\pdfoutput=1

\documentclass[aps,prstab,reprint,groupedaddress,showpacs,floatfix,a4paper]{revtex4-1}

\usepackage{graphicx}
  \setkeys{Gin}{width=\linewidth}
\usepackage{amsmath,amsfonts,amssymb}

\usepackage{color}

\usepackage{epstopdf}

\begin{document}


\title{Electronic and Optical Properties of Graphene Nanoribbons in External Fields} 


\author{Hsien-Ching Chung}
\email[]{E-mail: hsienching.chung@gmail.com}
\affiliation{Department of Physics, National Cheng Kung University, Tainan 70101, Taiwan}
\affiliation{Center for Micro/Nano Science and Technology (CMNST), National Cheng Kung University, Tainan 70101, Taiwan}

\author{Cheng-Peng Chang}
\affiliation{Center for General Education, Tainan University of Technology, Tainan 701, Taiwan}

\author{Chiun-Yan Lin}
\email[]{E-mail: l28981084@mail.ncku.edu.tw}
\affiliation{Department of Physics, National Cheng Kung University, Tainan 70101, Taiwan}


\author{Ming-Fa Lin}
\email[]{E-mail: mflin@mail.ncku.edu.tw}
\affiliation{Department of Physics, National Cheng Kung University, Tainan 70101, Taiwan}



\date{\today}

\begin{abstract}


A review work is done for electronic and optical properties of graphene nanoribbons in magnetic, electric, composite, and modulated fields.
Effects due to the lateral confinement, curvature, stacking, non-uniform subsystems and hybrid structures are taken into account.
The special electronic properties, induced by complex competitions between external fields and geometric structures, include many one-dimensional parabolic subbands, standing waves, peculiar edge-localized states, width- and field-dependent energy gaps, magnetic-quantized quasi-Landau levels, curvature-induced oscillating Landau subbands, crossings and anti-crossings of quasi-Landau levels, coexistence and combination of energy spectra in layered structures, and various peak structures in the density of states.
There exist diverse absorption spectra and different selection rules, covering edge-dependent selection rules, magneto-optical selection rule, splitting of the Landau absorption peaks, intragroup and intergroup Landau transitions, as well as coexistence of monolayer-like and bilayer-like Landau absorption spectra.
Detailed comparisons are made between the theoretical calculations and experimental measurements.
The predicted results, the parabolic subbands, edge-localized states, gap opening and modulation, and spatial distribution of Landau subbands, have been identified by various experimental measurements.

\end{abstract}


\maketitle
\tableofcontents

\newpage

\noindent\includegraphics[width=4cm, keepaspectratio]{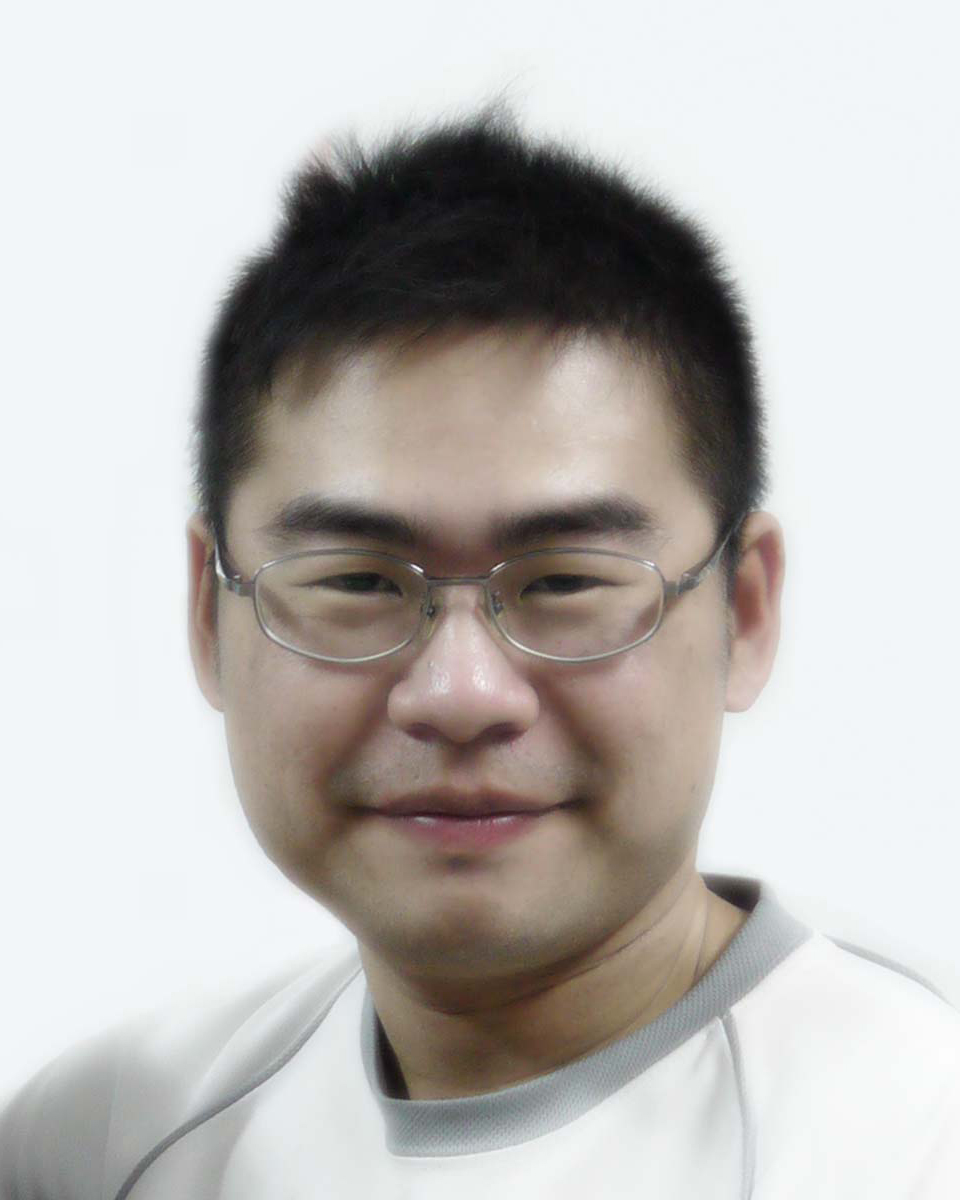}
\textbf{Hsien-Ching Chung}\\
Hsien-Ching Chung received his PhD in 2011 in physics from the National Cheng Kung University (Tainan, Taiwan). From 2011--2012, he was a postdoctoral fellow in the National Center for Theoretical Science (south), Tainan, Taiwan. Currently, H. C. Chung is a postdoctoral researcher in the Department of Physics as well as the Center for Micro/Nano Science and Technology in the National Cheng Kung University. His main scientific interests in condensed matter physics include electronic and optical properties of carbon-related materials and low-dimensional systems. H. C. Chung is a member of American Physical Society and American Chemical Society.

\noindent\includegraphics[width=4cm, keepaspectratio]{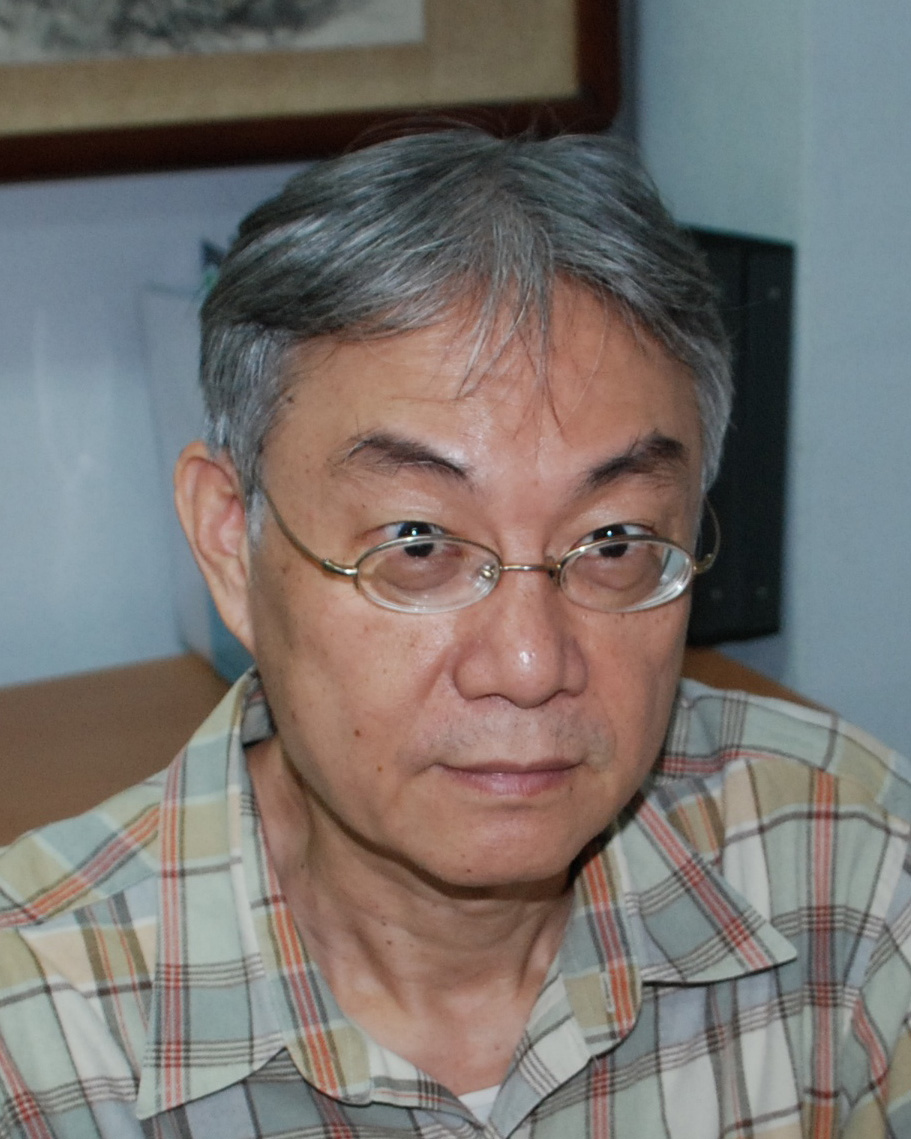}
\textbf{Cheng-Peng Chang}\\
Cheng-Peng Chang received his PhD in 1994 in solid state physics from the Department of Physics at the National Cheng Kung University (Tainan, Taiwan). He became professor, in 2005, at Tainan University of Technology. His main scientific interests focus on the theoretical studies of electronic structure, transport, and magneto-optical properties of solids (\emph{e.g.} quantum wells and graphene systems).

\noindent\includegraphics[width=4cm, keepaspectratio]{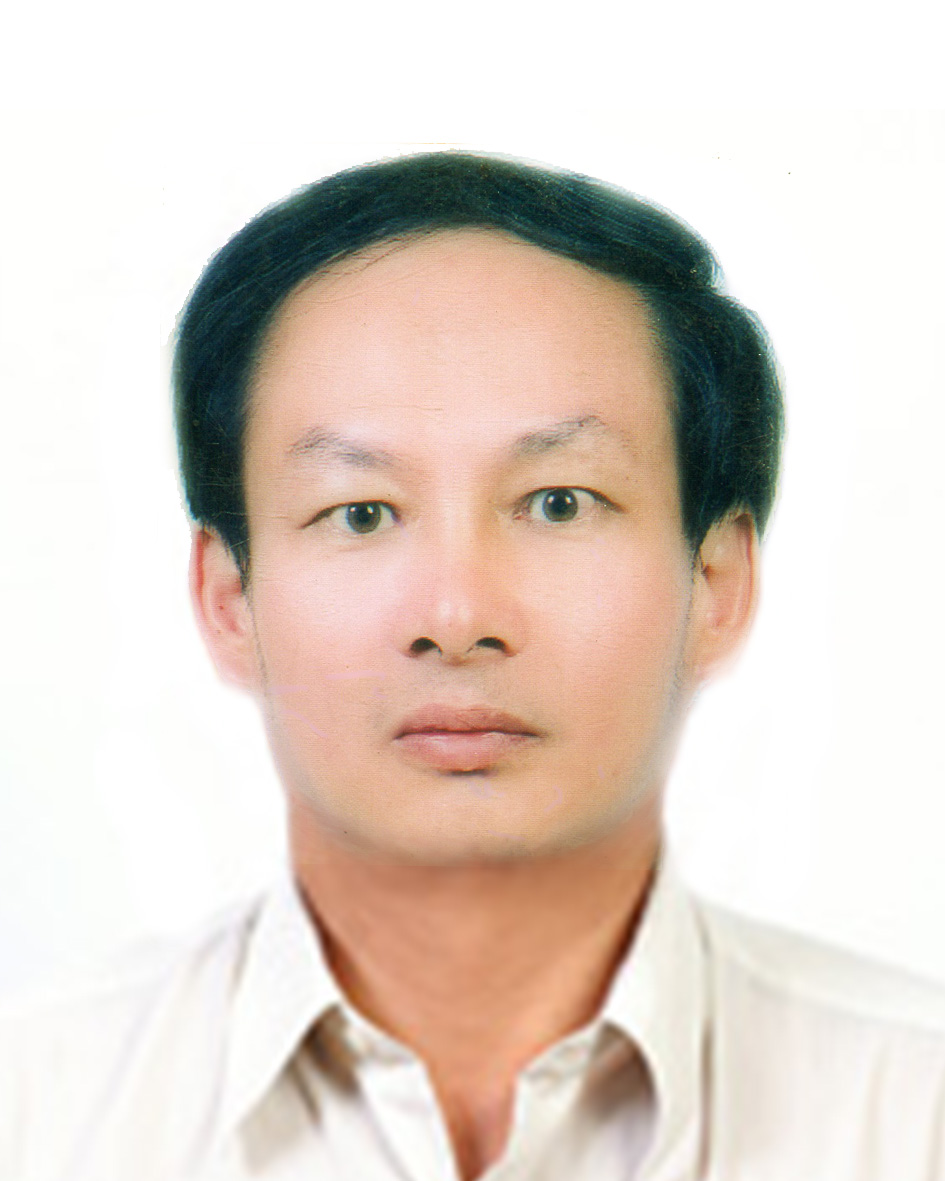}
\textbf{Ming-Fa Lin}\\
Ming-Fa Lin received his PhD in 1993 in physics from the National Tsing-Hua University (Hsinchu, Taiwan) where he stayed until 1995. After three years in the National Chiao Tung University, 1995--1997, he became professor in the National Cheng Kung University. Currently, M. F. Lin is distinguished professor in the Department of Physics in the National Cheng Kung University. His main scientific interests focus on essential properties of carbon-related materials and low-dimensional systems. M. F. Lin is a member of American Physical Society, American Chemical Society, and Physical Society of Republic of China (Taiwan).

\section{Introduction}
\label{sec:Introduction}

Carbon is one of the mainstream elements in the periodic table, and it can generate many substances very appropriate for researches in physics, chemistry, materials, biology and geology.
In condensed-matter systems, each carbon atom has four half-filled orbitals and is capable of forming crystals with stable covalent bonds, including diamond, graphite, graphene, carbon nanotube (CNT), C$_{60}$-related fullerene, and linear acetylenic carbon.
Diamond is a bulk material with strong $sp^3$ bondings and often used in the industry due to high hardness\cite{Phys.Rev.125(1962)828J.N.Plendl} and thermal conductivity.\cite{Phys.Rev.B42(1990)1104T.R.Anthony}
The $sp^2$-hybridized carbon allotropes possess open/closed surfaces and structures with various dimensionalities.
Graphite, which consists of stacked carbon sheets\cite{Proc.R.Soc.Lond.A106(1924)749J.D.Bernal} and can be easily found in nature, is applied as an electrical conductor,\cite{IntroductionToSyntheticElectricalConductorsJ.R.Ferraro} a moderator of nuclear reactor,\cite{Carbon20(1982)3B.T.Kelly} and a solid lubricant.\cite{BOOK_LubricantsAndLubrication}
Owing to the advance in nanotechnology, the low-dimensional carbon materials have been fabricated and observed.
Fullerene, a stable zero-dimensional carbon cluster with a closed structure, has been synthesized by Kroto \emph{et. al.} in 1985.\cite{Nature318(1985)162H.W.Kroto, Chem.Rev.91(1991)1213H.W.Kroto}
A CNT, a quasi-one-dimensional (quasi-1D) hollow cylinder with a nanoscale radius is discovered by Iijima~\emph{et. al.} in 1991.\cite{Nature354(1991)56S.Iijima, Nature363(1993)603S.Iijima, Nature363(1993)605D.S.Bethune}
In 2004, Geim~\emph{et al.} have demonstrated that graphene, a one-atom thick two-dimensional (2D) carbon sheet, can be isolated and transferred to another substrate.\cite{Science306(2004)666K.S.Novoselov}
Linear acetylenic carbon, found in 1995, is a 1D chain based on $sp$ bondings with alternating triple and single bonds.\cite{Science267(1995)362R.J.Lagow, Macromolecules28(1995)344J.Kastner}

A new scientific frontier has been explored since the discovery of graphene.
This nanomaterial attracts researchers, in both fundamental science and applied technology,\cite{Appl.Phys.Lett.99(2011)163102Y.H.Su, ACSSustainableChem.Eng.3(2015)1965Y.H.Su, ACSNano9(2015)8967H.C.Wu} for several reasons.
Graphene is the first truly 2D material,\cite{Science306(2004)666K.S.Novoselov, Proc.Natl.Acad.Sci.U.S.A.102(2005)10451K.S.Novoselov} which can serve as an ideal platform for realizing low-dimensional physics and applications.
It is a gapless semiconductor with cone-like energy spectrum near the Fermi energy.
The electronic structure of graphene gives rise to many incredible essential properties, such as high carrier mobility at room temperature ($>$ 200000 cm$^2$/Vs),\cite{Science312(2006)1191C.Berger, SolidStateCommun.146(2008)351K.I.Bolotin, Phys.Rev.Lett.100(2008)016602S.V.Morozov} superior thermoconductivity (3000--5000 W/mK),\cite{Phys.Rev.Lett.100(2008)016602S.V.Morozov, NanoLett.8(2008)902A.A.Balandin} extremely high modulus ($\sim$1 TPa) and tensile strength ($\sim$100 GPa),\cite{Science321(2008)385C.Lee} high transparency to incident light over a broad range of wavelength (97.7 $\%$),\cite{Science320(2008)1308R.R.Nair, Nat.Nanotechnol.5(2010)574S.Bae} and anomalous quantum Hall effect.\cite{Nature438(2005)197K.S.Novoselov, Nature438(2005)201Y.B.Zhang, Nat.Phys.2(2006)177K.S.Novoselov, Science315(2007)1379K.S.Novoselov}

Based on its superior electronic, thermal, mechanical, and optical properties,  graphene is considered to have high potential in future electronic and optical devices.
However, the gapless feature gives a low on/off ratio in graphene-based field effect transistors (FETs) and hinders the development of graphene nanoelectronics.
One of the most promising approaches to controlling electronic and optical properties is to make 1D strips of graphene, usually referred to as graphene nanoribbons (GNRs).
To achieve large-scale production of GNRs, numerous fabrication strategies, including both top-down and bottom-up schemes, have been proposed.
From the geometric point of view, cutting graphene seems to be the most simple and intuitive strategy to produce GNRs, and the available routes include lithographic patterning and etching of graphene,\cite{Phys.Rev.Lett.98(2007)206805M.Y.Han, PhysicaE40(2007)228Z.H.Chen, Nat.Nanotechnol.3(2008)397L.Tapaszto, NanoLett.9(2009)2083J.W.Bai} sonochemical breaking of graphene,\cite{Science319(2008)1229X.L.Li, Phys.Rev.Lett.100(2008)206803X.R.Wang, NanoRes.3(2010)16Z.S.Wu} metal-catalyzed cutting of graphene,\cite{NanoLett.8(2008)1912S.S.Datta, NanoRes.1(2008)116L.Ci, NanoLett.9(2009)457N.Severin, NanoRes.2(2009)695F.Schaffel, Phys.StatusSolidiB246(2009)2540F.Schaffel, NanoLett.9(2009)2600L.C.Campos, Adv.Mater.21(2009)4487L.Ci} and oxidation cutting of graphene.\cite{J.Am.Chem.Soc.132(2010)10034S.Fujii, Chem.Mater.19(2007)4396M.J.McAllister}
A much more imaginative strategy is unzipping of CNTs, since a nanotube can be viewed as a folded or zipped GNR.
The available routes for the reverse process cover chemical attack,\cite{Nature458(2009)872D.V.Kosynkin, Carbon48(2010)2596F.Cataldo}
laser irradiation,\cite{Nanoscale3(2011)2127P.Kumar}
plasma etching,\cite{Nature458(2009)877L.Jiao, NanoRes.3(2010)387L.Jiao}
metal-catalyzed cutting,\cite{NanoLett.10(2010)366A.LauraElias, Nanoscale3(2011)3876U.K.Parashar}
hydrogen treatment and annealing,\cite{ACSNano5(2011)5132A.V.Talyzin}
unzipping functionalized nanotubes by scanning tunneling microscope (STM) tips,\cite{NanoLett.10(2010)1764M.C.Pavia}
electrical unwrapping by transmission electron microscopy (TEM),\cite{ACSNano4(2010)1362K.Kim}
intercalation and exfoliation,\cite{NanoLett.9(2009)1527A.G.Cano-Marquez, ACSNano5(2011)968D.V.Kosynkin}
electrochemical unzipping,\cite{J.Am.Chem.Soc.133(2011)4168D.B.Shinde}
and sonochemical unzipping.\cite{Nat.Nanotechnol.5(2010)321L.Jiao, J.Am.Chem.Soc.133(2011)10394L.Xie}
Other strategies involve chemical vapor deposition (CVD)\cite{NanoLett.8(2008)2773J.Campos-Delgado, J.Am.Chem.Soc.131(2009)11147D.C.Wei, Nat.Nanotechnol.5(2010)727M.Sprinkle} and chemical synthesis.\cite{J.Am.Chem.Soc.130(2008)4216X.Y.Yang, Nature466(2010)470J.M.Cai, ACSNano3(2012)2020S.Blankenburg, Appl.Phys.Lett.105(2014)023101Y.Zhang}
The former is much compatible with semiconductor industry, and the latter is piecewise linking of molecular precursor monomers.
Although there are many routes for the synthesis of GNRs, the main obstacles and disadvantages precluding GNRs from real applications are to control the width and edge structure, substrate effect, edge termination, and defect.
Electronic properties, especially the energy gap, are sensitive to the ribbon width and edge structure.
They are also altered by charge transfer between substrate and GNRs.
During some synthesis processes, the ribbon edges might be partly terminated with hydrogen, oxygen, and other functional groups; therefore, the edge-related properties will be changed.
Non-hexagonal defects (often pentagonal and heptagonal structures) on the ribbon plane cause drastic changes in electronic properties.
Up to now, the researchers and engineers are continuously finding other routes of high yield to precisely control the nanoscale width and perfect edge structure and trying to overcome the technical problems for real applications.

The electronic and optical properties of GNRs are dominated by the ribbon width and edge structure.
Armchair and zigzag GNRs (AGNRs and ZGNRs) are two typical systems chosen for model studies.
The lateral quantum confinement is responsible for a lot of 1D parabolic subbands, where the electronic states are characterized by the regular standing waves.
Especially, the energy gaps induced in AGNRs scale inversely with the ribbon width.\cite{NanoLett.6(2006)2748V.Barone, Phys.Rev.Lett.97(2006)216803Y.W.Son}
Furthermore, ZGNRs present partial flat subbands near the Fermi energy with peculiar edge states localized at the ribbon boundaries.\cite{Phys.Rev.B54(1996)17954K.Nakada, J.Phys.Soc.Jpn.65(1996)1920M.Fujita}
The former have been observed by the experimental measurements on the electric conductance\cite{Phys.Rev.Lett.98(2007)206805M.Y.Han, Science319(2008)1229X.L.Li} and tunneling current,\cite{Nat.Nanotechnol.3(2008)397L.Tapaszto} and the latter are confirmed by the STM image.\cite{Phys.Rev.B71(2005)193406Y.Kobayashi, Phys.Rev.B73(2006)125415Y.Kobayashi}
As for the optical properties, the edge-dependent absorption selection rules are predicted to satisfy $|\Delta J| = odd$ for ZGNRs and $\Delta J = 0$ for AGNRs, where $J$ is the subband index\cite{J.Phys.Soc.Jpn.69(2000)3529M.F.Lin, Phys.Rev.B76(2007)045418H.Hsu, Opt.Express19(2011)23350H.C.Chung, Phys.Rev.B84(2011)085458K.Sasaki}  (Section~\ref{sec:ElectronicStructureOfGrapheneInNoBandE}).
The essential properties will be enriched by the magnetic and electric fields, curvature, stacking configurations, non-uniform subsystems, and hybrid structures, as discussed later.

A uniform perpendicular magnetic field can flock the neighboring electronic states and thus induce highly degenerate Landau levels (LLs) with quantized cyclotron orbitals.\cite{Z.Physik64(1930)629L.Landau}
The magnetic quantization will be seriously suppressed by the lateral confinement.
Their competition diversifies the magneto-electronic structures, covering partly dispersionless quasi-Landau levels (QLLs), 1D parabolic subbands, and partial flat subbands.\cite{Phys.Rev.B59(1999)8271K.Wakabayashi, Phys.Rev.B73(2006)195408L.Brey, J.Appl.Phys.103(2008)073709Y.C.Huang}
Such energy dispersions induce two kinds of peaks, symmetric and asymmetric ones, in the density of states (DOS).\cite{Nanotechnol.18(2007)495401Y.C.Huang, PhysicaE42(2010)711H.C.Chung}.
The QLLs can be formed for sufficiently wide GNRs, and each Landau wave function possesses a localized symmetric/antisymmetric distribution with a specific number of zero points (a quantum number $n$).
The magneto-optical spectra exhibit lots of symmetric and asymmetric absorption peaks.
The former ones come from the inter-QLL transitions and obey the magneto-optical selection rule of $| \Delta n | = 1$.
However, the latter ones originate from the transitions among parabolic subbands and abide by the edge-dependent selection rules.

A transverse electric field can generate an extra potential energy in GNRs.
The charge carriers experience different site energies so that the electronic and optical properties are dramatically changed.
There exist oscillatory energy subbands, more extra band-edge states, gap modulations with semiconductor-metal transitions, and irregular standing waves.\cite{Carbon44(2006)508C.P.Chang}
Meanwhile, the intensity, number, and frequency of asymmetric DOS peaks are obviously modified.
The absorption peaks associated with the edge-dependent selection rules are inhibited, and more extra peaks appear in the optical spectra (Section~\ref{EffectsOfElectricFieldsOfMonolayerABS}).
The different potential energies in GNRs hinder the formation of Landau orbitals, and the QLLs would tilt, become oscillatory, or exhibit crossings and anti-crossings.
Moreover, the inter-QLL optical transitions would be severely altered or even destroyed thoroughly.

Geometric configurations play an important role in modifying the fundamental properties.
The curved GNRs or the unzipped CNTs\cite{Nature458(2009)872D.V.Kosynkin} are the suitable systems for studying the magnetic quantization in curved surfaces.
The electrons experience a non-uniform effective magnetic field instead of the applied perpendicular uniform one, and the effects of magnetic quantization will gradually reduce from the ribbon center to either edge.
With the increase of curvature, the Landau subbands become oscillatory with more band-edge states, and the spatial distributions of Landau wave functions are distorted.
A detailed comparison between the curved GNRs and CNTs show the influences of curvature and the interplay between different boundary conditions.
The QLLs hardly survive in a cylindrical CNT with a periodic boundary condition except for a very large tube diameter and magnetic field, mainly owing to the vanishing of net magnetic flux.
The magneto-absorption peaks are dominated by the angular-momentum-dependent selection rules.\cite{J.Phys.Soc.Jpn.62(1993)1255H.Ajiki, J.Phys.Soc.Jpn.66(1997)3294M.F.Lin, Phys.Rev.B62(2000)16092S.Roche, Phys.Rev.B67(2003)045405F.L.Shyu}
On the other hand, the QLL-dominated and the curvature-induced extra absorption peaks coexist in the curved GNRs.

The electronic and optical properties are very sensitive to the changes in the number of layers and stacking configurations.
Few-layer and bilayer GNRs can be fabricated by cutting graphene\cite{NanoLett.9(2009)2083J.W.Bai, Science319(2008)1229X.L.Li} and unzipping multi-wall CNTs.\cite{Nature458(2009)872D.V.Kosynkin}
There are two typical stackings, namely, AB and AA stackings.
The bilayer GNRs possess two groups of parabolic subbands (or two groups of QLLs), and each group contains conduction and valence subbands.
The initial energies of subband groups and the wave functions are closely related to the stacking configurations.
For the AA stacking with higher symmetry, the interlayer atomic interactions induce the strong overlap between the valence subbands of the first group and the conduction subbands of the second group.
The metallic property keeps the same even in the presence of perpendicular and transverse electric fields.
On the other hand, the AB stacking has only a weak overlap between the conduction and valence subbands of the first group, where the metal-semiconductor transition will happen by the electric fields.
As to the optical properties, only the intragroup transitions are presented in the AA stacking, while in the AB stacking, the intergroup transitions are also available.
This important difference originates from that the wave functions of the former are the symmetric or antisymmetric superpositions of the tight-binding functions on different layers.

In the non-uniform bilayer GNRs and GNR-CNT hybrids, the complex interlayer atomic interactions between the constituent subsystems enrich the essential properties.
The former consisting of two GNRs with different widths are successfully fabricated by mechanical exfoliation from bulk graphite.\cite{Phys.Rev.B79(2009)235415C.P.Puls, NanoLett.10(2010)562X.Xu, Phys.Rev.B88(2013)125410J.Tian}
The relative positions of two subsystems can determine whether the QLLs will survive.
The modulation of their widths will generate various kinds of magneto-electronic spectra, including monolayer-like, bilayer-like, and coexistent QLL spectra, as well as distorted energy dispersions.
The feature-rich electronic properties are directly reflected on the magneto-absorption spectra.
Moreover, the couplings in GNR-CNT hybrids distort the QLL spectra with more band-edge states.\cite{Nanotechnology19(2008)105703T.S.Li}
The QLL absorption peaks and the additional nearby subpeaks are suppressed and introduced, respectively.

\begin{figure}
\begin{center}
  \includegraphics[width=\linewidth]{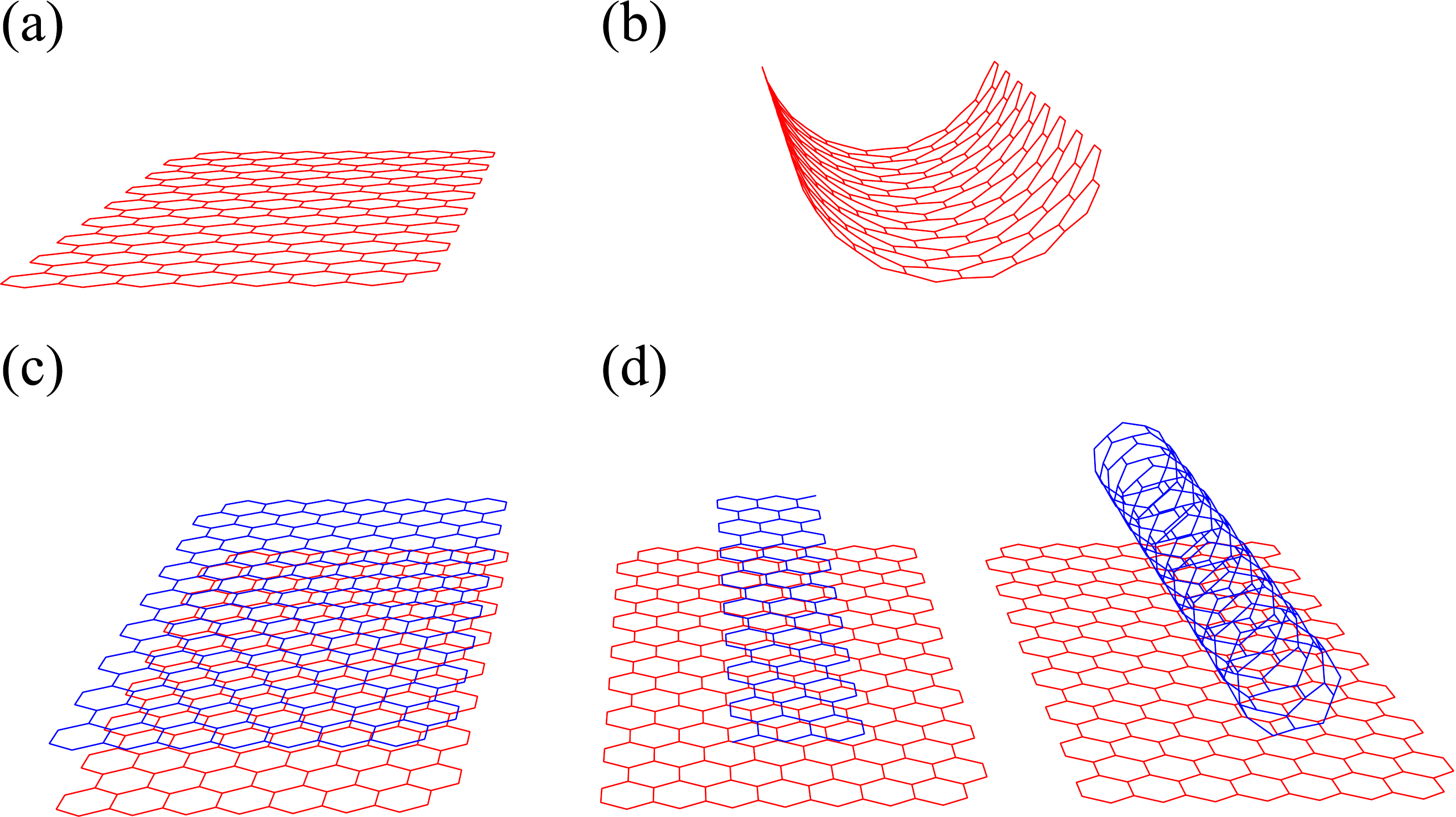}\\
  \caption[]{
  GNR-related systems: (a) monolayer GNRs, (b) curved GNRs, (c) bilayer GNRs, and (d) non-uniform GNRs and GNR-CNT hybrids.
  }
  \label{fig:Mono_Bi_Curve_Hybrid}
\end{center}
\end{figure}

This review is focused on GNR-related systems: (1) monolayer GNRs, (2) curved GNRs, (3) bilayer GNRs, and (4) non-uniform GNRs and GNR-CNT hybrids, as illustrated in Fig.~\ref{fig:Mono_Bi_Curve_Hybrid}.
Many effects are taken into account, including edge structures, ribbon widths, magnetic quantization, electric potential energies, curvatures and boundary conditions, stacking configurations, and complex couplings between subsystems.
A systematic understanding is given on the electronic properties and optical spectra.
Detailed comparisons are made between theoretical predictions and experimental observations, showing that most results are in agreement with each other, and some others require further experimental verifications.

\section{Monolayer systems}
\label{sec:MonolayerSystem}

A 2D graphene is considered to be a promising candidate for future nanoelectronics due to its unique electronic properties.\cite{Nat.Photonics4(2010)611F.Bonaccorso, Nat.Photonics6(2012)749A.N.Grigorenko}
Unfortunately, the graphene FETs cannot be turned off effectively under a gapless condition, in which an on/off current ratio typically around 5 in top-gated graphene FETs.\cite{NanoLett.10(2010)715F.Xia}
The gapless limitation can be overcome through the successful fabrications of GNRs, including the graphene cutting, CNT unzipping, CVD, and linkage of molecular precursor monomers (Section~\ref{sec:Introduction}).
Electrical transport experiments show that the lateral-confinement-induced gap renders the FET with a high on/off ratios of about $10^7$ at room temperature.\cite{Science319(2008)1229X.L.Li, Phys.Rev.Lett.100(2008)206803X.R.Wang}
In monolayer GNRs, the energy gap can be controlled by the ribbon width, electric and magnetic fields.
The predicted energy gap is inversely proportional to the ribbon width,\cite{NanoLett.6(2006)2748V.Barone, Phys.Rev.Lett.97(2006)216803Y.W.Son, Appl.Phys.Lett.88(2006)142102B.Obradovic} and this fact has been confirmed by the conductance\cite{Phys.Rev.Lett.98(2007)206805M.Y.Han, Science319(2008)1229X.L.Li} and tunneling current measurements.\cite{Nat.Nanotechnol.3(2008)397L.Tapaszto}
A metal-semiconductor transition will take place depending on the strength of transverse electric fields.
As magnetic fields are exerted, gap shrinkage happens and the GNR eventually undergoes a semiconductor-metal transition during the formation of Landau subbands.
The gap modulation opens the possibility for the design of side-gated GNR-field-effect devices,\cite{Phys.Rev.B76(2007)245426F.Molitor, Small4(2008)716J.F.Dayen, Appl.Phys.Lett.101(2012)093504B.Hahnlein, IEEEElectronDeviceLett.33(2012)330C.T.Chen, IEEETrans.ElectronDevices61(2014)3329L.T.Tung, Sci.Rep.4(2014)5581V.Panchal, IEEEJ.Electr.Dev.Soc.3(2015)144L.T.Tung} where the side gate offering a better alternative to top-gating scheme will prevent dielectric breakdown\cite{J.Electrochem.Soc.119(1972)591C.M.Osburn, J.Electrochem.Soc.119(1972)603C.M.Osburn, J.Appl.Phys.62(1987)2360S.Suyama, Phys.StatusSolidiA207(2010)286X.Li, NanoLett.12(2012)1165B.Standley, Sci.Rep.4(2014)5893K.Ho} and electrical hysteresis caused by top-gate dielectrics.\cite{Appl.Phys.Lett.88(2006)113507M.H.Yang, NanoLett.8(2008)3092K.Kim, NanoLett.9(2009)388D.B.Farmer, NanoLett.9(2009)1973T.Lohmann, Appl.Phys.Lett.95(2009)202101S.Unarunotai, NanoLett.9(2009)1472Y.Dan, NanoLett.10(2010)1149M.Lafkioti, Appl.Phys.Lett.95(2009)242104S.S.Sabri, Appl.Phys.Lett.96(2010)082114V.Geringer, J.Phys.-Condens.Matter22(2010)334214P.Joshi, J.Chem.Phys.133(2010)044703Z.M.Liao, ACSNano4(2010)7221H.Wang}
The geometric configurations and external fields not only modulate the gap but also alter the energy spacings among subbands.
These changes are reflected on the variations of DOS, and can be confirmed by means of measurements on tunneling current.\cite{Phys.Rev.B31(1985)805J.Tersoff, Phys.Rev.Lett.82(1999)1225P.Kim, Nat.Phys.3(2007)623G.Li, Phys.Rev.Lett.102(2009)176804G.Li, Nat.Commun.4(2013)1744G.Li}

Wave functions, which present the spatial information of electronic states, are very significant for realizing fundamental physical properties, such as charge densities,\cite{Phys.Rev.28(1926)1049E.Schrodinger, Phys.Rev.B78(2008)235311M.W.Y.Tu, Phys.Rev.B89(2014)121401P.M.PerezPiskunow, Phys.Rev.B90(2014)115423G.Usaj} state mixing,\cite{J.Phys.Soc.Jpn.80(2011)044602H.C.Chung} and optical selection rules.\cite{NanoLett.6(2006)2748V.Barone, J.Appl.Phys.103(2008)073709Y.C.Huang}
At zero field, the electronic states confined by the transverse boundaries display  regular standing waves throughout the whole nanoribbon.
The spatial distributions and nodal structures are dependent on the wave vectors, state energies, sublattices, ribbon widths, and edge structures.
Magnetic fields can quantize the neighboring electronic states with the formation of Landau orbitals and regular nodal structures.
The Landau states are strongly localized, and they will be distorted and truncated as the wave functions encounter either boundary of the ribbon.
On the other hand, electric fields can alter the spatial distributions, destroy the nodal structures, and cause the state mixings.
These aforementioned features can be verified by the spectroscopic-imaging STM, which is a potent tool for direct mapping of the wave functions and has been implemented to study the surface states in various systems.\cite{Phys.Rev.Lett.71(1993)1071Y.Hasegawa, Phys.Rev.Lett.73(1994)910G.Hormandinger, Nature363(1993)524M.F.Crommie, Phys.Rev.B60(1999)7792V.Meunier, Phys.Rev.Lett.82(1999)3520A.Rubio, Appl.Phys.A68(1999)275A.Rubio, Phys.Rev.B65(2002)245418J.Jiang, NanoLett.6(2006)1439A.vanHouselt, J.Phys.-Condens.Matter25(2013)014014R.Heimbuch}

The special relations between conduction and valence wave functions, which strongly depend on the geometric configurations and external fields, can enrich the absorption spectra.
At zero field, the absorption peaks obey the edge-dependent selection rules, where the number, structure, energy, and intensity are determined by the edge structures and ribbon widths.
The low-lying edge-dependent absorption peaks are changed into the QLL-dependent ones in the presence of magnetic fields, while the coexistence of two kinds of peaks at higher energy reflects the competition between the magnetic quantization and lateral confinement.
On the other hand, a weak electric field induces the coupling of the electronic states in adjacent subbands and modifies the selection rules.
In other words, the extra selection rules bring more absorption peaks.
However, under a large electric field, many low-intensity peaks are revealed due to the breaking of selection rules.
These feature-rich optical spectra and selection rules can be verified through infrared transmission experiments.\cite{Phys.Rev.Lett.97(2006)266405M.L.Sadowski, Phys.Rev.Lett.98(2007)197403Z.Jiang, Phys.Rev.Lett.100(2008)087401P.Plochocka, Phys.Rev.Lett.110(2013)246803J.M.Poumirol}

The organization of this section is stated as follows.
The first subsection describes the generalized tight-binding model, where the magnetic and electric fields are considered.
In the next subsection, the discussions are focused on the electronic properties,  including the energy dispersions, DOS, and wave functions, under the influence of external fields.
The lateral confinement, magnetic quantization, electric-field-induced collapse of QLLs, and gap modulation are further discussed.
The last subsection presents the feature-rich optical spectra with the edge-dependent, QLL-dependent, and electric-field-modified selection rules.
Also included are the detailed comparisons between the theoretical results and experimental observations.

\subsection{The tight-binding model}
\label{Sec:TightBindingModel}

\begin{figure}
\begin{center}
  \includegraphics[width=\linewidth]{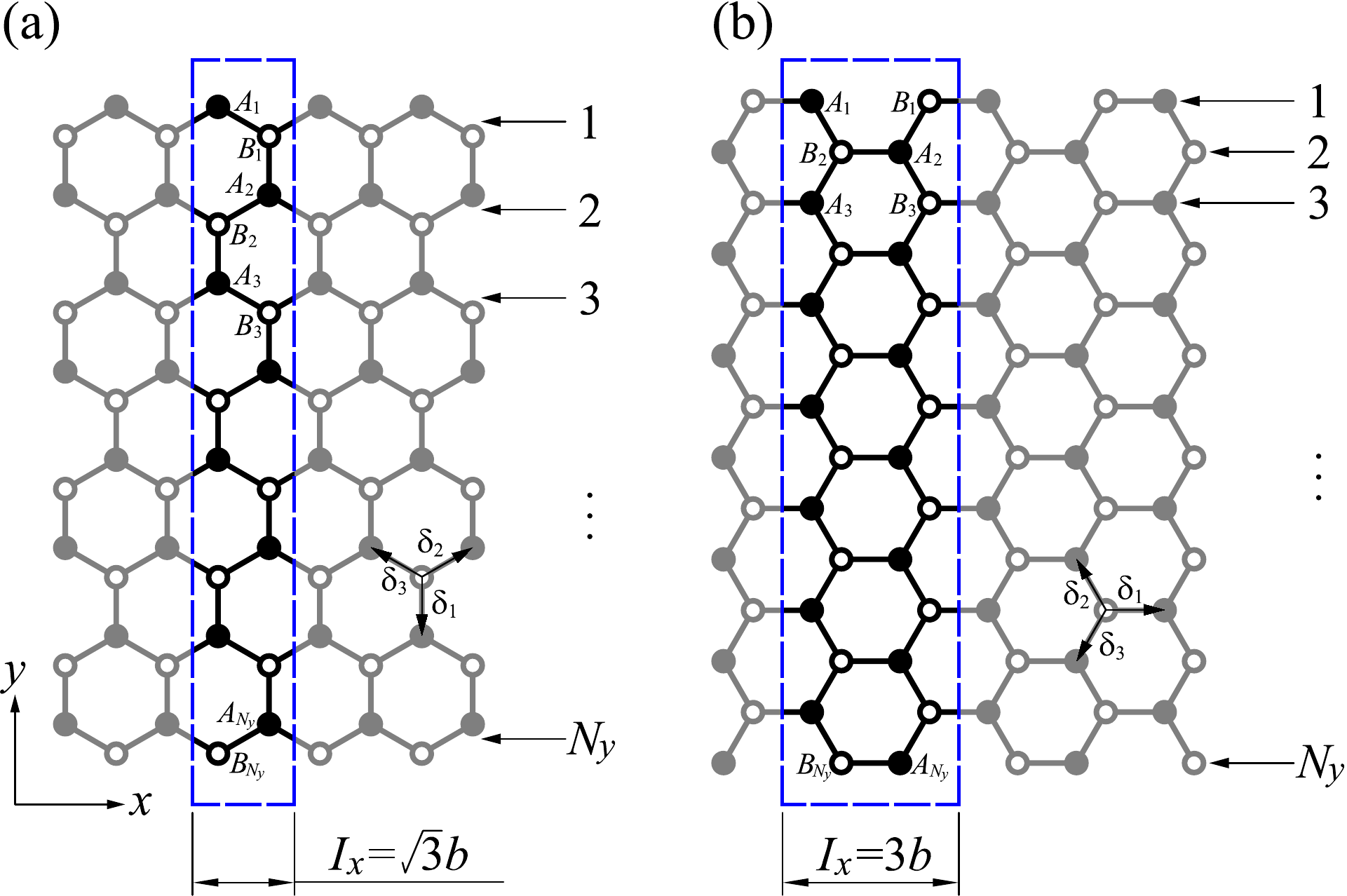}\\
  \caption[]{
  Geometric structures of the monolayer (a) ZGNRs and (b) AGNRs.
  }
  \label{fig:PrimitiveUnitCellofMonoArmZigGNRs}
\end{center}
\end{figure}

GNRs are quasi-1D strips made up of hexagonally structured carbon atoms.
The ZGNRs and AGNRs with the zigzag and armchair edges are shown in Fig.~\ref{fig:PrimitiveUnitCellofMonoArmZigGNRs}(a) and (b), respectively.
The primitive unit cells enclosed by the dashed rectangles have the periodic lengths: $I_x = \sqrt{3}b$ for ZGNRs and $I_x = 3b$ for AGNRs, where $b=1.42$ \AA~is the C--C bond length.
The first Brillouin zone is within the region $-\pi /I_x < k_x \leq \pi /I_x$.
The carbon atoms on the \emph{m}th zigzag line or dimer line (armchair line) are denoted by $A_m$'s or $B_m$'s.
$2N_y$ carbon atoms are included in a unit cell, where $N_y$ is the number of zigzag or dimer lines.
The ribbon widths of ZGNR and AGNR are characterized by $W_{zig} = b(3N_y/2 - 1)$ and $W_{arm} = \sqrt{3}b(N_y-1)/2$, respectively.
${\delta}_i$'s ($i=1,2,3$) are the nearest-neighbor vectors.

The low-energy electronic properties are dominated by the tight-binding functions of $2p_z$ orbitals.
Based on the $2N_y$ tight-binding functions, the Hamiltonian is presented by
\begin{equation}
\mathcal{H} = -t \sum_{m,n} \sum_{|{\delta}_i|}
|\mathbf{R}^A_m\rangle \langle \mathbf{R}^B_n| + \mathrm{H.c.},
\label{eq:TightBindingHamiltonianOfGNRs}
\end{equation}
with the Bloch wave function
\begin{equation}
| \Psi^{c,v} \rangle = \sum_m^{N_y}
A_m |\mathbf{R}^A_m\rangle + B_m |\mathbf{R}^B_m\rangle ,
\label{eq:WFOfGNRs}
\end{equation}
where $|\mathbf{R}^A_m\rangle$ ($|\mathbf{R}^B_n\rangle$) represents the tight-binding functions due to the $2p_z$ orbitals on the $m$th ($n$th) site of the sublattice $A$ ($B$) with the position vector $\mathbf{R}^A_m$ ($\mathbf{R}^B_n$).
Only the interactions between the nearest neighbors indicated by $|{\delta}_i| = |\mathbf{R}^A_m-\mathbf{R}^B_n| = b$ are considered, and $t \sim 2.6$ eV is the hopping energy.\cite{Phys.Rev.B43(1991)4579J.C.Charlier}
$A_m$ and $B_m$ are the site amplitudes of sublattices $A$ and $B$, respectively.
The diagonalization of the Hamiltonian matrix $\mathcal{H}$ yields the energy spectrum $E^{c,v}(k_x)$ and wave functions $| \Psi^{c,v} \rangle$, where the superscripts $c$ and $v$ denote the conduction and valence subbands, respectively.

When a uniform static magnetic field, $\mathbf{B}=B \hat{z}$, is applied perpendicularly to the ribbon plane, each tight-binding wave function has an extra phase term owing to the vector potential $\mathbf{A}$.\cite{Z.Phys.80(1933)763R.Peierls, J.Phys.Radium8(1937)397F.London, Phys.Rev.84(1951)814J.M.Luttinger}
That is to say, each Hamiltonian matrix element is given by the product of the zero-field element and a phase factor $\exp (i2\pi \theta _{mn})$.
The Peierls phase is a line integral of the vector potential $\mathbf{A}$ from the $m$th site to the $n$th site as defined by
\begin{equation}
\theta_{mn} = \frac{1}{\phi_{0}} \int_{\mathbf{R}_m}^{\mathbf{R}_n}\mathbf{A}\cdot d\mathbf{l} ,
\label{eq:PeierlsPhase}
\end{equation}
where $\mathbf{R}_m$ ($\mathbf{R}_n$) is the position vector of the $m$th ($n$th) site, $\mathbf{A}=(-B y,0,0)$ is chosen to preserve the translation invariance along the $x$-direction under the Landau gauge, and $\phi_0=h/e$ is the magnetic flux quantum.

In the presence of an external electric field $\mathbf{E}$, the field can be treated as perturbation (for $|\mathbf{E}| \ll t/b$) and the Hamiltonian becomes
\begin{equation}
\mathcal{H} = \mathcal{H}_0 + \mathcal{H}_1 ,
\end{equation}
where $\mathcal{H}_0$ is the unperturbed Hamiltonian (eqn~(\ref{eq:TightBindingHamiltonianOfGNRs})), and $\mathcal{H}_1 = -e V(\mathbf{r})$.
Here $e$ is the charge of an electron, and $V(\mathbf{r})$ is the electrostatic potential of the external electric field.
For a uniform field, the potential at the position $\mathbf{R}_m$ is
\begin{equation}
V(\mathbf{r}) = -\mathbf{E} \cdot \mathbf{R}_m .
\end{equation}
In other words, the electric field modifies the on-site energies of carbon atoms and thus alters the essential properties.

\subsection{Electronic properties}
\label{sec:ElectronicProperties_MonolayerSystem}

\subsubsection{Low-energy electronic structure.}
\label{sec:ElectronicStructureOfGrapheneInNoBandE}

\begin{figure}
\begin{center}
  \includegraphics[width=\linewidth, keepaspectratio]{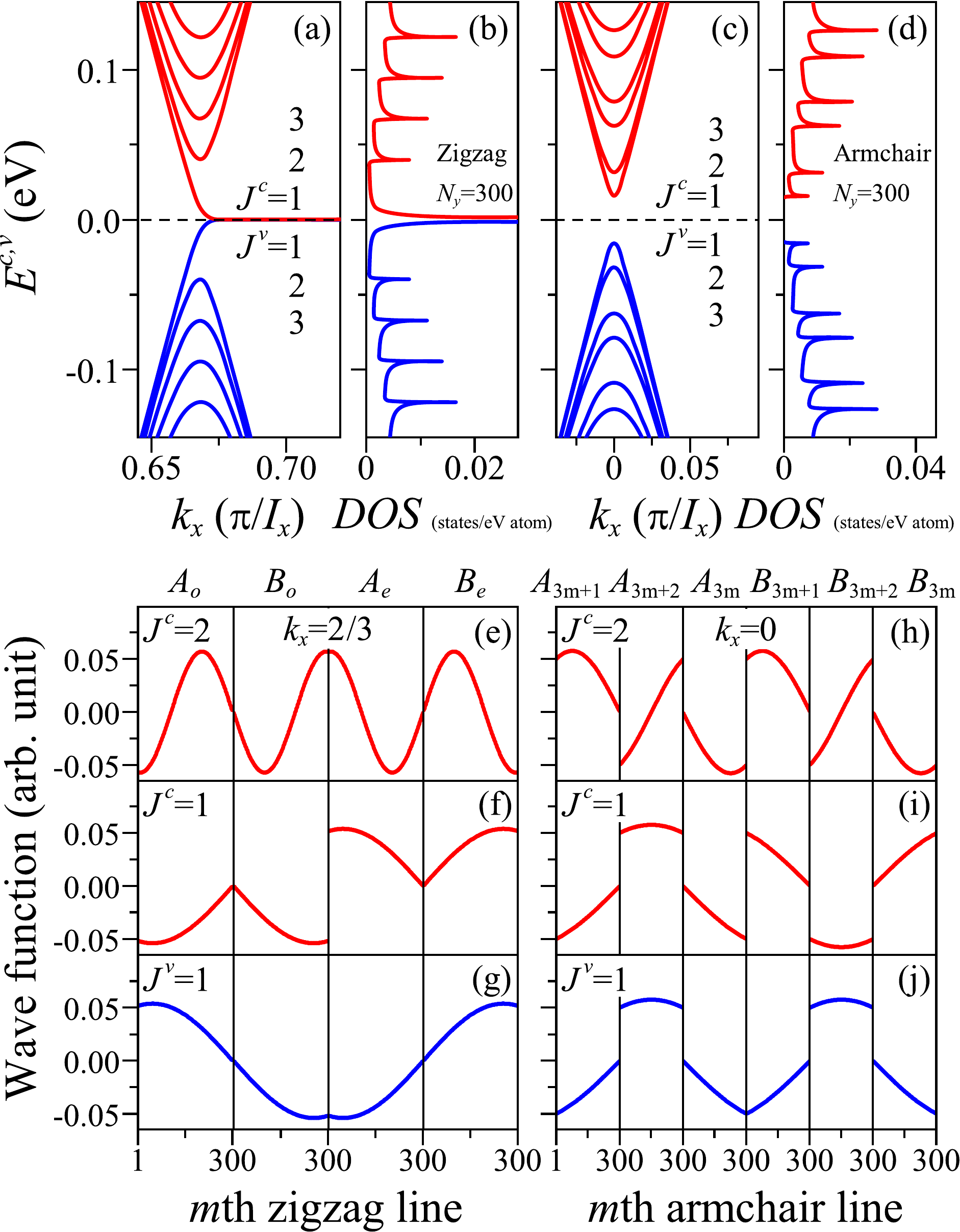}\\
  \caption[]{
  Low-energy band structures, DOS, and wave functions of the $N_y=300$ ZGNR and AGNR.
  }
  \label{fig:BS_DOS_WFofMonoGNRsNy300B0E0}
\end{center}
\end{figure}

The electronic properties are quite different between ZGNRs and AGNRs.
The low-energy band structure of the $N_y=300$ ZGNR in the absence of external fields is ploted in Fig.~\ref{fig:BS_DOS_WFofMonoGNRsNy300B0E0}(a).
All the energy subbands are 1D parabolic energy dispersions except for the two partial flat subbands lying on the Fermi level ($E_F = 0$), where $J^{c,v} = 1$, $2$, $3$,... are the subband indices accounting from $E_F=0$.
The parabolic subbands embody the lateral confinement, and their band-edge states are at $k_{x} \sim 2/3$ (in unit of $\pi$ over lattice constant).
The energy spacings are almost uniform and will shrink as the ribbon width becomes larger.
On the other hand, the states on the two partial flat subbands are strongly localized on the ribbon edges, and the charge density decays exponentially as a function of the distance from the edge.\cite{J.Phys.Soc.Jpn.65(1996)1920M.Fujita, Phys.Rev.B54(1996)17954K.Nakada, Phys.Rev.B59(1999)8271K.Wakabayashi}
For AGNRs, there are many 1D parabolic subbands, whose band-edge states are at $k_x=0$ as illustrated in Fig.~\ref{fig:BS_DOS_WFofMonoGNRsNy300B0E0}(c).
The spacing between band-edge states are non-uniform.
More importantly, there is a direct energy gap, which is inversely proportional to the ribbon width and approaches to zero for a very large $N_y$.\cite{Phys.Rev.B59(1999)8271K.Wakabayashi, Phys.Rev.B59(1999)9858Y.Miyamoto, Phys.Rev.B73(2006)045432M.Ezawa, Phys.Rev.B73(2006)235411L.Brey, Appl.Phys.Lett.89(2006)203107Y.Ouyang, Phys.Rev.Lett.97(2006)216803Y.W.Son, NanoLett.6(2006)2748V.Barone, NanoLett.7(2007)204D.A.Areshkin, Phys.Rev.B77(2008)245434H.Raza}
The size-dependent band gap originates from the lateral confinement and can also be found in semiconductor quantum structures.\cite{J.Appl.Phys.97(2005)073706G.Pellegrini, Phys.Rev.B72(2005)125325J.B.Li, Mater.Lett.60(2006)2526M.Li}

Part of the theoretical results have been verified by experimental measurements.
The angle-resolved photoemission spectroscopy (ARPES) can be utilized to investigate the actual distribution of electrons in the energy-momentum space of  low-dimensional systems.\cite{Phys.Rev.Lett.79(1997)467V.N.Strocov, Phys.Rev.Lett.81(1998)4943V.N.Strocov, Rev.Mod.Phys.75(2003)473A.Damascelli, Phys.Scr.T109(2004)61A.Damascelli, NewJ.Phys.7(2005)97F.Reinert, LowTemp.Phys.40(2014)286A.A.Kordyuk}
From the measurements on the unidirectionally aligned well-defined AGNRs, the 1D parabolic energy dispersions have been directly observed.\cite{ACSNano6(2012)6930P.Ruffieux}
The edge-localized states, which are also predicted by other theoretical methods,\cite{Phys.Rev.B59(1999)9858Y.Miyamoto, Phys.Rev.B73(2006)235411L.Brey} have been identified by STM\cite{Appl.Phys.A66(1998)S129Z.Klusek, Appl.Surf.Sci.161(2000)508Z.Klusek, Phys.Rev.B71(2005)193406Y.Kobayashi, Phys.Rev.B73(2006)125415Y.Kobayashi, Appl.Surf.Sci.241(2005)43Y.Niimi, Phys.Rev.B73(2006)085421Y.Niimi, Nat.Commun.5(2014)4311Y.Y.Li} and high-resolution ARPES.\cite{Phys.Rev.B73(2006)045124K.Sugawara}
It should be noted that these states are presented in ZGNRs, but absent in AGNRs.
Some theoretical works\cite{J.Am.Chem.Soc.109(1987)3721S.E.Stein, Synth.Met.17(1987)143K.Tanaka, FullereneSci.Technol.4(1996)565M.Fujita} on the finite-size graphite systems, either as macro molecules or as 1D systems, have also displayed such edge-dependent localizations.
In addition, recent STM measurements have revealed the edge-localized states in GNRs with chiral edges,\cite{Nat.Phys.7(2011)616C.Tao, NanoLett.12(2012)1928M.Pan} and demonstrated that these states can survive on the boundary with partial zigzag edges.
As to the lateral-confinement-induced energy gap, the inverse relation to the ribbon width is confirmed by the STM\cite{Nat.Nanotechnol.3(2008)397L.Tapaszto} and conductance experiments.\cite{Phys.Rev.Lett.98(2007)206805M.Y.Han, Science319(2008)1229X.L.Li}
Although in earlier studies, the edge roughness limits the accuracy of measurements and causes difficulties in obtaining a GNR with a specific gap.
Recent synthesis techniques have shown that the roughness of the ribbon edges can be controlled within the atomic accuracy,\cite{J.Am.Chem.Soc.130(2008)4216X.Y.Yang, Nature466(2010)470J.M.Cai, Sci.Rep.2(2012)983H.Huang, ACSNano7(2013)6123Y.C.Chen, Appl.Phys.Lett.105(2014)023101Y.Zhang} as an indication that the width-dependent energy gap could be well-controlled for applications.

The DOS, defined as $\sum_{\sigma, k_x, J^{c,v}} \delta [\omega - E^{c,v}(k_x, J^{c,v})]$ ($\sigma$ indicates the spin), directly reflects the main features of energy subbands.\cite{Sci.Rep.5(2015)9423P.Y.Lo}
Many states accumulate at the local maxima and minima of 1D energy subbands, and hence, a lot of peaks are presented in the DOS.
Such special singular structures are called van Hove singularities (vHSs),\cite{Phys.Rev.89(1953)1189L.VanHove} which play significant roles in optical transitions.
For both types of GNRs, there are many asymmetric peaks as shown in Fig.~\ref{fig:BS_DOS_WFofMonoGNRsNy300B0E0}(b) and (d).
These peaks are divergent in the square-root form $1/\sqrt{|\omega - E^{c,v}_{be}|}$, where the band-edge state energy $E^{c,v}_{be}$ correspond to the peak energy.
The peak height, which is inversely proportional to the square-root of subband curvature, grows as the state energy increases.
It is remarkable that the peak spacings are uniform in ZGNRs, while non-uniform in AGNRs.
Besides the asymmetric peaks, the symmetric delta-function-like peak at the Fermi level, associated with the band-edge states of partial flat subbands, is the highest one in the DOS of ZGNRs.
The symmetric peak near $E_F = 0$ in ZGNRs has been confirmed by scanning tunneling spectroscopy (STS),\cite{Phys.Rev.B71(2005)193406Y.Kobayashi, Appl.Surf.Sci.241(2005)43Y.Niimi, Phys.Rev.B73(2006)085421Y.Niimi, Nat.Commun.5(2014)4311Y.Y.Li} whereas the edge-dependent features of the asymmetric peaks, including the number, frequency, and height, still need further investigations.

The electrons confined in the finite-size GNRs exhibit regular standing waves.
Their oscillatory patterns, which can be characterized by the number of nodes (zero points), are closely related to the edge structures, sublattices, energies, and wavevectors.
Especially, in ZGNRs, the wave functions can be decomposed into the subenvelope functions on the $A$ and $B$ sublattices at the odd and even zigzag lines (Fig.~\ref{fig:BS_DOS_WFofMonoGNRsNy300B0E0}(e)--(g));\cite{J.Appl.Phys.103(2008)073709Y.C.Huang, Phys.Rev.B78(2008)115422Y.C.Huang} while in AGNRs, the subenvelope functions are on the $A$ and $B$ sublattices at the ($3m$)th, ($3m+1$)th, and ($3m+2$)th dimer lines ($m$ is an integer; Fig.~\ref{fig:BS_DOS_WFofMonoGNRsNy300B0E0}(h)--(j)).\cite{Phys.Rev.B.75(2007)165414H.Zheng, J.Appl.Phys.107(2010)083712S.C.Chen}
The wave functions will oscillate with more nodes in the increase of $|E^{c,v}|$; furthermore, there exist only a phase-shift difference for various wavevectors.
Here, the wave functions at $k_x = 2/3$ for ZGNRs and $k_x = 0$ for AGNRs are presented, in which the band-edge states possess high DOS and optical response.
It is significant that the relations in the wave functions are also edge-dependent and will directly determine the available optical transitions.
For ZGNRs, the spatial distributions of wave functions alternate between symmetric and anti-symmetric patterns as the subband index increases or decreases by one.\cite{Phys.Rev.B76(2007)045418H.Hsu, Opt.Express19(2011)23350H.C.Chung}
For example, the wave functions for the $J^v = 1$ and $J^c = 1$ \& $2$ subbands are symmetric, anti-symmetric, and symmetric, respectively (Fig.~\ref{fig:BS_DOS_WFofMonoGNRsNy300B0E0}(e)--(g)).
On the other hand, there are two kinds of relations for AGNRs.
One is that for a specific subband, the subenvelope functions of $A$ sublattices ($| \Psi_A \rangle$) and those of $B$ sublattices ($| \Psi_B \rangle$) are in phase or 180$^\circ$ out of phase ($| \Psi_A \rangle = | \Psi_B \rangle$ or $| \Psi_A \rangle = - | \Psi_B \rangle$), \emph{e.g.} $| \Psi_A (J^c=1) \rangle = - | \Psi_B (J^c=1) \rangle$ (Fig.~\ref{fig:BS_DOS_WFofMonoGNRsNy300B0E0}(i)).
The other is that, for the conduction and valence subbands with the same subband index, their subenvelope functions of $A$ ($B$) sublattices are in phase or out of phase by $\pi$, \emph{i.e.} $| \Psi^c_{A(B)} (J^c = J^v) \rangle = \pm | \Psi^v_{A(B)} (J^v) \rangle$.\cite{Opt.Express19(2011)23350H.C.Chung}
For instance, the subenvelope functions of $A$ sublattices for the $J^c = 1$ and $J^v = 1$ subbands are identical (Fig.~\ref{fig:BS_DOS_WFofMonoGNRsNy300B0E0}(i) and (j)).
The relations of the spatial distributions are distinct for ZGNRs and AGNRs; therefore, the optical selection rules are expected to depend on the edge structures (Section~\ref{sec:EdgeDependentAbsorptionSpectraOfMonoGNRs}).

The spectroscopic-imaging STM,\cite{Science262(1993)218M.F.Crommie, Phys.Rev.Lett.91(2003)196804T.Maltezopoulos, Nature427(2004)328S.Ilani, Science319(2008)782C.R.Moon, Nat.Phys.4(2008)454C.R.Moon, Science327(2010)665A.Richardella} which provides the information of wave functions through the local DOS, is a powerful tool to measure the standing waves on surfaces of various condensed-matter systems.
The direct observations on the regular oscillations have been done at Au(111) and Cu(111) surface steps.\cite{Phys.Rev.Lett.71(1993)1071Y.Hasegawa, Phys.Rev.Lett.73(1994)910G.Hormandinger, Nature363(1993)524M.F.Crommie, Science262(1993)218M.F.Crommie}
The theoretical predictions\cite{Phys.Rev.B60(1999)7792V.Meunier, Phys.Rev.Lett.82(1999)3520A.Rubio, Appl.Phys.A68(1999)275A.Rubio, Phys.Rev.B65(2002)245418J.Jiang} on the standing waves in a metallic CNT with a finite length are verified by the experimental measurements on the spatial distributions.\cite{Science283(1999)52L.C.Venema, Nature412(2001)617S.G.Lemay, Phys.Rev.Lett.93(2004)166403J.Lee}
Also, some related works have presented the spatial mapping of the electronic states in the troughs between self-organized Pt nanowires on Ge(001).\cite{NanoLett.6(2006)1439A.vanHouselt, J.Phys.-Condens.Matter25(2013)014014R.Heimbuch}
The spectroscopic-imaging STM can be used to identify the feature-rich standing waves in GNRs, \emph{e.g.} the edge-dominated special relations in the wave functions.

\subsubsection{Magnetic quantization.}
\label{sec:MagneticQuantizationInMonolayer}

Magnetic quantization constrains carrier motions in real space, brings neighboring electronic states together, and creates highly degenerate Landau states.\cite{Z.Physik64(1930)629L.Landau}
The low-energy electronic structures are determined by the competition between the magnetic quantization and finite-size boundary.
The 1D parabolic subbands evolve into the composite subbands with  dispersionless QLLs, and the standing waves become the localized Landau wave functions with a regular spatial symmetry.
The dispersionless Landau states can survive only when the ribbon width is comparable to the magnetic length ($\sqrt{\hbar/eB}$), and more QLLs come to exist under the extension of ribbon width.
It is remarkable that the QLL energies are proportional to the square-root of  magnetic field and subband index ($n^{c,v}$), \emph{i.e.} $|E^{c,v}| \propto \sqrt{n^{c,v}B}$.
This identifiable relation can be a criterion in locating where the electronic structures are dominated by the magnetic quantization.

\begin{figure}
\begin{center}
  \includegraphics[width=\linewidth, keepaspectratio]{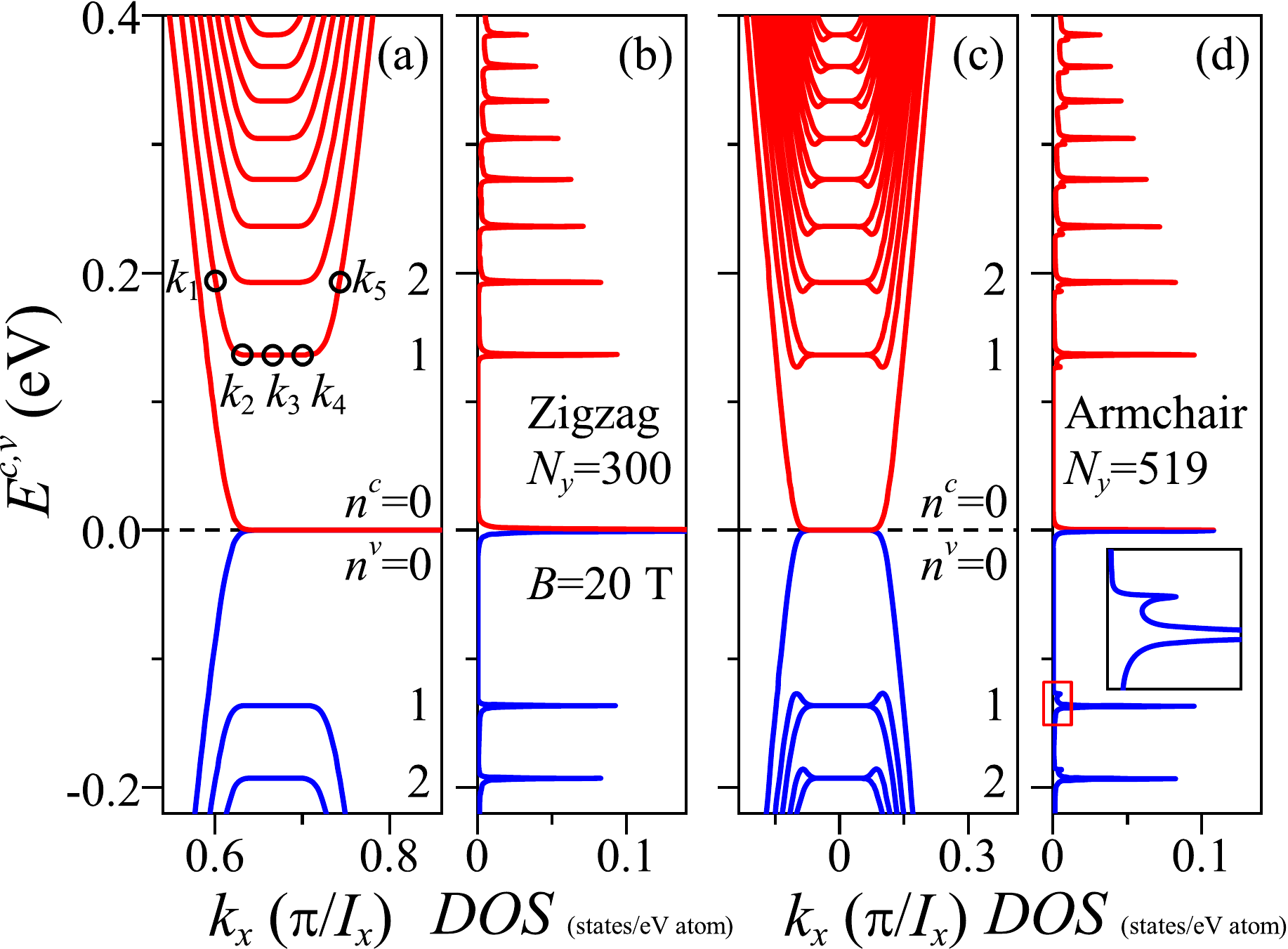}\\
  \caption[]{
  Low-energy magneto-electronic structures and DOS of the $N_y=300$ ZGNR and $N_y=519$ AGNR at $B=20$ T.
  The ZGNR and AGNR are with the same width for comparison.
  $n^{c,v}$ is the subband index accounting from $E_F = 0$.
  The circles mark the states of the $n^c = 1$ subband at $k_1$--$k_5$.
  }
  \label{fig:BSandDOSofGNRsInB}
\end{center}
\end{figure}

The low-energy electronic structures are dramatically changed by the magnetic fields.
There exist a lot of Landau subbands composed of parabolic subbands and dispersionless QLLs.
For example, the $n^{c} = 1$ Landau subband in ZGNR is a combination of two parabolic dispersions and one QLL (Fig.~\ref{fig:BSandDOSofGNRsInB}(a)).
The widths of QLLs gradually shrink as the state energy grows, and the QLL energies proportional to $\sqrt{n^{c,v}}$ are independent of the edge structures.\cite{Phys.Rev.B59(1999)8271K.Wakabayashi, Phys.Rev.B73(2006)195408L.Brey, J.Appl.Phys.103(2008)073709Y.C.Huang}
It is notable that the band structures are edge-dependent.
Especially, in ZGNRs, the formation center of QLLs is at $k_x=2/3$, and the partial flat subbands remain lying on the Fermi level.
The conjunction of the $n^{c,v} = 0$ QLLs and flat subbands indicates the coexistence of Landau and edge-localized states.
The former mostly survive near the formation center, while the latter mainly reside at the zone boundaries.\cite{Phys.Rev.Lett.98(2007)116802V.Lukose, J.Phys.Soc.Jpn.80(2011)044602H.C.Chung}
As to AGNRs (Fig.~\ref{fig:BSandDOSofGNRsInB}(c)), all the QLLs are formed around $k_x=0$ and each QLL is doubly degenerate.
At the ends of a QLL, the splitting of degeneracy takes place, \emph{i.e.} one of the two subbands has monotonic $k_x$-dependence and the other exhibits non-monotonic dependence with extra band-edge states.\cite{Phys.Rev.B73(2006)195408L.Brey, Phil.Trans.R.Soc.A368(2010)5431O.Roslyak}
This degeneracy splitting has been identified through the magneto-transport measurements.\cite{Phys.Rev.Lett.107(2011)086601R.Ribeiro, J.Low.Temp.Phys.170(2013)541J.Beard}
In addition, the gradual merging of the $n^{c,v} = 0$ QLLs causes the shrinkage of energy gap, and the GNR undergoes a semiconductor-metal transition.\cite{Nanotechnol.18(2007)495401Y.C.Huang, J.Appl.Phys.108(2010)033709S.B.Kumar}

Landau subbands induce many special structures in the DOS.
The primary structures are the symmetric peaks corresponding to the QLLs and  partial flat subbands; the secondary structures are the asymmetric peaks originating from the extra band-edge states of Landau subbands (Fig.~\ref{fig:BSandDOSofGNRsInB}(b) and (d)).
The features of the symmetric Landau peaks is independent of the edge structure.
The $\sqrt{n^{c,v}}$-dependence in the positions of QLLs remains the same.
The peak height is proportional to the QLL width and gradually reduces as the state energy grows.
Moreover, the edge-dependent energy dispersions are reflected on the DOS.
In ZGNRs (Fig.~\ref{fig:BSandDOSofGNRsInB}(b)), the highest symmetric peak at $E_F = 0$ is associated with the partial flat subbands; while in AGNRs, the adjacent asymmetric subpeaks (inset in Fig.~\ref{fig:BSandDOSofGNRsInB}(b)) correspond to the extra band-edge states.

\begin{figure}
\begin{center}
  \includegraphics[width=\linewidth, keepaspectratio]{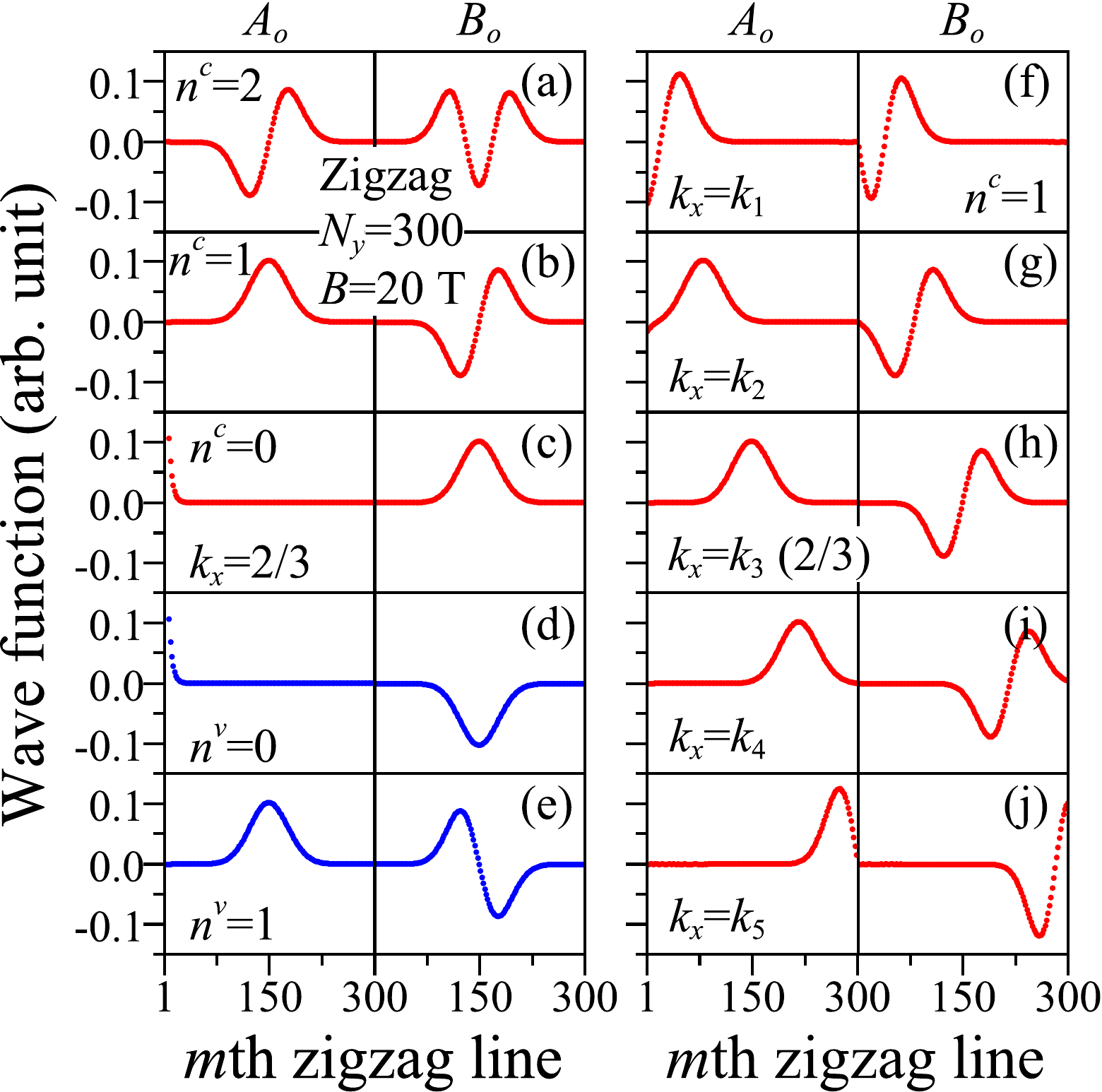}\\
  \caption[]{
  Landau wave functions of (a)--(e) the QLLs for $n^c=0$--$2$ and $n^v=0$ \& $1$ at $k_x = 2/3$ and (f)--(j) the $n^c = 1$ Landau subband at various wavevectors.
  $k_x = k_1$--$k_5$ are remarked to those in Fig.~\ref{fig:BSandDOSofGNRsInB}(a).
  }
  \label{fig:LandauWFofZGNRs}
\end{center}
\end{figure}

Magnetic wave functions in the GNRs present peculiar spatial distributions, where the localization center, waveform, and node number are very sensitive to the state energy and wavevector.
The Landau wave functions at the formation center of the $N_y = 300$ ZGNR are well behaved and localized at the ribbon center, as illustrated in Fig.~\ref{fig:LandauWFofZGNRs}(a)--(e).
For the $n^{c,v}=n$ QLL at $k_x > 0$ ($k_x < 0$), the wave functions of the $A$ and $B$ sublattices can be, respectively, described by $\phi_{n-1}$ and $\phi_n$ ($\phi_n$ and $\phi_{n-1}$), where $\phi_n$ is the wave function of harmonic oscillator, a product of the $n$th-order Hermite polynomial and a Gaussian distribution function.
Such functions are either symmetric or anti-symmetric, and the node number of the $B$ sublattice is larger than ($k_x > 0$) or smaller than ($k_x < 0$) that of the $A$ sublattice by one.
For instance, there are, respectively, one and two nodes on the $A$ and $B$ sublattices for the $n^c = 2$ QLL (Fig.~\ref{fig:LandauWFofZGNRs}(a)).
Further, the node number on the $A$ ($B$) sublattice grows with the increase of state energy, that is, the node numbers on the $B$ sublattice of the QLLs for $n^c = 1$, $2$, and $3$ are $1$, $2$, and $3$, respectively.
On the other hand, the spatial distributions of the magnetic wave functions are strongly dependent on the wavevector and reflect the competition between the lateral confinement and magnetic quantization.
When the wavevector shifts from $k_3 = 2/3$ to $k_2$ or $k_4$ (Fig.~\ref{fig:BSandDOSofGNRsInB}(a)), the $n^c = 1$ QLL remains highly degenerate, and the wave localization center moves to the ribbon edges with a consistent waveform (Fig.~\ref{fig:LandauWFofZGNRs}(g) and (i)).
However, for a larger wavevector shift, the wave functions are distorted and truncated by the ribbon edges, \emph{e.g.} $k_x = k_1$ and $k_5$ (Fig.~\ref{fig:LandauWFofZGNRs}(f) and (j)).
The electronic states are no longer dispersionless and correspond to the parabolic dispersion.

\begin{figure}
\begin{center}
  \includegraphics[width=\linewidth, keepaspectratio]{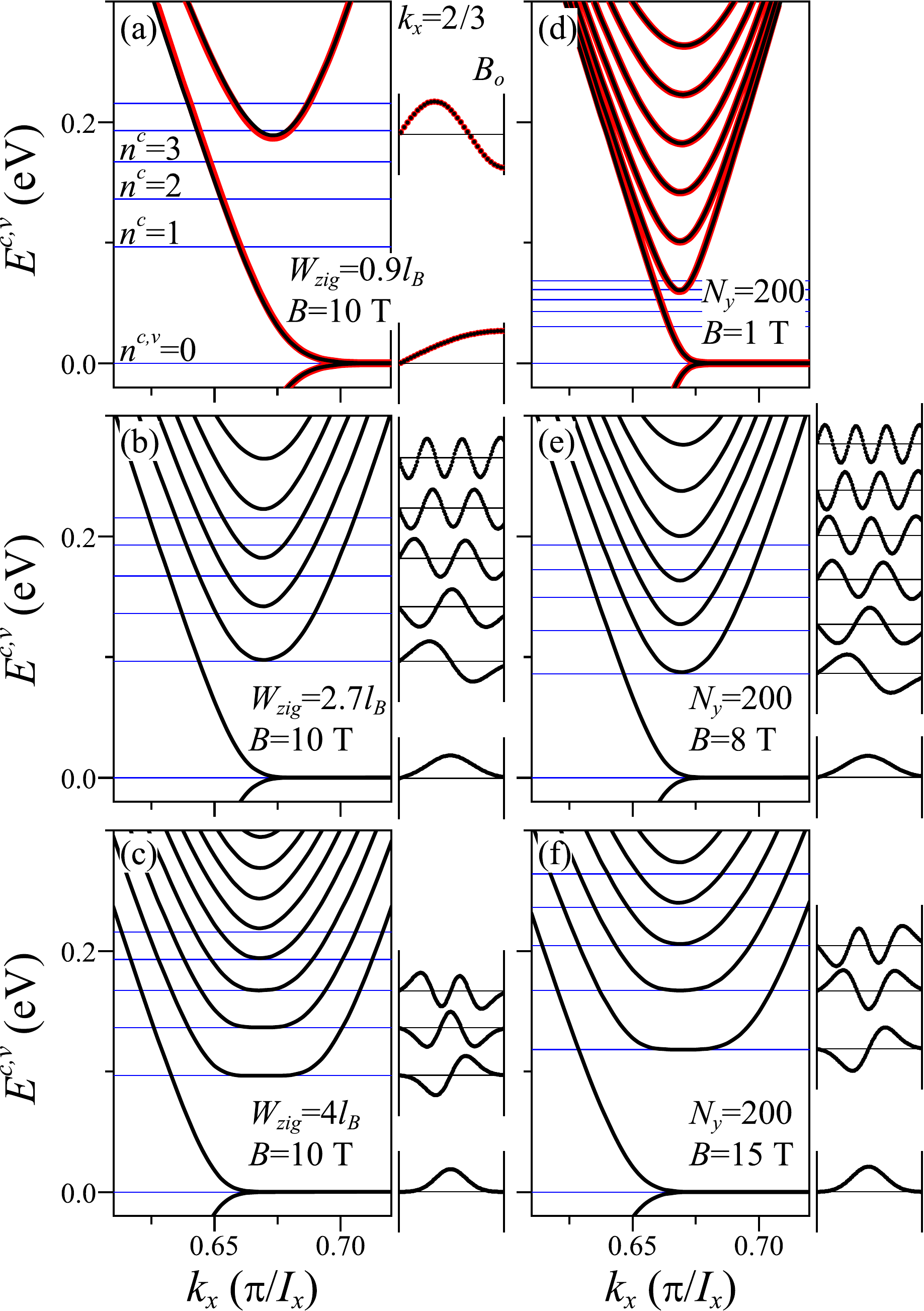}\\
  \caption[]{
  Energy spectra and wave functions of (a)--(c) the ZGNRs with different widths at $B=10$ T and (d)--(f) the $N_y=200$ ZGNRs under various magnetic fields.
  The first five QLL energies are indicated (solid blue lines).
  The heavy red curves and dots illustrate the zero-field spectra and wave functions, respectively.
  }
  \label{fig:MagneticAndQuantumConfinementsWithWF}
\end{center}
\end{figure}

The competition responsible for the evolution of electronic structures is worthy of further investigations.
The appearance of QLLs demonstrates where the magnetic fields dominate the electronic properties.
The magnetic length $l_B=\sqrt{\hbar/eB}$, which is a criterion in determining the formation of Landau states, can characterize the distribution width of the Landau wave functions.
For a narrower GNR, the energy dispersions are almost identical to the zero-field one (heavy red curves), meaning there are no QLLs and no changes on the standing waves (heavy red dots) as shown in Fig.~\ref{fig:MagneticAndQuantumConfinementsWithWF}(a).
Once the ribbon width is larger than the magnetic length, the QLLs are initiated from the low-lying electronic states.
In the case of $W_{zig} = 2.7 l_B$ (Fig.~\ref{fig:MagneticAndQuantumConfinementsWithWF}(b)), the energy spacings between subbands shrink, and the QLLs for $n^{c,v} = 0$ and 1 are revealed.
The $n^{c,v} = 0$ Landau subbands have a lot of dispersionless Landau states, and the $n^{c} = 1$ subband has only one at the formation center.
Meanwhile, the band-edge state energy of the parabolic subband is the same as that of the $n^c = 1$ QLL, \emph{i.e.} the parabolic subband evolves into the $n^c=1$ Landau subband.
The $n^c = 1$ Landau wave function entirely resides in the nanoribbon, while the higher-energy ones are distorted and truncated.
But for $l_B \ll W_{zig}$  (Fig.~\ref{fig:MagneticAndQuantumConfinementsWithWF}(c)), many QLLs are formed with more well-behaved Landau wave functions.
The magnetic quantization dominates the low-energy band structure around the formation center.
Also, the magneto-electronic properties can be tuned by the field strength.
The energy dispersions and electronic states are almost unchanged for a weaker magnetic field (Fig.~\ref{fig:MagneticAndQuantumConfinementsWithWF}(d)).
When the field strength increases and $l_B$ is comparable to $W_{zig}$, the QLLs make their appearances (Fig.~\ref{fig:MagneticAndQuantumConfinementsWithWF}(e)).
As to a much stronger magnetic field ($l_B \ll W_{zig}$), the magnetic quantization determines the electronic structure, and hence, there are many QLLs with well-behaved spatial distributions (Fig.~\ref{fig:MagneticAndQuantumConfinementsWithWF}(f)).

\begin{figure}
\begin{center}
  \includegraphics[width=\linewidth, keepaspectratio]{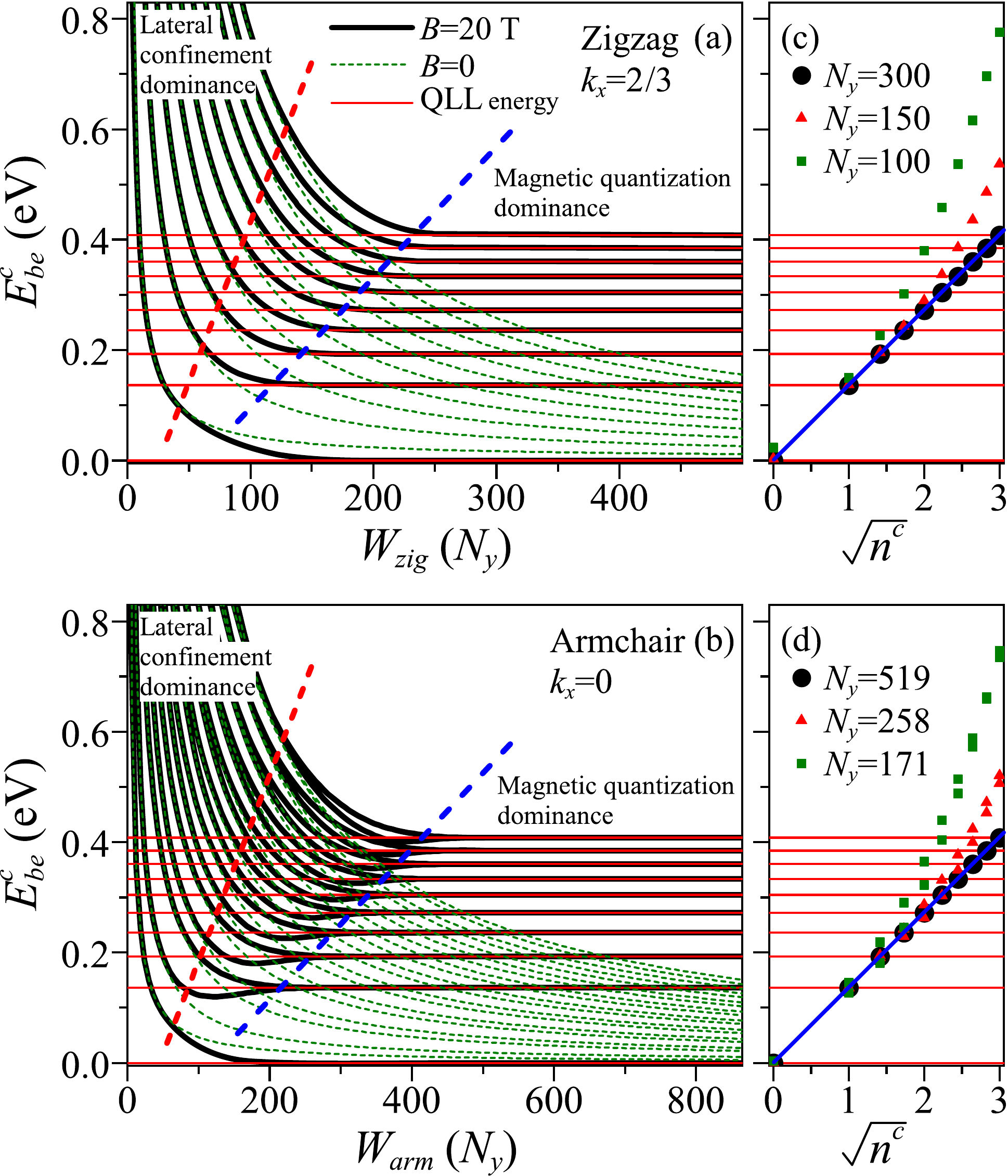}\\
  \caption[]{
  Dependence of band-edge state energies on the ribbon width and on the square-root of subband index for the low-lying conduction subbands.
  The length scale of $W_{arm}$ and $W_{zig}$ are identical.
  The solid blue lines in (c) and (d) indicate the linear relations between $E_{be}^c$ and $\sqrt{n^c}$.
  }
  \label{fig:EnergiesOfQLLsVSWidthAndSqrtNc}
\end{center}
\end{figure}

\begin{figure}
\begin{center}
  \includegraphics[width=\linewidth, keepaspectratio]{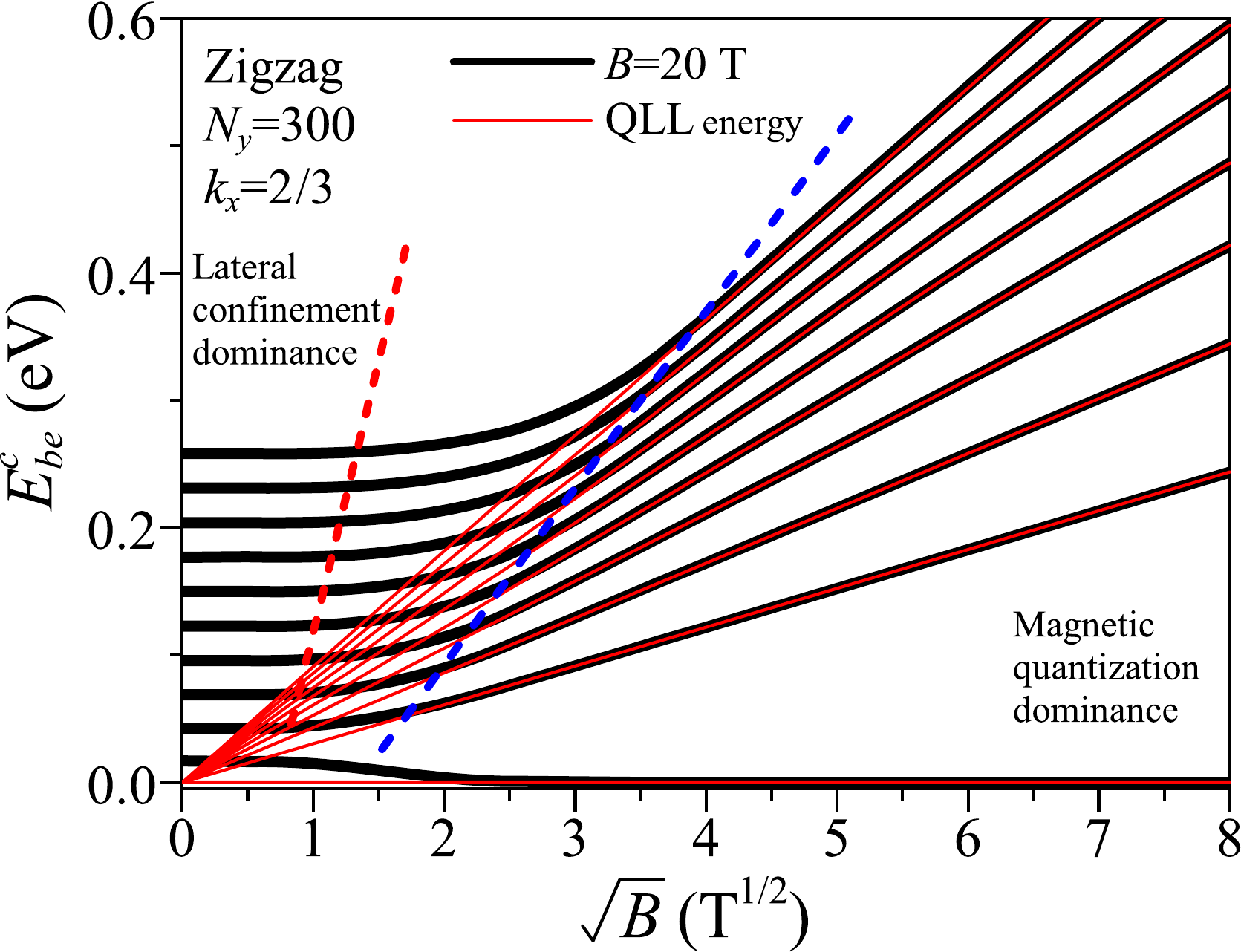}\\
  \caption[]{
  Dependence of band-edge state energies on $\sqrt{B}$ for the low-lying conduction subbands (heavy black curves) and the linear relations between $E_{be}^c$ and $\sqrt{B}$ (solid red lines).
  }
  \label{fig:EnergiesOfQLLsVSsqrtB}
\end{center}
\end{figure}

The competition-induced evolution of magneto-electronic properties, including the subband structure, energy spacing, band gap, as well as the energy dependence on the ribbon width, magnetic field strength, and subband index, is explored in detail and can be directly verified by experimental measurements.
For ZGNRs under a specific magnetic field, the band-edge state energies (heavy black curves) deviate from the zero-field ones (dashed green curves) and then approach to the QLL ones (solid red lines) with the increase of ribbon width (Fig.~\ref{fig:EnergiesOfQLLsVSWidthAndSqrtNc}(a)).
In other words, the energy dispersions evolve from the parabolic ones to the partially dispersionless ones.
The dominant effect can be distinguished by the deviations and approaches, as shown by the dashed blue and red lines, respectively.
On the left side of the dashed red line, the state energies are the same as the zero-field ones, indicating that the electronic structures are dominated by the lateral confinement; while on the right side of the dashed blue line, they are identical to the QLL ones, meaning that the magnetic quantization is the dominant effect.
The region between these two lines clearly illustrates the strong competition of two quantum effects; therefore, the Landau wave functions are truncated and no QLLs are formed.
As to AGNRs (Fig.~\ref{fig:EnergiesOfQLLsVSWidthAndSqrtNc}(b)), the deviation and approach lines are the same with those in ZGNRs, \emph{i.e.} the width-dependent dominant regions are independent of the edge structures.
Apparently, the lowest magneto-electronic state (the lowest heavy black curve), being lower than the zero-field one (the lowest dashed green curve), rapidly approaches to zero, indicating that the magnetic quantization accelerates the gap shrinkage and results in the semiconductor-metal transition.
Moreover, the state energies exhibit the $\sqrt{n^c}$-dependence (solid blue lines) for sufficiently wide GNRs (Fig.~\ref{fig:EnergiesOfQLLsVSWidthAndSqrtNc}(c) and (d)).
Similarly, the band-edge state energies (heavy black curves) undergo the deviations and approaches with the increase of magnetic fields, and the dominant effects can be clearly identified (Fig.~\ref{fig:EnergiesOfQLLsVSsqrtB}).
After the QLL formation, the state energies are very close to those of monolayer graphene, and they can be expressed by $E^{c,v} (n^{c,v}) = \pm \sqrt{2e \hbar v_F^2 B n^{c,v}}$, where $v_F = 3tb/2$ is the Fermi velocity and $+$ ($-$) stands for the conduction (valence) states.\cite{Phys.Rev.104(1956)666J.W.McClure, Phys.Rev.Lett.95(2005)146801V.P.Gusynin}
The $\sqrt{n^{c,v}B}$-dependence, being different from the linear $B$-dependence in the two-dimensional electron gas (2DEG), has been confirmed via infrared transmission experiments.\cite{Phys.Rev.Lett.98(2007)197403Z.Jiang, Phys.Rev.Lett.100(2008)087401P.Plochocka}

The main characteristics and the competition-induced evolution of prominent Landau peaks in the DOS can be verified by STS.
It is an extension of STM\cite{Phys.Rev.Lett.49(1982)57G.Binnig, Helv.Phys.Acta55(1982)726G.Binnig, Phys.Rev.Lett.50(1983)120G.Binnig} and provides detailed information about the DOS on a sample surface, such as CNTs\cite{Nature391(1998)59J.W.G.Wildoer, Phys.Rev.Lett.82(1999)1225P.Kim, Science283(1999)52L.C.Venema, Phys.Rev.B62(2000)5238L.C.Venema} and silicon.\cite{Phys.Rev.Lett.56(1986)1972R.J.Hamers, Surf.Sci.181(1987)295R.M.Feenstra, Phys.Rev.B48(1993)17892M.N.Piancastelli}
The tunneling conductance ($dI/dV$), which is proportional to the DOS,\cite{Phys.Rev.B31(1985)805J.Tersoff} apparently reveals the main characteristics, \emph{i.e.} the structures, positions, and intensities of the peaks.
In very wide GNRs, the low-lying symmetric Landau peaks with the $\sqrt{n^{c,v}B}$-dependence have been observed.\cite{Science324(2009)924D.L.Miller, Nature467(2010)185Y.J.Song}
Recently, the spatial evolution of Landau states towards the ribbon edges has been reported,\cite{Nat.Commun.4(2013)1744G.Li} and the gradual disappearance of the Landau peaks clearly exhibit the competition.

The competition is also reflected in the magneto-transport responses.
For a 2D graphene,\cite{Nature438(2005)197K.S.Novoselov, Nat.Mater.10(2011)357D.Waldmann, Phys.Rev.B84(2011)115429Z.Tan} the magneto-resistance as a function of the inverse magnetic field shows a typical Shubnikov-de Haas oscillation (SdHO).\cite{LeidenCommun.207a(1930)3L.Shubnikov}
When the magnetic length becomes comparable to the width of nanoribbon, the competition has been unveiled by the anomalous SdHO, a strong departure from the $1/B$ periodicity.\cite{Phys.Rev.B73(2006)241403N.M.R.Peres, Science312(2006)1191C.Berger}
Furthermore, the breakdown of Hall quantization in narrow GNRs gives another experimental verification.
The $2G_0$ Hall plateau (with $G_0 = 2 e^2/h$) at the lowest filling factor of 2  vanishes, and the seriously distorted structures in conductance are generated for any filling factors.\cite{Phys.Rev.Lett.107(2011)086601R.Ribeiro, J.Low.Temp.Phys.170(2013)541J.Beard}

Experimental measurements have made significant progress in the direct observations on Landau wave functions.
Since 2DEG is usually deeply buried in semiconducting heterostructures, the experimental observations on the nodal structures of Landau wave functions are elusive.
However, the advent of surface 2DEG in graphene,\cite{Phys.Rev.Lett.102(2009)026803Y.Niimi} doped semiconductors,\cite{Phys.Rev.Lett.90(2003)056804M.Morgenstern, Phys.Rev.B81(2010)155308S.Becker} and topological insulators\cite{Phys.Rev.Lett.105(2010)076801P.Cheng, Phys.Rev.B82(2010)081305T.Hanaguri} has opened the way for the experimental verifications.
Recently, the observations about the Landau orbitals without zero point are revealed in the STM real-space imaging of electronic probability density,\cite{Phys.Rev.Lett.101(2008)256802K.Hashimoto, Nat.Phys.6(2010)811D.L.Miller} while at the later time the further observations on the concentric-ring-like nodal structures have also been realized.\cite{Phys.Rev.Lett.109(2012)116805K.Hashimoto, Nat.Phys.10(2014)815Y.S.Fu}
In GNRs, the Landau wave functions are affected by the geometric configurations and external fields.
The lateral confinements and curvatures (Section~\ref{Sec:CurvedSystem}) result in the distortions and truncations in spatial distributions.
Bilayer GNRs with different stackings possess various nodal structures (Section~\ref{BilayerSystem}).
Electric fields cause severe mixings of Landau orbitals in the collapsed QLLs.
The QLL-related mixings in the non-uniform GNRs and GNR-CNT hybrids lead to the discontinuous and kink-form distributions (Section~\ref{NonUniformGNRs}).
The diverse spatial distributions in Landau wave functions can be examined by the spectroscopic-imaging STM measurements on nodal structures.\cite{Phys.Rev.Lett.101(2008)256802K.Hashimoto, Nat.Phys.6(2010)811D.L.Miller, Phys.Rev.Lett.109(2012)116805K.Hashimoto, Nat.Phys.10(2014)815Y.S.Fu}

\subsubsection{Electric-field-induced rich properties.}
\label{sec:EffectsOfElectricFieldsOfMonolayerSystem}

A transverse electric field can induce the different site energies and the destruction of the mirror symmetry about the $xz$-plane, so that the electronic properties are significantly diversified.
The low-lying electronic structures are sensitive to the field strength and reflected on the DOS, including the splitting of partial flat subbands, mixing of conduction and valence subbands, extra band-edge states, gap opening, and semiconductor-metal transition.
The gate-voltage-controllable gap, which can be verified by the measurements on conductance\cite{Phys.Rev.Lett.98(2007)206805M.Y.Han, Science319(2008)1229X.L.Li} and absorption spectrum,\cite{NanoLett.2(2002)155M.E.Itkis} provides the flexibility in the design of GNR-based nanoelectronic devices.
The main characteristics of wave functions, the symmetry, node number, and localization center of the spatial distribution, are drastically altered and can be further examined by the spectroscopic-imaging STM.\cite{Nature363(1993)524M.F.Crommie, Science283(1999)52L.C.Venema, NanoLett.6(2006)1439A.vanHouselt}

\begin{figure}
\begin{center}
  \includegraphics[width=\linewidth, keepaspectratio]{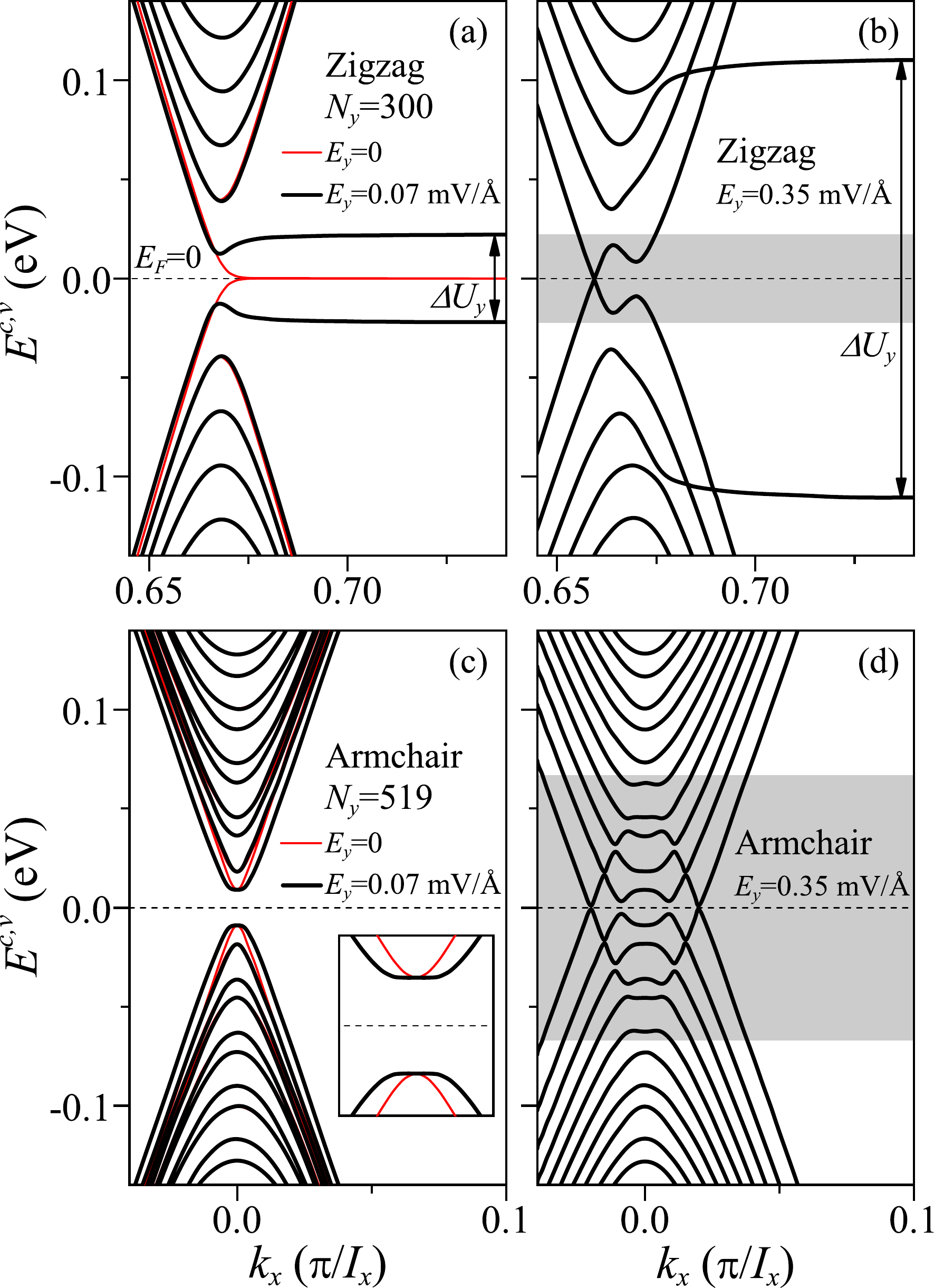}\\
  \caption[]{
  Electronic structures of the $N_y = 300$ ZGNR and $N_y = 519$ AGNR under various transverse electric fields.
  Inset illustrates the details of widened parabolic subbands, and gray zones mark the extra band-edge states.
  }
  \label{fig:BSofGNRsB0Evarious}
\end{center}
\end{figure}

The low-lying energy subbands are violently changed by the transverse electric field $E_y \hat{y}$.\cite{Carbon44(2006)508C.P.Chang, Phys.Rev.Lett.99(2007)056802D.S.Novikov, Fuller.Nanotub.CarbonNanostruct.21(2013)183R.Alaei}
For ZGNRs, the doubly degenerate partial flat subbands are split with an energy spacing, $\Delta U_y = | e E_y W_{zig} |$, reflecting the edge localization of  electronic states.
An energy gap is opened and enlarged with the increasing field strength (Fig.~\ref{fig:BSofGNRsB0Evarious}(a)).
After reaching a maximum value, it shrinks and vanishes due to the mixing of valence and conduction subbands, \emph{e.g.} the mixed $J^c = 1$ and $J^v = 1$ parabolic subbands in Fig.~\ref{fig:BSofGNRsB0Evarious}(b).
Both the metal-semiconductor and semiconductor-metal transitions are revealed during the variation of field strength.
It is noticeable that the strong electric fields lead to the oscillatory subbands with more extra band-edge states.
As to AGNRs, the low-lying parabolic subbands are widened with more band-edge states (Fig.~\ref{fig:BSofGNRsB0Evarious}(c)).
The energy gap monotonically reduces and then approaches to zero with the subband mixing as $E_y$ grows, for instance, the $J^{c,v} = 1$--$5$ oscillatory subbands in Fig.~\ref{fig:BSofGNRsB0Evarious}(d).
There exists a semiconductor-metal transition in AGNRs.\cite{Phil.Trans.R.Soc.A368(2010)5431O.Roslyak, Phys.Lett.A374(2010)4061O.Roslyak}
Approximately, the $E_y$-dependent maximum gap is inversely proportional to the ribbon width, regardless of the edge structures.\cite{Carbon44(2006)508C.P.Chang, Phys.Rev.B77(2008)245434H.Raza, Appl.Phys.Lett.98(2011)263105S.B.Kumar}

\begin{figure}
\begin{center}
  \includegraphics[width=\linewidth, keepaspectratio]{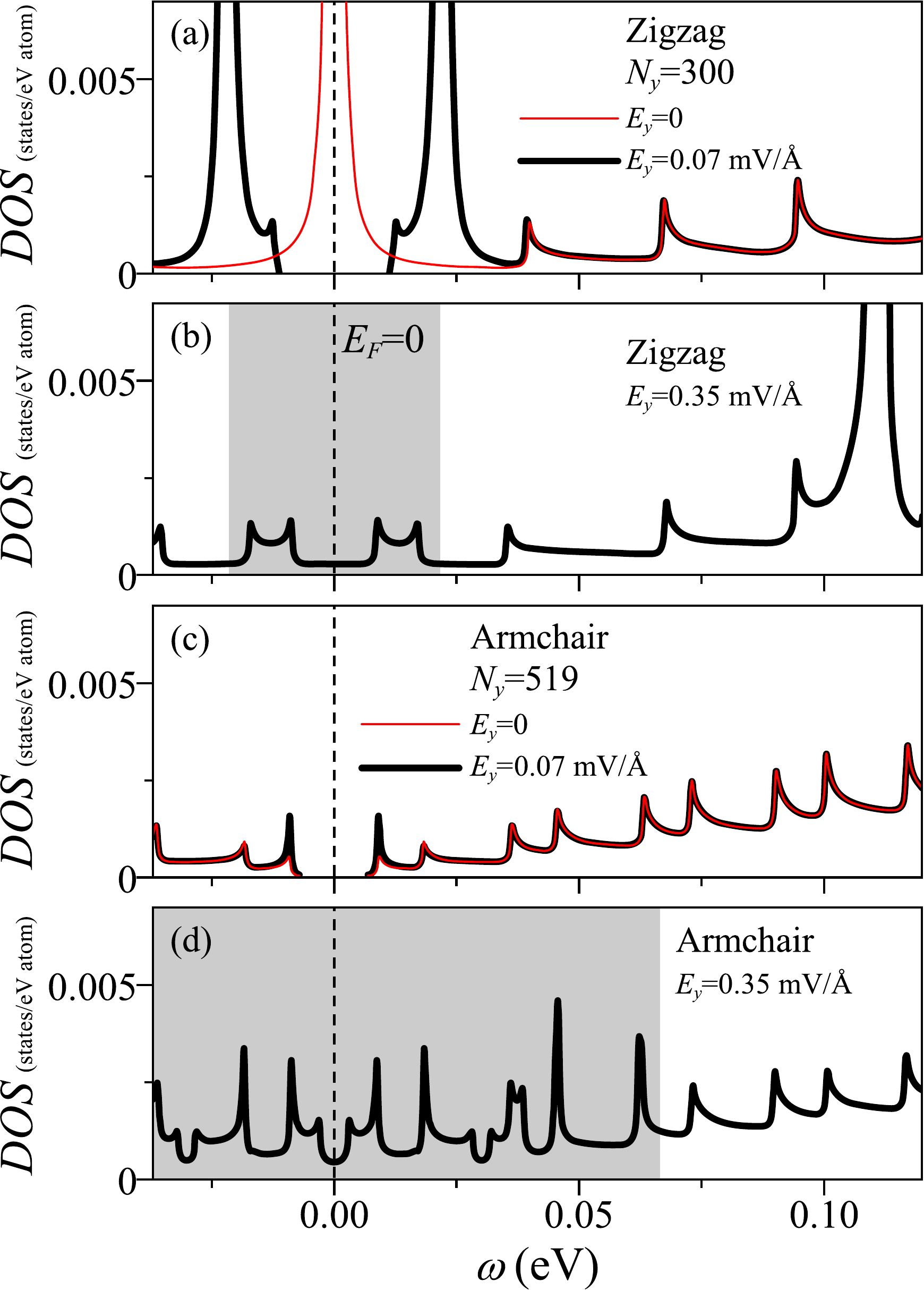}\\
  \caption[]{
  DOS of the $N_y = 300$ ZGNR and $N_y = 519$ AGNR under various electric fields.
  The gray zones indicate the additional peaks.
  }
  \label{fig:DOSofGNRsB0Evarious}
\end{center}
\end{figure}

The special structures in the DOS are enriched by the transverse electric field.
In ZGNRs, the highest symmetric peak at $\omega = 0$ corresponding to the partial flat subbands is split into two ones with a spacing of $\Delta U_y$ (Fig.~\ref{fig:DOSofGNRsB0Evarious}(a)).
Such peaks are rapidly far away from the Fermi level and appear at higher energies.
Meanwhile, the low-lying peak structures are replaced by pairs of asymmetric ones (Fig.~\ref{fig:DOSofGNRsB0Evarious}(b)).
While for AGNRs, only the low-lying two peaks are enhanced at small $E_y$ (Fig.~\ref{fig:DOSofGNRsB0Evarious}(c)).
Many additional peaks are revealed due to the wide range of subband mixing in the further increase of $E_y$ (Fig.~\ref{fig:DOSofGNRsB0Evarious}(d)).
The variations in number, position, height, and structure of prominent peaks near the Fermi level can be tested by STS measurements.\cite{Science324(2009)924D.L.Miller, Nature467(2010)185Y.J.Song, Nat.Commun.4(2013)1744G.Li}

\begin{figure}
\begin{center}
  \includegraphics[width=\linewidth, keepaspectratio]{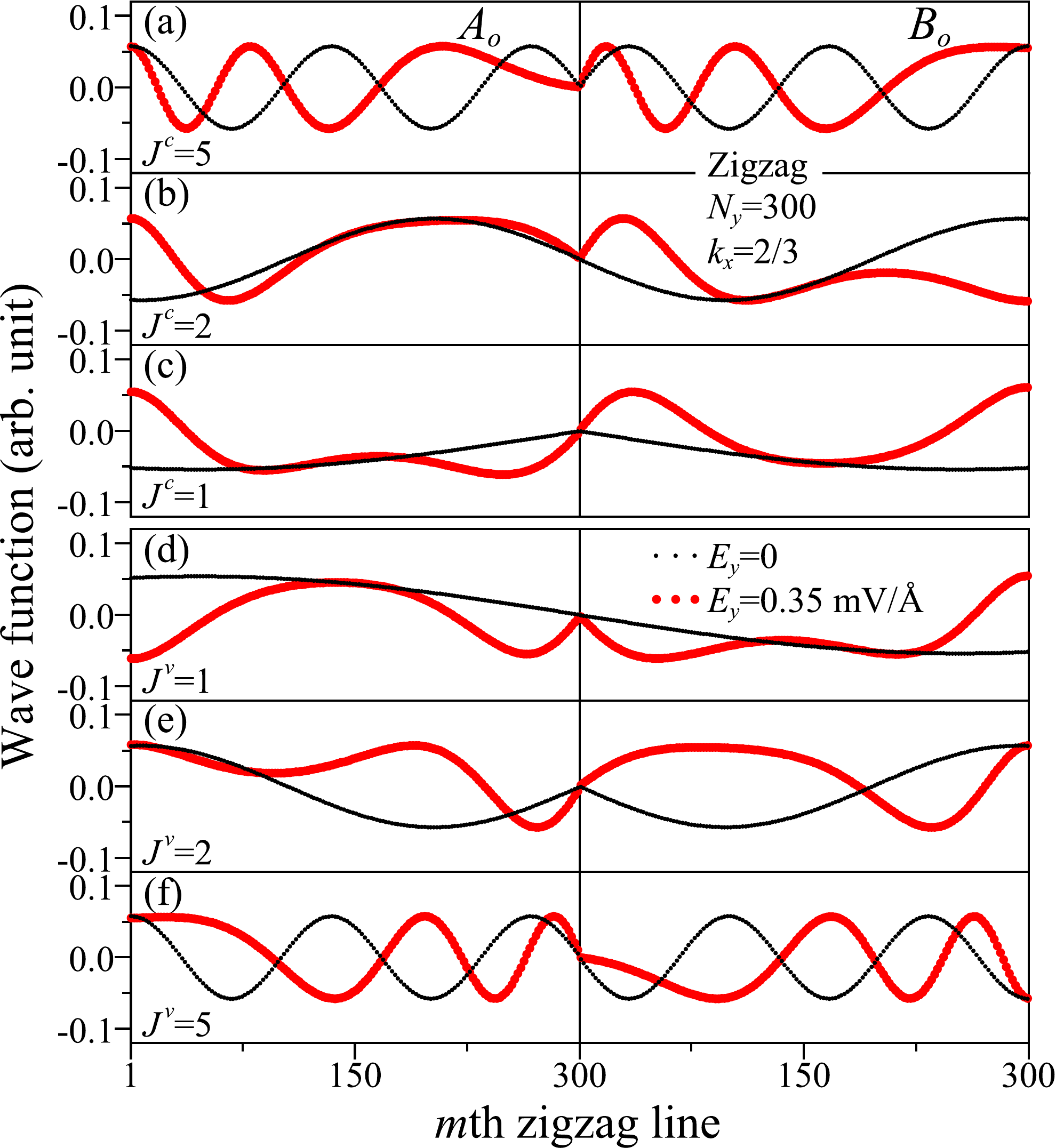}\\
  \caption[]{
  Wave functions of the $N_y = 300$ ZGNR at $k_x = 2/3$ under $E_y = 0$ (light black dots) and $E_y = 0.35$ mV/{\AA} (heavy red dots).
  }
  \label{fig:WFofZGNRInE}
\end{center}
\end{figure}

The spatial distributions of standing waves are severely distorted by $E_y$.
The symmetric and anti-symmetric features no longer exist (heavy red dots in Fig.~\ref{fig:WFofZGNRInE}).
The regularities of node numbers are destroyed, especially for the low-lying energy subbands.
For example, instead of one, there are two and three nodes in the sublattices $A$ and $B$ of the $J^c=1$ wave function, respectively (Fig.~\ref{fig:WFofZGNRInE}(c)).
While for the higher-energy electronic states, the effects of the electric field are greatly reduced, \emph{e.g.} the $J^{c,v} = 5$ wave functions (Fig.~\ref{fig:WFofZGNRInE}(a) and (f)).
Their spatial distributions are shifted to the ribbon edges without the variations in node numbers and amplitudes.
Moreover, the conduction and valence ones shift oppositely to each other.
By means of the spectroscopic-imaging STM,\cite{Nature363(1993)524M.F.Crommie, Science283(1999)52L.C.Venema, NanoLett.6(2006)1439A.vanHouselt} the destructions of zero-point regularity and distribution symmetry and the changes of amplitude and localization center can be examined.

\subsubsection{Diverse properties in a composite field.}
\label{sec:Effects_of_Composite_Fields_Monolayer}

A composite field, which consists of a perpendicular magnetic field and a transverse electric field, can diversify the magneto-electronic properties.
The strong competition between these two fields will be obviously revealed in the electronic properties.
The different site energies hinder the formation of Landau states with the same energy.
The symmetry of energy spectrum is dramatically changed.
The QLLs are tilted, and their spacings are gradually reduced during the variations of electric and magnetic fields.
There exists a critical electric field with a specific ratio to the magnetic one, in which the tilted QLLs are collapsed with zero energy spacing and the Landau wave functions are thoroughly destroyed.

\begin{figure}
\begin{center}
  \includegraphics[width=\linewidth, keepaspectratio]{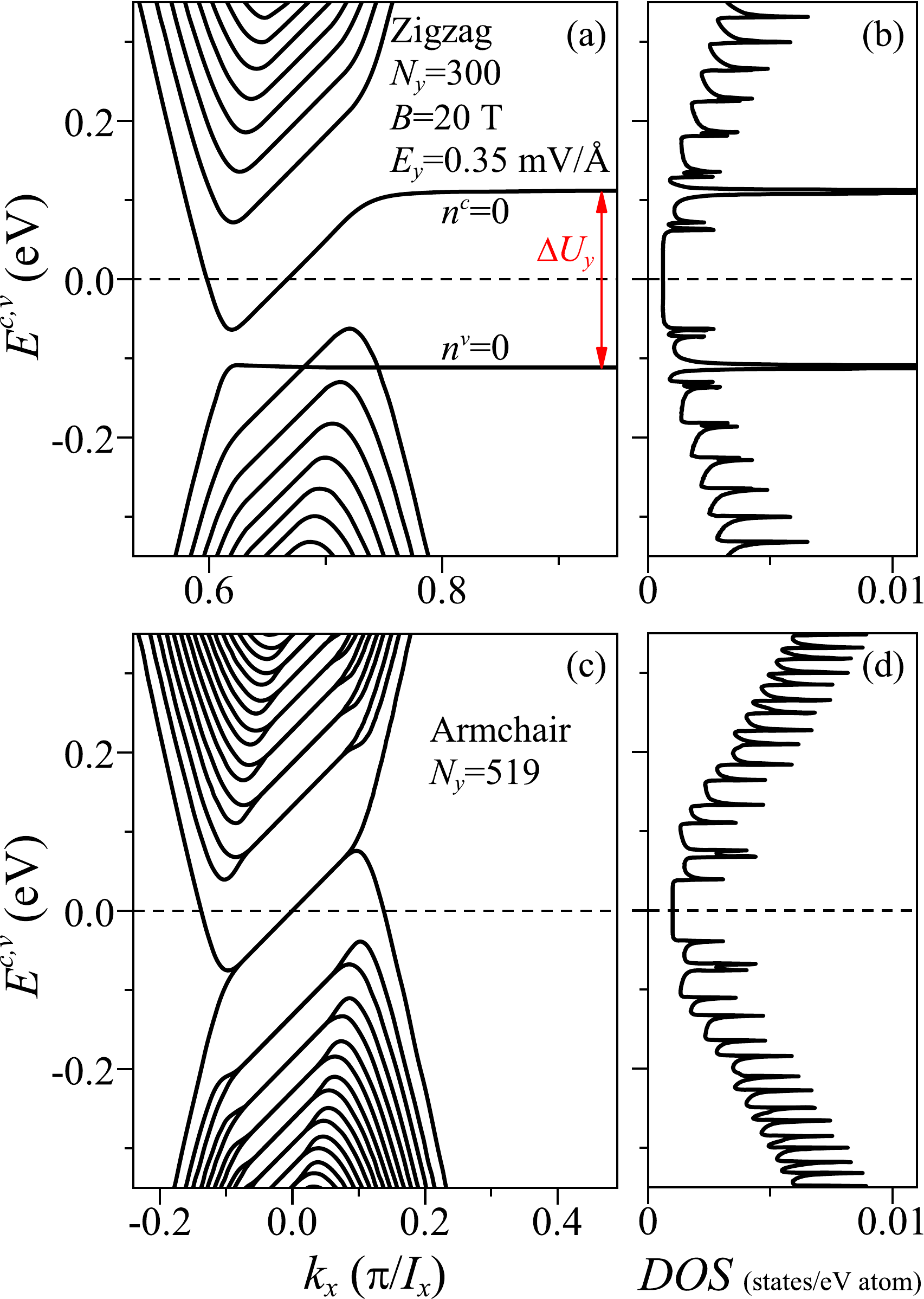}\\
  \caption[]{
  Magneto-electronic structures and DOS with respect to the $N_y=300$ ZGNR and $N_y=519$ AGNR in the presence of a composite field.
  }
  \label{fig:BSandDOSofGNRsInBandE}
\end{center}
\end{figure}

The zero-field energy subbands exhibit the symmetric energy spectrum, $E^{c,v} (k_x) = E^{c,v} (-k_x)$ and $E^{c} (k_x) = -E^{v} (k_x)$.
The former and the latter result from the inversion symmetry along the $x$-axis and the equivalence of $A$ and $B$ sublattices, respectively.
Under the magnetic or electric field, these symmetric properties keep the same.
However, a composite field breaks the inversion invariance and makes the energy dispersion change into anti-symmetric spectrum $E^{c} (k_x) = -E^{v} (-k_x)$ (Fig.~\ref{fig:BSandDOSofGNRsInBandE}(a) and (c)).
When the magnetic field is fixed at the $z$-axis, $x \rightarrow -x$ and $y \rightarrow -y$ are operated simultaneously.
The variation of the transverse electric potential from $V(y)$ to $V(-y)=-V(y)$ means that the electron spectrum will become the hole spectrum, or vice versa.
Although the energy spectrum is anti-symmetric, the DOS remains symmetric about $\omega = 0$ (Fig.~\ref{fig:BSandDOSofGNRsInBandE}(b) and (d)).

The QLLs are tilted with the same angle for various $E_y$'s.
The tilting pivots are at $k_x = 2/3$ and $k_x = 0$ for ZGNRs and AGNRs, respectively (Fig.~\ref{fig:BSandDOSofGNRsInBandE}(a) and (c)).
The low-lying conduction and valence subbands coexist within a specific energy range,\cite{J.Phys.Soc.Jpn.80(2011)044602H.C.Chung, Phys.Lett.A376(2012)1215M.Yang, Phys.Rev.B91(2015)155409B.Ostahie} for instance, the $n^{c,v} = 0$ subbands of the AGNRs (Fig.~\ref{fig:BSandDOSofGNRsInBandE}(c)).
With the increase of $E_y$, the tilting angle is getting larger, and more subbands overlap simultaneously.
It is remarkable that, at the zone boundary of ZGNR, the partial flat subbands split with an energy spacing of $\Delta U_y$, which features the edge-localized states.
Near the tilting pivots, the $n^{c} = 0$ subband possesses only the Landau states and behaves like a QLL, but the $n^{v} = 0$ subband has only the edge-localized states and preserves the partial flat energy dispersion.\cite{J.Phys.Soc.Jpn.80(2011)044602H.C.Chung}
As a result, the mixed Landau and edge-localized states can be separated by the electric field.
Additionally, the large overlap of conduction and valence subbands is predicted to induce the irregular oscillations in the ballistic conductance.\cite{Phil.Trans.R.Soc.A368(2010)5431O.Roslyak, Phys.Lett.A374(2010)4061O.Roslyak}

DOS under the electric field exhibits a lot of prominent structures, including asymmetric peaks, a finite plateau centered at $\omega = 0$, and a pair of symmetric peaks.
The asymmetric peaks arise from the band-edge states of composite subbands with tilted QLLs and parabolic dispersions.
The low-lying tilted QLLs with linear dispersions cross over the Fermi level, and thus induce the plateau structure.
As to ZGNRs, there is an energy spacing of $\Delta U_y$ between two quite strong symmetric peaks associated with the splitting subbands of $n^{c,v} = 0$.
The increase of $E_y$ will significantly reduces the asymmetric peak height, narrows the plateau width, and enlarges $\Delta U_y$.
These variations can be investigated by STS measurements,\cite{Science324(2009)924D.L.Miller, Nature467(2010)185Y.J.Song, Nat.Commun.4(2013)1744G.Li} so that the tilt of QLLs can be verified.

\begin{figure}
\begin{center}
  \includegraphics[width=\linewidth, keepaspectratio]{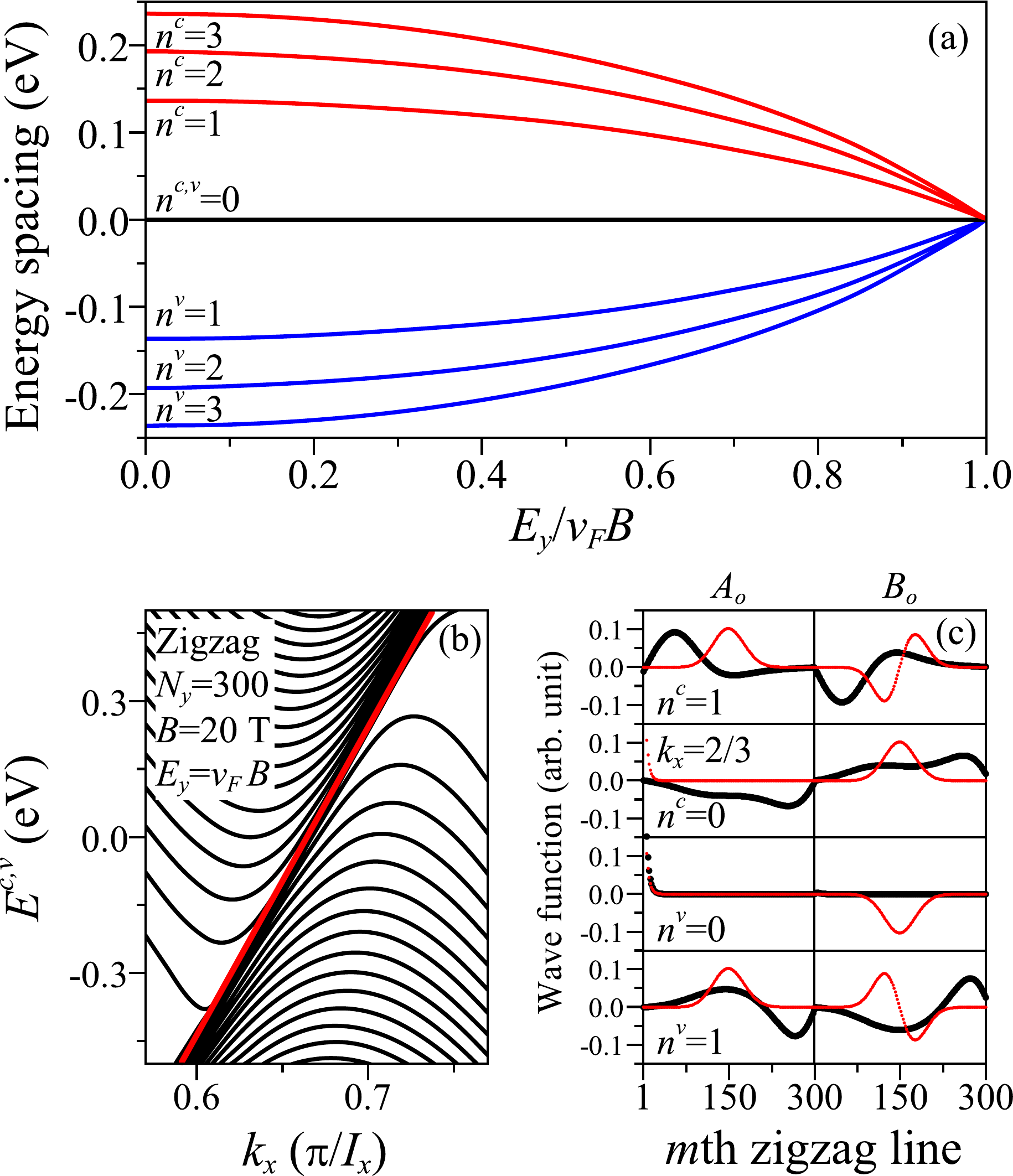}\\
  \caption[]{
  Collapse of QLLs by the transverse electric field.
  (a) The spacings between $n^{c,v} = 1$--$3$ QLLs and the lowest tilted QLL as a function of $v_FB$.
  (b) The magneto-electronic structure at the critical electric field of $|E_y| = v_F|B|$.
  (c) Large black dots: the wave functions of collapsed QLLs at $k_x = 2/3$.
  Small red dots: the regular Landau wave functions.
  }
  \label{fig:TheCollapseOfQLLsInBandE}
\end{center}
\end{figure}

The collapse of QLLs can be identified from the energy spacings and wave functions.
The energy spacings between the $n^{c,v}=1$--$3$ and $n^{c,v}=0$ QLLs at $k_x = 2/3$ gradually decline in the increase of $E_y$ (Fig.~\ref{fig:TheCollapseOfQLLsInBandE}(a)).
The main characteristics of QLLs are absent at a critical field of $|E_y| = v_F|B|$.
In other words, the tilted QLLs are piled up on a straight line, and there are no spacings between them (Fig.~\ref{fig:TheCollapseOfQLLsInBandE}(a) and (b)).
The wave functions are severely mixed and distorted, as evidenced by the entire destructions in the node regularity and spatial symmetry of the $n^{c,v} = 0$ and $1$ wave functions at $k_x = 2/3$ (large black dots in Fig.~\ref{fig:TheCollapseOfQLLsInBandE}(c)).
The gradual collapse of QLLs indicates an intense competition between the transverse centrifugal force and the longitudinal Coulomb force.
As the former is fully suppressed by the latter, the cyclotron motion is destroyed and the dispersionless QLLs can not be formed.

\subsubsection{Effects of modulated electric fields on magneto-electronic properties.}

Magneto-electronic properties under modulated electric fields are an intriguing field of less exploration.\cite{Phys.Rev.B75(2007)125429A.Matulis, Phys.Rev.B76(2007)195416M.Tahir, Phys.Rev.B77(2008)115446M.Barbier, Phys.Rev.B79(2009)115427J.H.Ho}
A modulated electric potential can be created by a periodic array of electrostatic gates,\cite{Phys.Rev.Lett.62(1989)1177R.W.Winkler, J.Vac.Sci.Technol.B10(1992)2904C.G.Smith, Phys.Rev.Lett.78(1997)705A.Messica, Phys.Rev.B55(1997)4482A.Soibel, Phys.Rev.B85(2012)075427S.Goswami, NewJ.Phys.17(2015)043035M.Staab} thin films of multiferroic BiFeO$_3$ with ordered domain strips,\cite{NanoLett.9(2009)1726Y.H.Chu, Nat.Nanotechnol.5(2010)143S.Y.Yang,  Phys.Rev.Lett.107(2011)126805J.Seidel, Adv.Mater.23(2011)1530Y.P.Chiu, NanoLett.11(2011)828C.T.Nelson} and two interfering laser beams.\cite{Appl.Phys.Lett.45(1984)663K.Tsubaki, Europhys.Lett.8(1989)179D.Weiss, Phys.Rev.B39(1989)13020D.Weiss, Phys.Rev.Lett.62(1989)2020C.W.J.Beenakker, Phys.Rev.Lett.64(1990)1473R.R.Gerhardts}
The magnetoresistance measurements on the GaAs/AlGaAs heterojunctions have been done to realize the Weiss oscillations in the 2DEG systems.\cite{Europhys.Lett.8(1989)179D.Weiss, Phys.Rev.B39(1989)13020D.Weiss, Phys.Rev.Lett.62(1989)1173R.R.Gerhardts, Phys.Rev.Lett.62(1989)2020C.W.J.Beenakker}
Recently, graphene in the non-linear electric potentials can be achieved by applying one or several local top-gate voltages,\cite{Science317(2007)638J.R.Williams, Phys.Rev.Lett.98(2007)236803B.Huard, Phys.Rev.Lett.99(2007)166804B.Oezyilmaz, Nat.Phys.5(2009)222A.F.Young, Phys.Rev.Lett.102(2009)026807N.Stander, J.Phys.D-Appl.Phys.47(2014)094003K.Tsukagoshi} creating the charge inhomogeneities in the interactions with Ru(0001) substrate,\cite{Phys.Rev.B76(2007)075429S.Marchini, Phys.Rev.Lett.100(2008)056807A.L.Vazquez, Adv.Mater.21(2009)2777Y.Pan} and placing on thin films of ferroelectric BiFeO$_3$.\cite{Appl.Phys.Lett.99(2011)132904Y.Zang, J.Appl.Phys.112(2012)054103Y.Zang, Appl.Phys.Lett.105(2014)142902R.K.Katiyar}
The modulated electric field is relatively easy to reform the different site energies and severely suppresses the cyclotron motion in the magnetic field.
The main effect is to turn the QLLs into the oscillatory subbands with more extra band-edge states.
Moreover, the symmetry of energy spectra, energy spacing, and spatial distribution of Landau wave functions are drastically modified or entirely destroyed.

\begin{figure*}
\begin{center}
  \includegraphics[width=\linewidth, keepaspectratio]{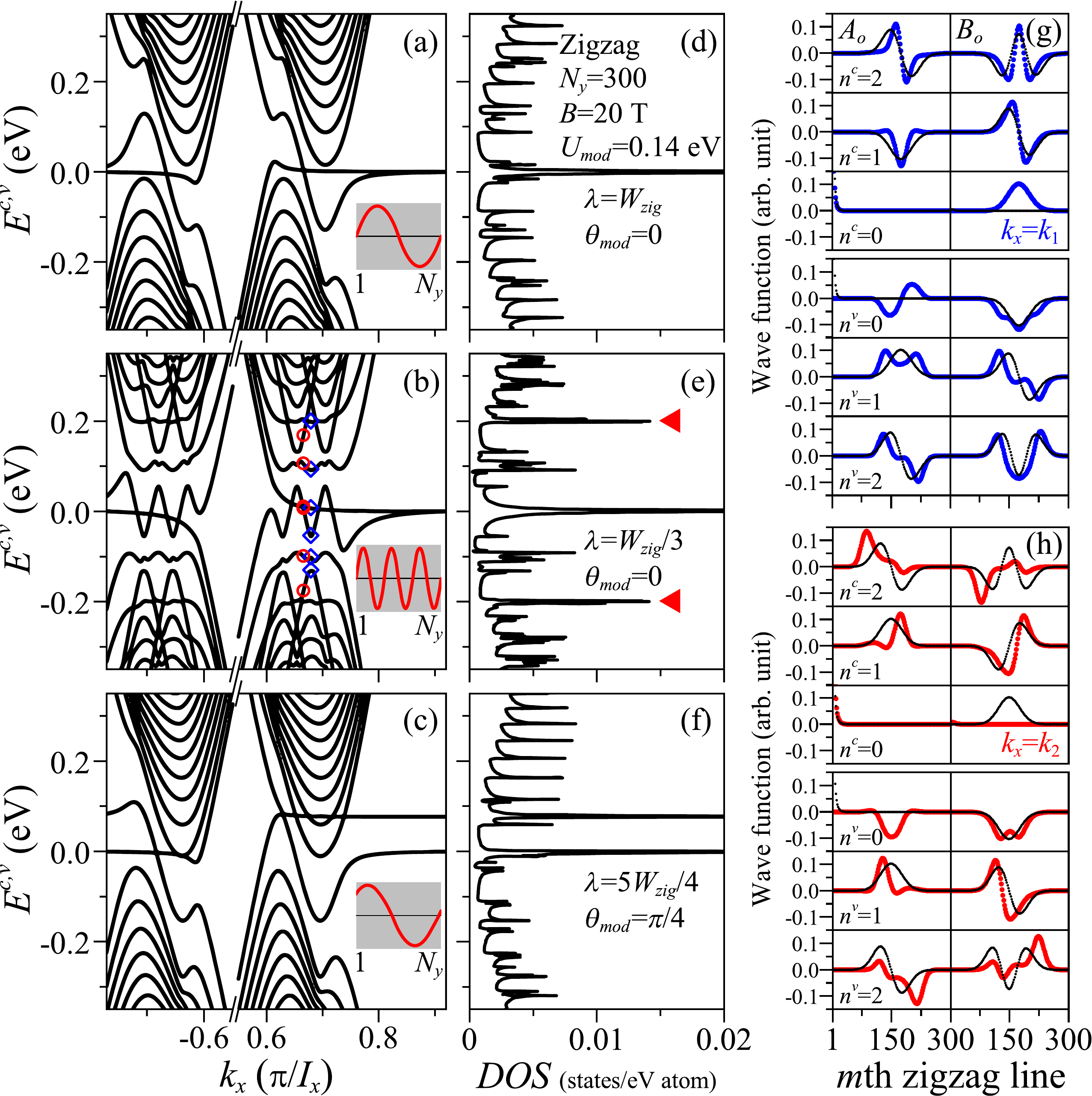}\\
  \caption[]{
  Band structures of the $N_y=300$ ZGNR under the uniform magnetic field $B = 20$ T and modulated electric potential with $U_{mod} = 0.14$ eV, (a) ($\lambda=W_{zig}$, $\theta_{mod}=0$); (b) ($\lambda=W_{zig}/3$, $\theta_{mod}=0$); (c) ($\lambda=W_{zig}$, $\theta_{mod}=\pi/4$).
  Each inset displays the modulated electric potential.
  (d)--(f) The corresponding DOS.
  (g) and (h) The $n^{c,v} = 0$--$2$ wave functions of band-edge states at $k_x = k_1$ (blue diamonds) and non-band-edge states at $k_2$ (red circles) in (b), respectively.
  The regular Landau wave functions are indicated by small black dots.
  }
  \label{fig:BS_DOS_WFofZGNR_Ny300inBandEmod}
\end{center}
\end{figure*}

A modulated electric potential is assumed to be in the form of $U_{mod} \sin[2\pi (y/\lambda) + \theta_{mod})]$, where $U_{mod}$, $\lambda$, and $\theta_{mod}$ are the amplitude, modulation period, and extra phase shift, respectively.\cite{Phys.Lett.A372(2008)5999S.C.Chen}
Since the localization center of Landau wave function moves with the variation of $k_x$, the energy dispersions of oscillatory subbands are closely related to the modulation form of electric potential.
The modulated electric field dominates the $k_x$-dependent energy dispersions so that the subbands oscillate in the form similar to the potential.
Each low-lying subband oscillates once with two extra band-edge states under the sinusoidal potential of one period (Fig.~\ref{fig:BS_DOS_WFofZGNR_Ny300inBandEmod}(a)).
These subbands have a wide-range energy overlap due to the large oscillation amplitudes.
The modulation effect is reduced for the higher-energy subbands.
The amplitudes of the $n^{c,v} = 1$ subbands become smaller than that of the lowest subband; furthermore, the $n^{c,v} \geq 2$ subbands possess only slightly distorted parabolic dispersions.
The increase of potential period reflects on the oscillations of low-lying subbands and brings severe changes to the higher-energy dispersions.
For a period of $\lambda = W_{zig}/3$, the lowest subband oscillates three times with more extra band-edge states (Fig.~\ref{fig:BS_DOS_WFofZGNR_Ny300inBandEmod}(b)).
The higher-energy dispersions exhibit irregular oscillations and energy spacings owing to the mixings of neighboring oscillatory subbands.
In general, the symmetries of electronic properties are mainly determined by the spatial distribution of modulated potential.
For example, the energy spectra in Fig.~\ref{fig:BS_DOS_WFofZGNR_Ny300inBandEmod}(a) \& (b) and Fig.~\ref{fig:BS_DOS_WFofZGNR_Ny300inBandEmod}(c), respectively, illustrate the anti-symmetric and asymmetric band structures.
In addition, the different potential energies at two edges will induce the splitting of degenerate partial flat subbands (Fig.~\ref{fig:BS_DOS_WFofZGNR_Ny300inBandEmod}(c)).

The prominent structures in DOS are sensitive to the modulation period and extra phase shift.
Numerous asymmetric peaks arising from the oscillatory subbands are revealed in Fig.~\ref{fig:BS_DOS_WFofZGNR_Ny300inBandEmod}(d).
With the increase of $\lambda$, the number of peaks is largely enhanced, and some lower peaks merge into a higher one (red triangles in Fig.~\ref{fig:BS_DOS_WFofZGNR_Ny300inBandEmod}(e)).
The symmetric DOS about $\omega = 0$, which results from the anti-symmetric energy dispersion, is absent under the electric potential without spatial symmetry (Fig.~\ref{fig:BS_DOS_WFofZGNR_Ny300inBandEmod}(f)).
Moreover, the highest symmetric peak at $\omega = 0$ (Fig.~\ref{fig:BS_DOS_WFofZGNR_Ny300inBandEmod}(d) and (e)) is split with a potential difference between two edges (Fig.~\ref{fig:BS_DOS_WFofZGNR_Ny300inBandEmod}(f)).
The aforementioned variations of the prominent structures in DOS can be explored by STS measurements;\cite{Phys.Rev.B76(2007)075429S.Marchini, Phys.Rev.Lett.100(2008)056807A.L.Vazquez, Adv.Mater.21(2009)2777Y.Pan} therefore, the severe subband mixing, symmetry breaking in energy dispersions, and splitting of partial flat subbands will be identified.

Also, the wide-range subband mixing indicates drastic modifications in the magnetic wave functions and the severe suppression on cyclotron motion.
The spatial distributions of irregular wave functions (large dots in Fig.~\ref{fig:BS_DOS_WFofZGNR_Ny300inBandEmod}(g) and (h)) are much different from those of the regular ones (small black dots) in amplitudes and node numbers.
Each $n^{c,v} = n$ irregular wave function is the superposition of $n^{c,v} = n$, $n \pm 1$, $n\pm 2$,... Landau wave functions, as evidenced by the finite inner product between the former and the latter.
For instance, the wave function of $B$ sublattice for $n^v = 1$ subband (Fig.~\ref{fig:BS_DOS_WFofZGNR_Ny300inBandEmod}(g)), which possesses four local extrema instead of two, is the superposition of $n^{c,v} = n$ and $n \pm 1$ Landau wave functions.
Such band-mixing-induced modifications and wave function superpositions can be found in the band-edge states ($k_x = k_1$) and non-band-edge states ($k_x = k_2$) as shown in Fig.~\ref{fig:BS_DOS_WFofZGNR_Ny300inBandEmod}(g) and (h), respectively.
Especially, the wave functions of $k_2$ exhibit extreme spatial distributions, \emph{e.g.} the $n^{c} = 2$ ($n^{v} = 2$) wave function is mostly distributed on the left (right) side.

\subsection{Optical properties}

When a GNR is present in an electromagnetic field, with the electric polarization $\hat{\mathbf{e}} \Vert \hat{x}$, electrons are excited from the occupied to the unoccupied states.
The vertical optical transitions with the same wave vector ($\Delta k_x = 0$) are available since the photon momentum is almost zero.
Based on the Fermi's golden rule, the optical absorption function is given by
\begin{align}
\nonumber
& A( \omega ) \propto \\
\nonumber
& \sum_{h,h'=c,v} \sum_{n,n'} \int_{\text{B.Z.}} \frac{dk_{x}}{2\pi}
 \operatorname{Im} \bigg\{ \frac{f[E^{h'}(k_x,n')]-f[E^{h}(k_x,n)]}{E^{h'}(k_x,n')-E^{h}(k_x,n)-\omega-i\Gamma} \bigg\} \\
& \times
\left| \bigg\langle\Psi^{h'}(k_x,n')\bigg|\frac{\hat{\mathbf{e}}\cdot\mathbf{p}}{m_e}\bigg|\Psi^{h}(k_x,n)\bigg\rangle \right| ^2 ,
\label{eq:AbsFormulaForGNR}
\end{align}
where $m_{e}$ is the bare electron mass, $\mathbf{p}$ the momentum operator, $f[E^h(k_x,n)]$ the Fermi-Dirac distribution function, and $\Gamma$ (2 meV) the phenomenological broadening parameter.
The spectral intensity is determined by the joint density of states (the first term) and the velocity matrix element (the second term).
The latter is evaluated within the gradient approximation,\cite{Phys.Rev.160(1967)649G.Dresselhaus, Phys.Rev.B7(1973)2275L.G.Johnson, Phys.Rev.B19(1979)5019N.V.Smith, Phys.Rev.B47(1993)15500L.C.LewYanVoon} as successfully done for other carbon-related systems, such as layered graphites,\cite{Phys.Rev.B7(1973)2275L.G.Johnson, Phys.Rev.B67(2003)165402A.Gruneis} CNTs,\cite{Phys.Rev.B67(2003)165402A.Gruneis, Carbon42(2004)3169J.Jiang} and few-layer graphenes.\cite{NewJ.Phys.12(2010)083060C.W.Chiu, ACSNano4(2010)1465Y.H.Ho}
The velocity matrix element is expressed as
\begin{equation}
M^{hh'}(k_x) = \frac{1}{\hbar} \sum_{{l,l'}=1} C_l^{h'*}(k_x,n') C_{l'}^h(k_x,n) \frac{\partial H_{l,l'}(k_x)}{\partial k_x} ,
\label{eq:VelocityMatrixElementOfGNRs}
\end{equation}
where $C_{l'}^h(k_x,n)$ is the amplitude on lattice site of the subenvelope  function.
$\partial H_{l,l'}/\partial k_x$ related to the nearest-neighboring atomic interactions is non-vanishing for $\mathbf{k}_x \cdot {\delta}_i \neq 0$.
Whether the optical transitions can survive rely on the finite inner product among the initial and final state on different sublattices and $\partial H_{l,l'}/\partial k_x$.
The effects due to the geometric configurations and external fields that diversify the optical spectra will be thoroughly examined.

\subsubsection{Edge-dependent absorption spectra.}
\label{sec:EdgeDependentAbsorptionSpectraOfMonoGNRs}

The low-energy optical spectra of GNRs exhibit a lot of absorption peaks, while that of graphene is featureless due to the linear Dirac cone.\cite{NewJ.Phys.12(2010)083060C.W.Chiu}
Lateral confinement is responsible for the difference; that is, only the former have many vHSs in the DOS.
Also, the edge structures play a crucial role in optical selection rules.
ZGNRs and AGNRs are predicted to have different selection rules.\cite{Phys.Rev.B76(2007)045418H.Hsu, J.Phys.Soc.Jpn.69(2000)3529M.F.Lin}
The former are distinct from the armchair CNTs (ACNTs) in the selection rule, but the opposite is true between the latter and ACNTs.
The edge-dependent selection rules can be realized from the special relations between the regular standing waves of the initial and final states.\cite{Opt.Express19(2011)23350H.C.Chung, Phys.Rev.B84(2011)085458K.Sasaki}
The experimental measurements are required to verify the feature-rich absorption spectra.

The prominent structures of low-frequency absorption spectra at zero temperature are very sensitive to the edge structures and ribbon widths.
\begin{figure}
\begin{center}
  \includegraphics[width=\linewidth, keepaspectratio]{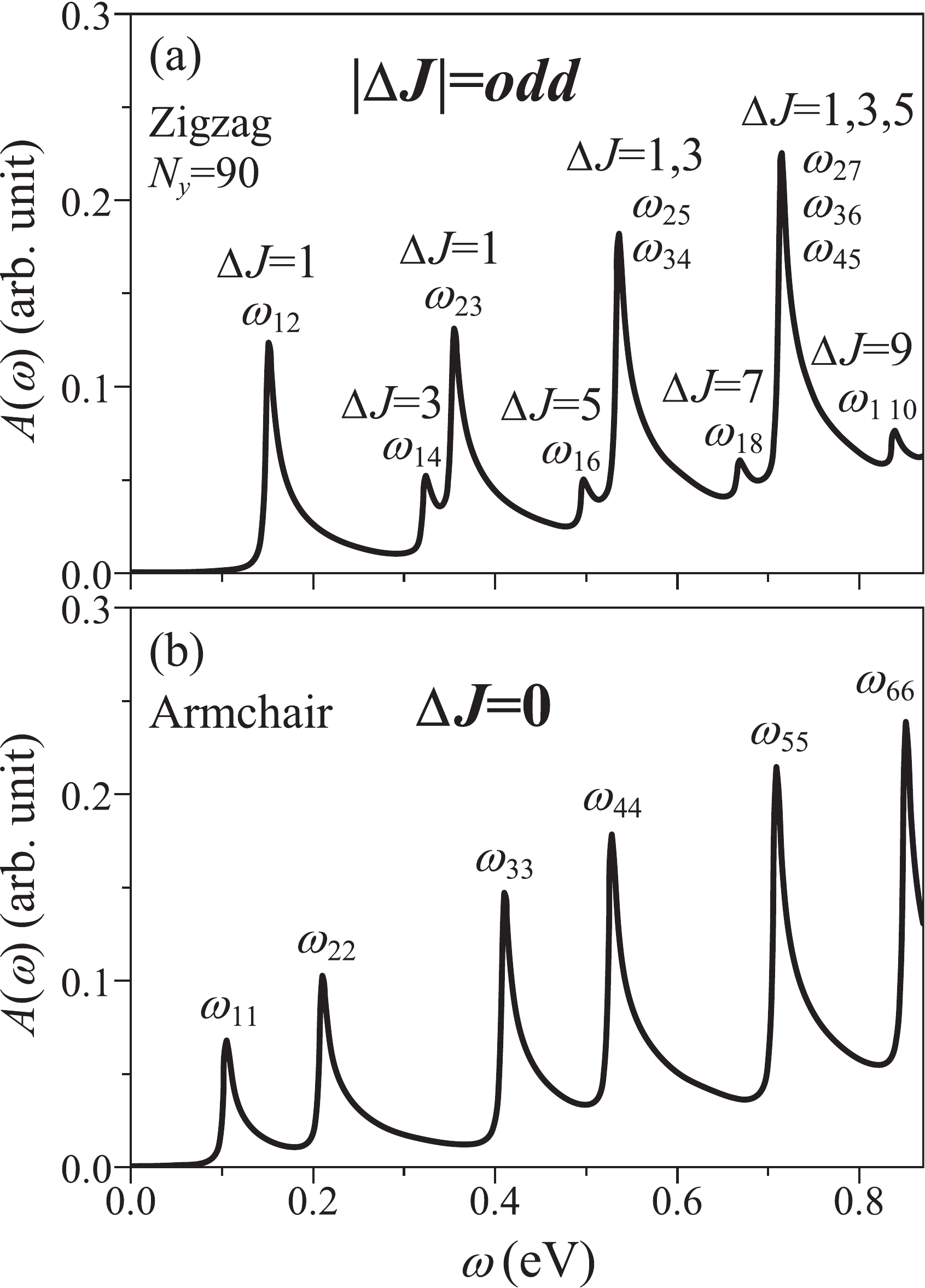}\\
  \caption[]{
  Low-energy absorption spectra of the $N_y=90$ (a) ZGNR and (b) AGNR.
  }
  \label{fig:ABSofGNR}
\end{center}
\end{figure}
A large number of divergent asymmetric peaks ($\omega_{J^vJ^c}$'s), which originate from the vertical interband transitions between the $J^v$th valence and the $J^c$th conduction parabolic subbands, are revealed in the optical spectra (Fig.~\ref{fig:ABSofGNR}).
For ZGNRs, the absorption peaks obey the selection rule, $|\Delta J |= |J^c - J^v| = odd$, \emph{i.e.} the optical transitions are available only between two subbands with an odd index difference (Fig.~\ref{fig:ABSofGNR}(a)).
The optical transition frequencies of $J^v=J \rightarrow J^c=J+1$ and $J^v=J+1 \rightarrow J^c=J$ are the same due to the symmetry of energy dispersions, so that only the transition channels of $J^v \leq J^c$ are displayed.
Moreover, such peaks can be further classified into the higher and lower ones.
The higher principal peaks become strong with the increasing frequency owing to the decrease of subband curvature (Fig.~\ref{fig:BS_DOS_WFofMonoGNRsNy300B0E0}) and the increase of transition channels.
For example, $\omega_{27}$, $\omega_{36}$, and $\omega_{45}$ merge into the fourth principal peak.
Meanwhile, there exists one lower subpeak between two adjacent principal peaks.
These subpeaks correspond to the transitions from the $J^v = 1$ valence subband to the $J^c = 4$, $6$, $8$,... conduction subbands.
That the initial subband has a very large curvature near $k_x = 2/3$ is the main reason for the weak spectral intensity.
As to AGNRs, the selection rule of $\Delta J = 0$, indicates that only the vertical excitations from the valence to conduction subbands of the same index are allowed (Fig.~\ref{fig:ABSofGNR}(b)).
The peak positions and the threshold frequency, which associate with the energy spacings between two available subbands, will grow quickly for narrow GNRs.
It is remarkable that the available optical excitations for GNRs and CNTs are quite different.
The selection rules of GNRs are edge-dependent, while all the CNTs possess an identical selection rule similar to that of AGNRs (Section~\ref{sec:MagnetoElectronicPropertiesOfCNTs}).

The peculiar relations in the spatial distributions of regular standing waves are responsible for the edge-dependent selection rules, especially the relations between two equivalent $A$ and $B$ sublattices as well as the relations between the conduction and valence states.
The available optical transitions are determined by the finite product among the initial state on the $A$ ($B$) sublattice, the final state on the $B$ ($A$) sublattice, and $\partial H_{l,l'}/\partial k_x$ (eqn~(\ref{eq:VelocityMatrixElementOfGNRs})).
For any $k_x$ states in ZGNRs, $\partial H_{l,l'}/\partial k_x$ due to the  nearest-neighbor interactions is a constant for ${\delta}_2$ and ${\delta}_3$, but vanishes for ${\delta}_1$.
Moreover, when the difference of subband index is an odd number, the symmetric properties of the spatial distribution are the same for the conduction $A$ ($B$) sublattice and valence $B$ ($A$) sublattice (Fig.~\ref{fig:BS_DOS_WFofMonoGNRsNy300B0E0}).
The selection rule of $|\Delta J| = odd$ can be derived by the detailed analytical calculations.\cite{Opt.Express19(2011)23350H.C.Chung, Phys.Rev.B84(2011)085458K.Sasaki}
On the other hand, as to the band-edge states of AGNRs, $\partial H_{l,l'}/\partial k_x$ for ${\delta}_1$ is twice that for ${\delta}_2$ and ${\delta}_3$.
Furthermore, the selection rule of $\Delta J = 0$ can be obtained with the following two special relations, including the in-phase/out-of-phase relations between $A$ and $B$ sublattices in a specific subband, and between the conduction and valence subenvelope functions of an identical subband index (Section~\ref{sec:ElectronicStructureOfGrapheneInNoBandE}).
The edge-dependent selection rules need further experimental verifications.

\subsubsection{Magneto-absorption spectra.}
\label{MagnetoAbsorptionSpectra}

Magnetic fields induce the highly degenerate QLLs and change the regular standing waves into the Landau wave functions.
The well-behaved Landau wave functions have the unique spatial symmetry, so that the available optical transitions associated with the symmetric absorption peaks are governed by the QLL-dependent selection rule.
However, the competition between the lateral confinement and magnetic quantization will be revealed at the higher-frequency asymmetric peaks.
The coexistence of the QLL-dependent and edge-dependent absorption peaks relys on the ribbon width and field strength.
On the experimental aspect, it is difficult to measure the optical properties from one single nanoscale GNR because of the low-intensity optical responses.
Recently, the array of GNRs can be fabricated lithographically on layered graphene via the optical method,\cite{Nat.Nanotechnol.6(2011)630L.Ju, Adv.Mater.25(2013)4723J.G.Son} gas phase etching,\cite{Nat.Chem.2(2010)661X.Wang} electron beam,\cite{Nat.Commun.4(2013)1951M.Freitag, Nat.Photonics7(2013)394H.Yan, Carbon61(2013)229C.H.Huang, Phys.Rev.Lett.110(2013)246803J.M.Poumirol} helium ion beam,\cite{ACSNano8(2014)1538A.N.Abbas} and self-assembly block copolymer template.\cite{ACSNano6(2012)9700X.Liang, ACSNano6(2012)6786G.Liu}
Its optical response is sufficiently strong for experimental measurements.
Part of the theoretical predictions on the magneto-optical properties have been verified.\cite{Phys.Rev.Lett.110(2013)246803J.M.Poumirol}

\begin{figure}
\begin{center}
  \includegraphics[width=\linewidth, keepaspectratio]{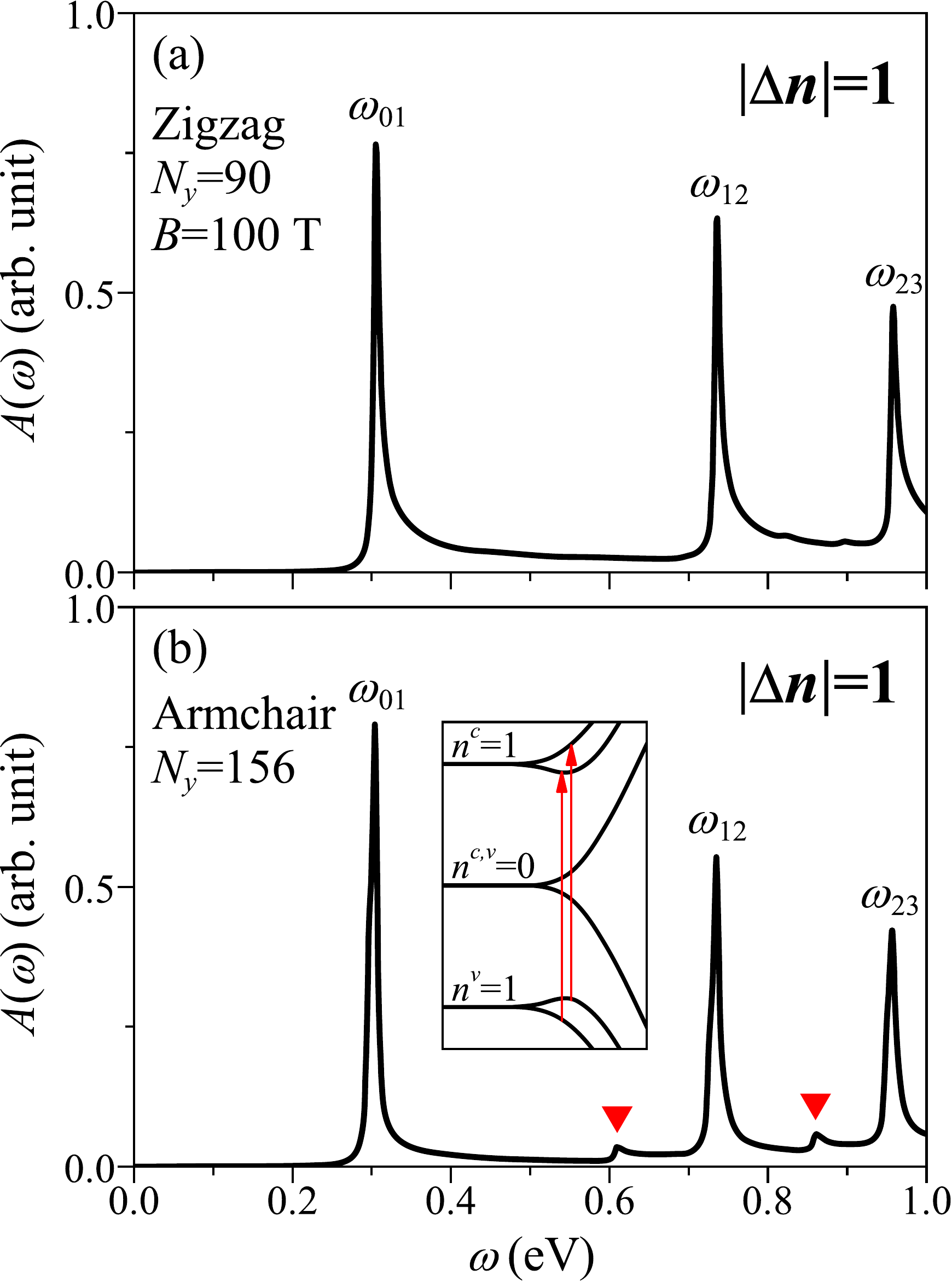}\\
  \caption[]{
  Low-frequency magneto-absorption spectra of the (a) $N_y = 90$ ZGNR and (b) $N_y = 156$ AGNR.
  Inset shows the optical transitions due to the extra band-edge states.
  }
  \label{fig:ABSofGNRinB}
\end{center}
\end{figure}

GNRs have an edge-independent selection rule in the low-frequency magneto-absorption spectra, as shown in Fig.~\ref{fig:ABSofGNRinB}.
There are many prominent symmetric peaks ($\omega_{n^v n^c}$'s) corresponding to the vertical transitions from the $n^v$ QLL to the $n^c = n^v \pm 1$ QLL.
The peak height, which is proportional to the width of QLL, declines with the increasing frequency.
The valence and conduction Landau wave functions have similar spatial distributions, and the oscillation modes on the $A$ and $B$ sublattices for any subbands differ by one (Section~\ref{sec:MagneticQuantizationInMonolayer}).
In other words, the Landau modes for the different sublattices of valence and conduction subbands with adjacent indices are the same.
The velocity matrix elements have finite values for the initial and final states with the same mode, so that the optical transitions can survive for the subbands satisfying the magneto-optical selection rule, $|\Delta n| = | n^c - n^v | = 1$.\cite{Nanotechnol.18(2007)495401Y.C.Huang, J.Appl.Phys.103(2008)073709Y.C.Huang}
It is obvious that the transition frequencies exhibit the simple relation $\sqrt{B}(\sqrt{n^{v}} + \sqrt{n^{v} \pm 1})$ based on the $\sqrt{n^{c,v}B}$ relation in state energy and the QLL-dependent selection rule of $|\Delta n| = 1$.
In addition, the lower additional absorption peaks in AGNRs (red triangles) originate from the vertical transitions associated with the extra band-edge states (inset in Fig.~\ref{fig:ABSofGNRinB}(b)).

\begin{figure}
\begin{center}
  \includegraphics[width=\linewidth, keepaspectratio]{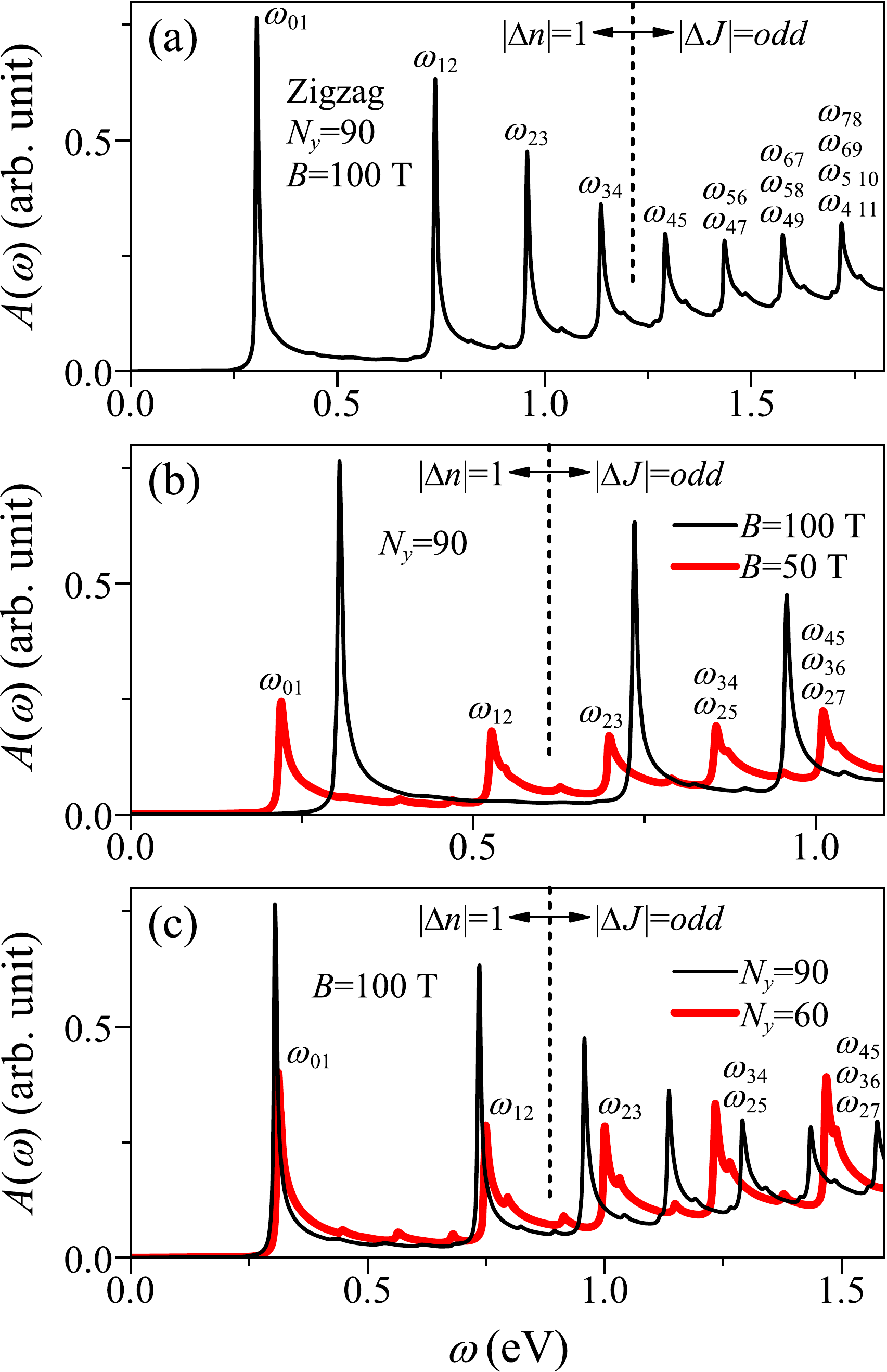}\\
  \caption[]{
  Coexistence of the QLL- and edge-dependent optical transitions.
  The magneto-absorption spectra of the ZGNR at (a) ($N_y = 90$, $B = 100$ T); (b) ($N_y = 90$, $B = 50$ T); (c) ($N_y = 60$, $B = 100$ T).
  The dashed lines separate the distinct selection rules.
  The transition channels are indicated.
  }
  \label{fig:SurvivalRegionOfMagnetoOpticalSelectionRule}
\end{center}
\end{figure}

The edge- and QLL-dependent selection rules govern the higher- and lower-frequency optical excitations, respectively.
With the increase of frequency, the spectral intensity gradually descends, and the symmetric peaks transform into the asymmetric ones.
The former arise from the transitions between QLLs of $|\Delta n| = 1$, while the latter composed of multiple transition channels correspond to parabolic subbands of $|\Delta J| = odd$, as shown respectively on the left and right sides of the dashed line in Fig.~\ref{fig:SurvivalRegionOfMagnetoOpticalSelectionRule}(a).
The governing regions for the two selection rules are affected by the field strength and ribbon width.
The lower the magnetic field strength is, the smaller the critical frequency is.
The number, height, and transition frequency of symmetric peaks on the left side of dashed line are reduced (Fig.~\ref{fig:SurvivalRegionOfMagnetoOpticalSelectionRule}(b)).
The higher QLL transition peaks are replaced by the multi-channel asymmetric ones, \emph{e.g.} the fourth peak of $\omega_{34}$ and $\omega_{25}$.
On the other hand, the critical frequency also exhibits the red shift with the decreasing ribbon width.
The peak intensities of $\omega_{01}$ and $\omega_{12}$ are lowered, while their frequencies remain the same (Fig.~\ref{fig:SurvivalRegionOfMagnetoOpticalSelectionRule}(c)).
In addition, the weak bump structures come from the vertical excitations related to the $n^{c,v} = 0$ subbands.

There exist certain important differences between 2D graphene and quasi-1D GNR in magneto-absorption spectra.
As to graphene, the Landau absorption peaks have the uniform intensity, $\sqrt{B}$-dependent transition energy, and magneto-optical selection rule of $|\Delta n| = 1$.\cite{Phys.Rev.Lett.98(2007)157402V.P.Gusynin, Phys.Rev.B77(2008)115313M.Koshino}
These unique features have been confirmed through the magneto-optical transmission\cite{Phys.Rev.Lett.97(2006)266405M.L.Sadowski, Phys.Rev.Lett.98(2007)197403Z.Jiang, Phys.Rev.B76(2007)081406R.S.Deacon, Phys.Rev.Lett.100(2008)087401P.Plochocka} and magneto-Raman spectroscopy.\cite{NanoLett.14(2014)4548S.Berciaud}
On the other hand, the main differences induced by the competition between magnetic quantization and lateral confinement include the structure and intensity of absorption peaks, deviation of $\sqrt{B}$-dependence, and coexistence of selection rules.
Part of the theoretical predictions in GNRs have been recently verified by the infrared transmission measurement on epitaxial GNR arrays,\cite{Phys.Rev.Lett.110(2013)246803J.M.Poumirol} mainly for the threshold absorption peak of $\omega_{01}$ in the magnetic-field strength.
For narrow GNRs, $\omega_{01}$ does not obey the $\sqrt{B}$-dependence at low magnetic field, but the opposite is true at high one.
Especially, the coexistent selection rules are worthy of further experimental measurements.

\subsubsection{Effects of electric fields.}
\label{EffectsOfElectricFieldsOfMonolayerABS}

The special competition among the electric field, lateral confinement, and magnetic quantization drastically modifies electronic properties and thus optical spectra.
For a sufficiently strong electric field, energy dispersions are highly distorted with more extra band-edge states or collapsed with zero energy spacings.
Moreover, the wave functions exhibit the mixing of neighboring electronic states and extremely irregular distributions.
During the variation of field strength, the diverse optical spectra are revealed, including the gradual suppression of prominent absorption peaks with edge- or QLL-dependent selection rules, the generation of lower peaks with extra selection rules, the coexistence of comparable peaks with different selection rules, the creation of numerous peaks without selection rules, and the almost vanishing optical transitions.
The electric-field-induced rich spectra can be explored via the optical experiments\cite{Phys.Rev.Lett.110(2013)246803J.M.Poumirol, Phys.Rev.Lett.97(2006)266405M.L.Sadowski, Phys.Rev.Lett.98(2007)197403Z.Jiang, Phys.Rev.B76(2007)081406R.S.Deacon, Phys.Rev.Lett.100(2008)087401P.Plochocka} on the side-gated GNRs.\cite{Phys.Rev.B76(2007)245426F.Molitor, Small4(2008)716J.F.Dayen, Appl.Phys.Lett.101(2012)093504B.Hahnlein, IEEEElectronDeviceLett.33(2012)330C.T.Chen, IEEETrans.ElectronDevices61(2014)3329L.T.Tung, Sci.Rep.4(2014)5581V.Panchal, IEEEJ.Electr.Dev.Soc.3(2015)144L.T.Tung}

\begin{figure}
\begin{center}
  \includegraphics[width=\linewidth, keepaspectratio]{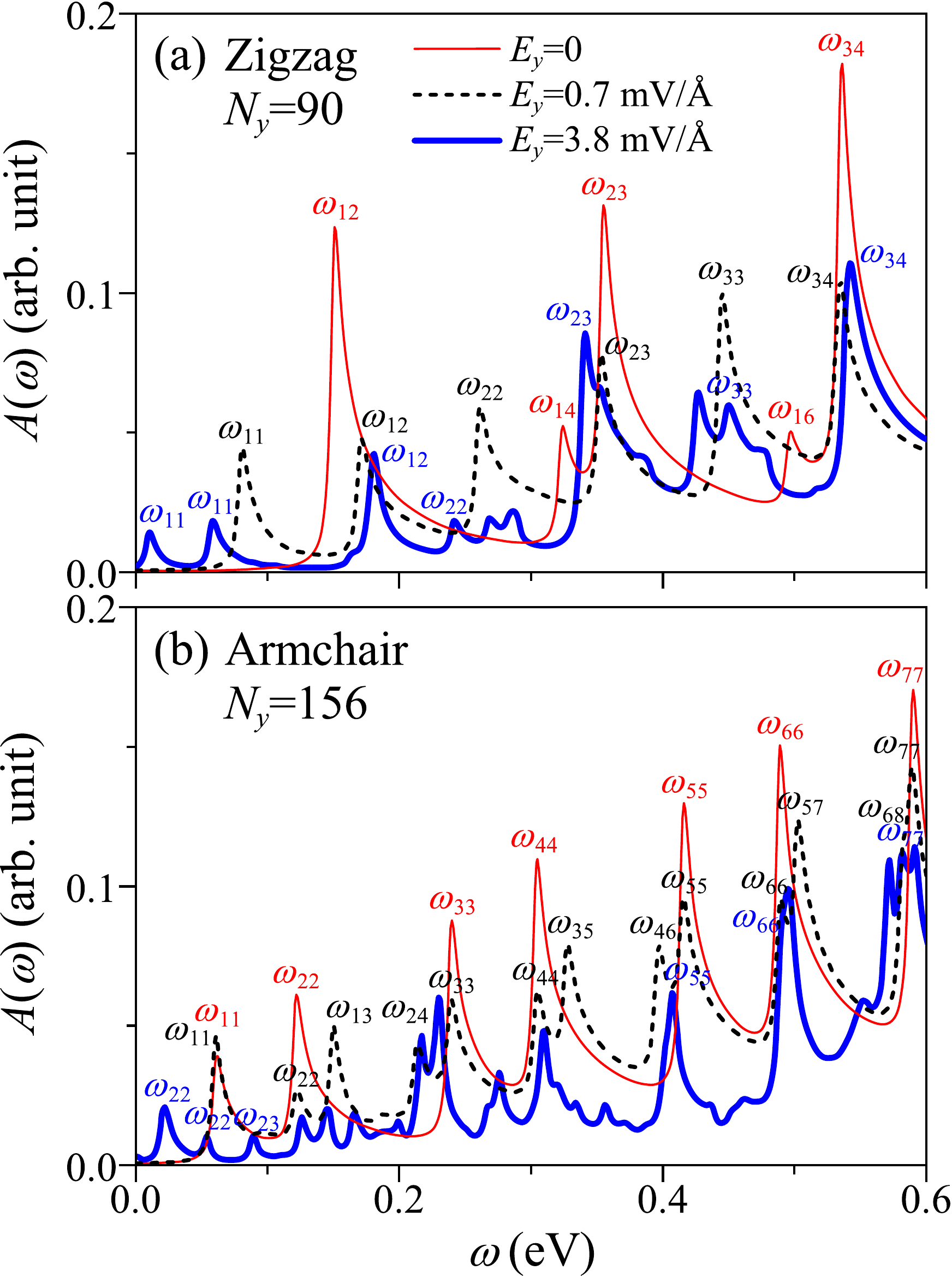}\\
  \caption[]{
  Low-energy absorption spectra of the ZGNR and AGNR in various electric fields $E_y = 0$ (light red curves), $E_y = 0.7$ mV/{\AA} (dashed black curves), and $E_y = 3.8$ mV/{\AA} (heavy bule curves).
  The transition channels for peaks obeying the selection rules are given.
  }
  \label{fig:ABSofMonoGNRsInE}
\end{center}
\end{figure}

Electric field can enrich absorption spectra by competing with the lateral confinement.\cite{Carbon44(2006)508C.P.Chang}
With the increase of electric field, the prominent peaks of the edge-dependent selection rules are lowered, and the subpeaks satisfying the extra selection rules come to exist.
Two kinds of peaks coexist and their intensities become comparable.
For ZGNRs, the absorption peaks obey the edge-dependent selection rule of $|\Delta J| = odd$ and the extra one of $\Delta J = 0$ (dashed black curve in Fig.~\ref{fig:ABSofMonoGNRsInE}(a)).
Similarly, AGNRs have the coexistent optical transitions corresponding to the edge-dependent and extra selection rules of $\Delta J = 0$ and $|\Delta J| = 2$, respectively (dashed black curve in Fig.~\ref{fig:ABSofMonoGNRsInE}(b)).
The electric field leads to the mixing of adjacent standing waves in the distorted wave functions so that the extra optical transitions are available.
Selection rules are thoroughly destroyed at a very large electric field (heavy bule curves in Fig.~\ref{fig:ABSofMonoGNRsInE}), in which many absorption peaks come from the band-edge states of oscillatory subbands (Fig.~\ref{fig:BSofGNRsB0Evarious}) associated with the serious mixing of several neighboring electronic states.

\begin{figure}
\begin{center}
  \includegraphics[width=\linewidth, keepaspectratio]{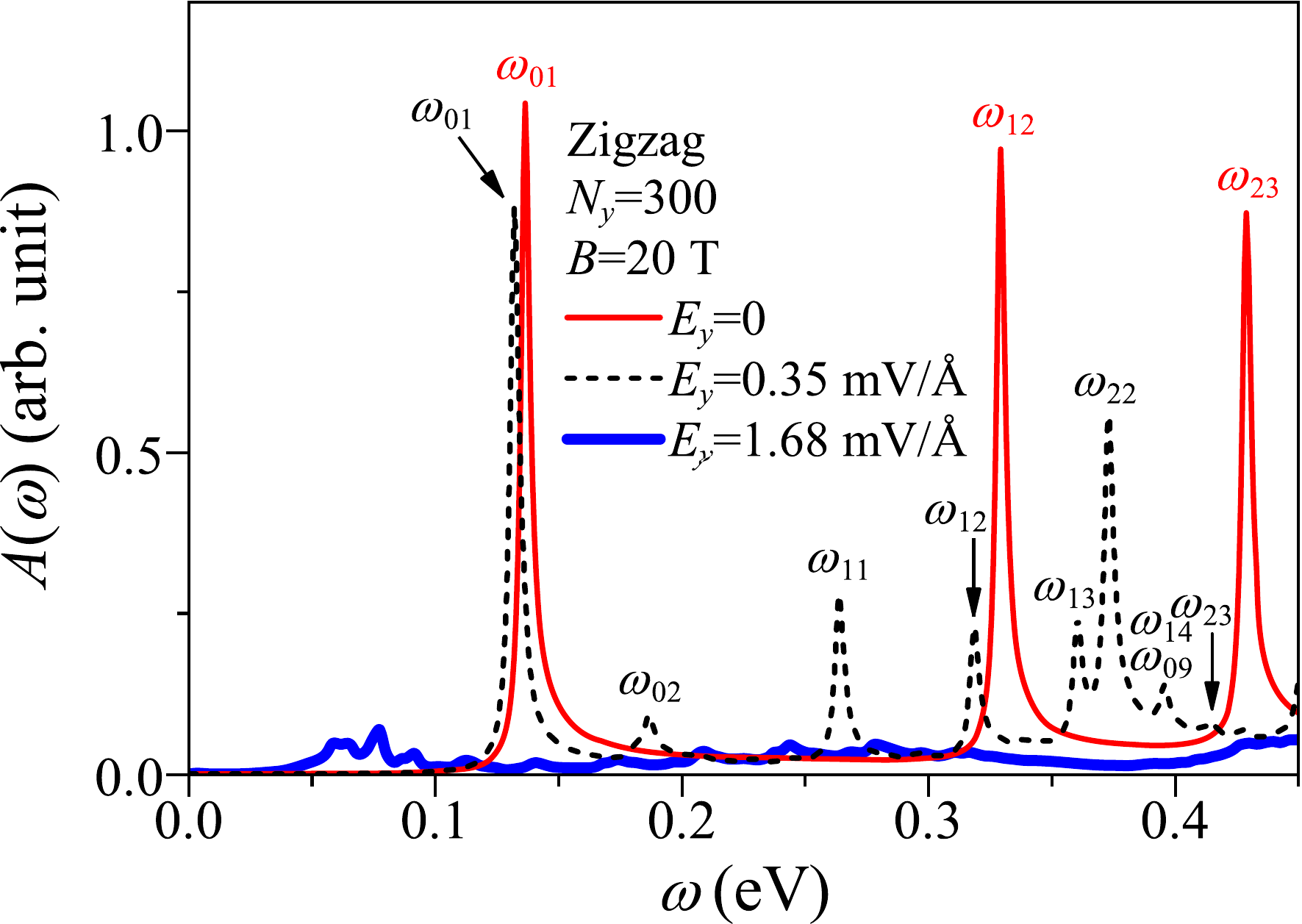}\\
  \caption[]{
  Low-energy magneto-absorption spectra of the ZGNR in $B = 20$ T and various electric fields $E_y = 0$ (light red curve), $E_y = 0.35$ mV/{\AA} (dashed black curve), and $E_y = 1.68$ mV/{\AA} (heavy bule curve).
  The arrows indicate the Landau absorption peaks.
  }
  \label{fig:ABSofMonoGNRInBandEcollapse}
\end{center}
\end{figure}

The electric field can suppress the magnetic quantization and diversify the magneto-optical spectra.
The intensity of Landau absorption peaks is reduced, especially for the higher-frequency ones.
Meanwhile, many additional optical transitions between the tilted QLLs are observable (dashed black curve in Fig.~\ref{fig:ABSofMonoGNRInBandEcollapse}).
The distorted Landau wave functions are responsible for the suppression of $|\Delta n| = 1$ Landau transitions, and the constant energy spacing between the tilted QLLs (Fig.~\ref{fig:BSandDOSofGNRsInBandE}) stands for the sufficiently strong extra peaks.
Moreover, the optical excitations from tilted QLLs are entirely suppressed at the critical electric field (heavy blue curve in Fig.~\ref{fig:ABSofMonoGNRInBandEcollapse}), where the spatial distribution of Landau wave functions are severely distorted and the collapsed QLLs have zero spacing (Fig.~\ref{fig:TheCollapseOfQLLsInBandE}).
The remaining low-intensity subpeaks are related to the band-edge states.

\section{Curved systems}
\label{Sec:CurvedSystem}

A flexible carbon membrane, composed of $sp^2$ bonding and exhibiting a high Young's modulus,\cite{Science321(2008)385C.Lee} can be easily bent without loosing its unique electronic and atomic structural properties.
These curled and folded $sp^2$-hybridized materials contain curved GNRs,\cite{J.Phys.Soc.Jpn.81(2012)064719C.Y.Lin} folded GNRs,\cite{Phys.Rev.Lett.105(2010)106802E.Prada} fullerenes,\cite{Nature318(1985)162H.W.Kroto, Chem.Rev.91(1991)1213H.W.Kroto} CNTs,\cite{Nature354(1991)56S.Iijima, Nature363(1993)603S.Iijima} carbon tori,\cite{Nature385(1997)780J.Liu, Science293(2001)1299M.Sano} as well as  folded,\cite{Phys.Rev.Lett.102(2009)015501Z.Liu, Phys.Rev.Lett.104(2010)166805J.Zhang,  NanoLett.12(2012)5207L.Ortolani, Nat.Commun.5(2014)4709C.Cong} rippled,\cite{Nat.Nanotechnol.4(2009)562W.Bao, NewJ.Phys.12(2010)093018B.Borca} bubbled,\cite{J.Appl.Phys.110(2011)064308S.Goler, Nat.Commun.3(2012)823J.Lu, Nat.Commun.4(2013)1556C.H.Y.X.Lim, Angew.Chem.Int.Ed.53(2014)215C.H.Y.X.Lim, NanoLett.15(2015)6162G.Zamborlini} and edge-scrolled\cite{Phys.Rev.Lett.108(2012)166602A.Cresti, Nature446(2007)60J.C.Meyer, Science299(2003)1361L.M.Viculis, NanoLett.9(2009)2565X.Xie} graphenes.
These systems possess specific geometric symmetries with dimensionalities ranging from zero- to two-dimension and curved surfaces with either open or closed boundary.

Magnetic quantization on curved surfaces is one of the mainstream research topics, and motivates many theoretical and experimental studies, such as the QLLs in curved GNRs,\cite{J.Phys.Soc.Jpn.81(2012)064719C.Y.Lin} periodical Aharonov-Bohm oscillations in CNTs,\cite{J.Phys.Soc.Jpn.62(1993)1255H.Ajiki, J.Phys.Soc.Jpn.65(1996)505H.Ajiki, Phys.Rev.B62(2000)16092S.Roche, Phys.Rev.B67(2003)045405F.L.Shyu, Science304(2004)1129S.Zaric} persistent currents in carbon tori,\cite{J.Phys.Soc.Jpn.67(1998)1094M.F.Lin, Phys.Rev.B57(1998)6731M.F.Lin, Phys.Rev.B70(2004)075411C.C.Tsai, J.Phys.-Condes.Matter18(2006)8313F.L.Shyu} quantum Hall effects in graphenes with scrolled edges,\cite{Phys.Rev.Lett.108(2012)166602A.Cresti} diamagnetism in $\mathrm{C}_{60}$ and $\mathrm{C}_{70}$,\cite{Chem.Phys.Lett.165(1990)79P.W.Fowler, Chem.Phys.Lett.169(1990)362R.C.Haddon, J.Phys.Chem.95(1991)3457R.S.Ruoff, Nature350(1991)46R.C.Haddon, J.Am.Chem.Soc.115(1993)7876M.Prato, Chem.Phys.Lett.238(1995)270R.Zanasi, Nature378(1995)249R.C.Haddon} pseudo-LLs in rippled graphenes,\cite{Phys.Rev.B77(2008)075422F.Guinea, Euro.Phys.Lett.84(2008)17003T.O.Wehling, SolidStateCommun.149(2009)1140F.Guinea, Phys.Rep.496(2010)109M.A.H.Vozmediano, Science329(2010)544N.Levy, Phys.Rev.B85(2012)035422H.Yan, Nat.Commun.3(2012)1068D.Guo, Phys.Rev.B87(2013)205405L.Meng} and QLLs in folded GNRs.\cite{Phys.Rev.Lett.105(2010)106802E.Prada}
Recently, the curved GNRs can be fabricated by the unzipping of CNTs\cite{Nature458(2009)872D.V.Kosynkin, Nature458(2009)877L.Jiao} (Section~\ref{sec:Introduction}).
In the two systems, the relationships among the curvatures, boundary conditions, and magnetic fields deserve a detailed investigation.
Curvatures change the magnetic flux passing through the curled honeycomb lattice, and the effects of the quantization become position-dependent.
Also, the non-parallel alignment of the $2p_z$ orbitals leads to the decrease (appearance) of $\pi$ ($\sigma$) bondings and the modification of hopping integrals.
The energy dispersions of Landau subbands and the spatial distributions of wave functions will be severely altered.
On the other hand, the electronic states of a cylindrical CNT with a periodic boundary condition are characterized by the quantized angular momenta, $J$'s.
A perpendicular magnetic field induces the coupling of angular momenta, and thus the energy subbands split and the circular standing waves drastically change.
Whether the QLLs will survive is mainly determined by the tube diameter and field strength.
Moreover, the prominent magneto-absorption peaks abide by the QLL-dependent and angular-momentum-dependent selection rules.
The main features of the peaks, including the position, height, number, and structure, will be notably diversified.

The organization of this section is stated as follows.
The first two subsections describe the magneto-electronic properties of curved GNRs and CNTs, respectively.
The effects of curvatures and boundary conditions on the energy dispersions, DOS, and wave functions are presented.
In the last subsection, the curvature- and boundary-induced variations in magneto-optical properties are provided.
Detailed comparisons are also made between the theoretical results and experimental observations.

\subsection{Magneto-electronic properties of curved GNRs}
\label{sec:MagnetoElectronicPropertiesOfCurvedGNR}

\begin{figure}
\begin{center}
  \includegraphics[width=\linewidth, keepaspectratio]{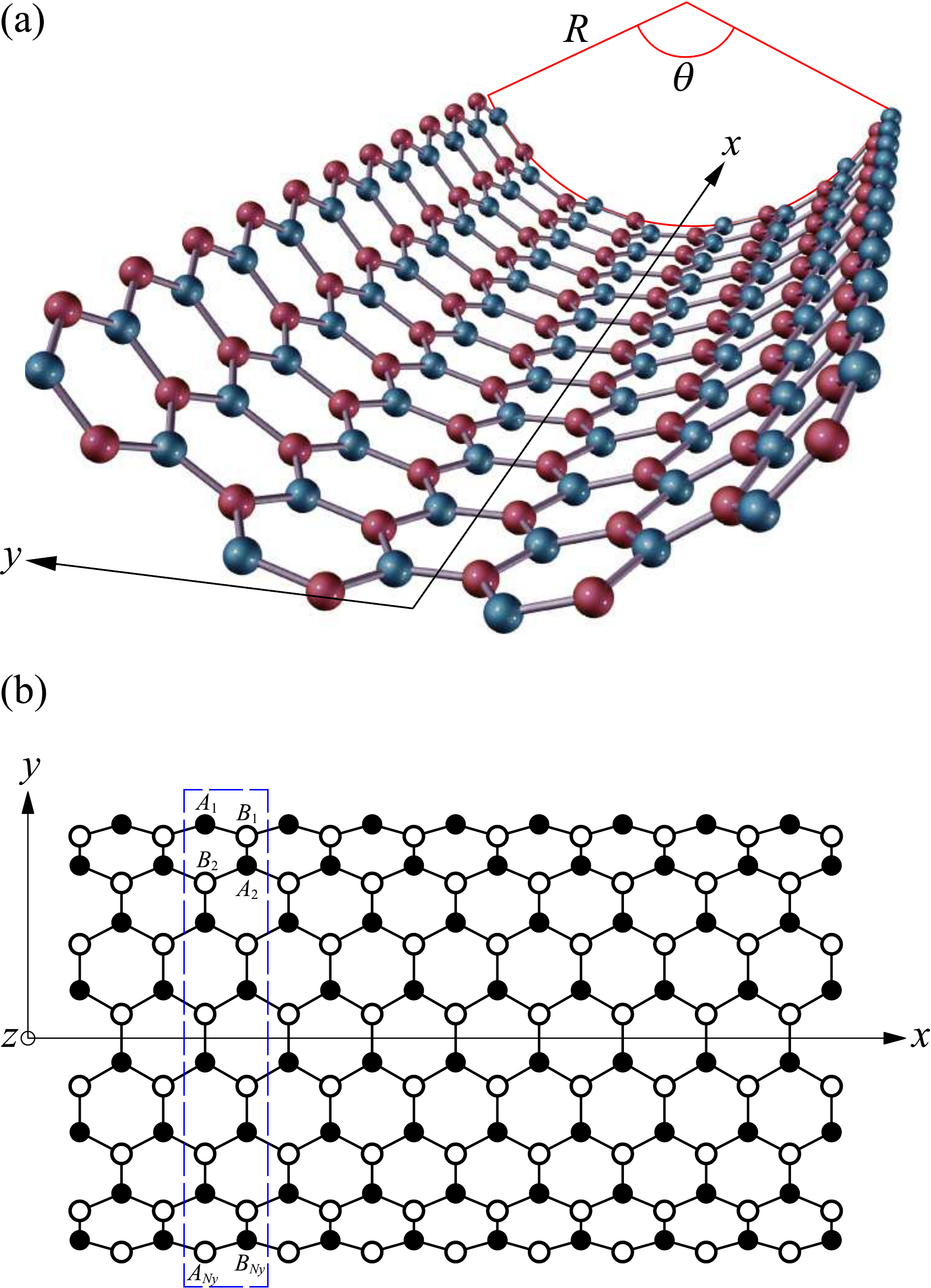}\\
  \caption[]{
  Geometric structure of a curved ZGNR: (a) side view and (b) top view. $R$ and $\theta$ represent the curvature radius and central angle, respetively. The primitive unit cell is enclosed by the dashed rectangle, indicating $A_1$, $B_1$,..., $A_{N_y}$, and $B_{N_y}$ atoms.
  }
  \label{fig:GeometricStructureOfCurvedZGNR}
\end{center}
\end{figure}

A curved GNR can be characterized by the curvature radius ($R$) and the central angle ($\theta$), as depicted in Fig.~\ref{fig:GeometricStructureOfCurvedZGNR}(a).
The misorientation of $2p_z$ orbitals in the curved surface leads to the change of hopping integral between two nearest-neighbor atoms as described
\begin{align}
\nonumber
\gamma_{\alpha}(\theta_{\alpha})=
V_{pp\pi }\{\cos (\theta_{\alpha})+4[(R/b)\sin ^{2}({\theta_{\alpha}}/{2})]^{2}\} \\
-4V_{pp\sigma}[(R/b) \sin ^{2}({\theta_{\alpha}}/{2})]^{2} ,
\end{align}
where $\alpha$ ($= 1$, $2$, $3$) corresponds to the three nearest neighbors, and $\theta_\alpha$ ($\theta_1 = b/R$; $\theta_2 = \theta_3 =b/2R$) is the arc angle between two nearest-neighbor atoms.\cite{Carbon69(2014)151C.Y.Lin}
The first and the second terms indicate $\pi$ and $\sigma$ hopping energies, respectively.
The Slater-Koster parameters $V_{pp\pi }=-2.66$ eV and $V_{pp\sigma}=6.38$ eV,\cite{Phys.Rev.B57(1998)15037J.C.Charlier} and $\theta_{\alpha}$ indicates the arc angle between orbitals.
The curvature effect has been confirmed to modify the electronic properties of CNTs,\cite{Carbon33(1995)893J.W.Mintmire, Phys.Rev.Lett.78(1997)1932C.L.Kane, J.Phys.Soc.Jpn.71(2002)1820F.L.Shyu} as well as been adapted to explain the gap opening in zigzag CNTs with small diameters\cite{Science292(2001)702M.Ouyang}.
When a uniform magnetic field $\mathbf{B}=B\hat{z}$ is applied to the curved GNR, the vector potential is expressed as $\mathbf{A} = B R \sin(\Phi) \hat{x}$ in terms of the cylindrical coordinates ($r, \Phi, x$) and the Peierls phase (eqn~(\ref{eq:PeierlsPhase})) that is added to the wave functions.
An effective magnetic field $B_{eff}=B\cos(\Phi)$, defined as the perpendicular $\mathbf{B}$ component on the curved surface, is introduced to explore the electronic properties, which are enriched and diversified as a result of the competition between the magnetic field and geometric curvature.
In other words, curving a GNR provides an ideal means to study the effects of a spatially modulated magnetic field.

\begin{figure}
\begin{center}
  \includegraphics[width=\linewidth, keepaspectratio]{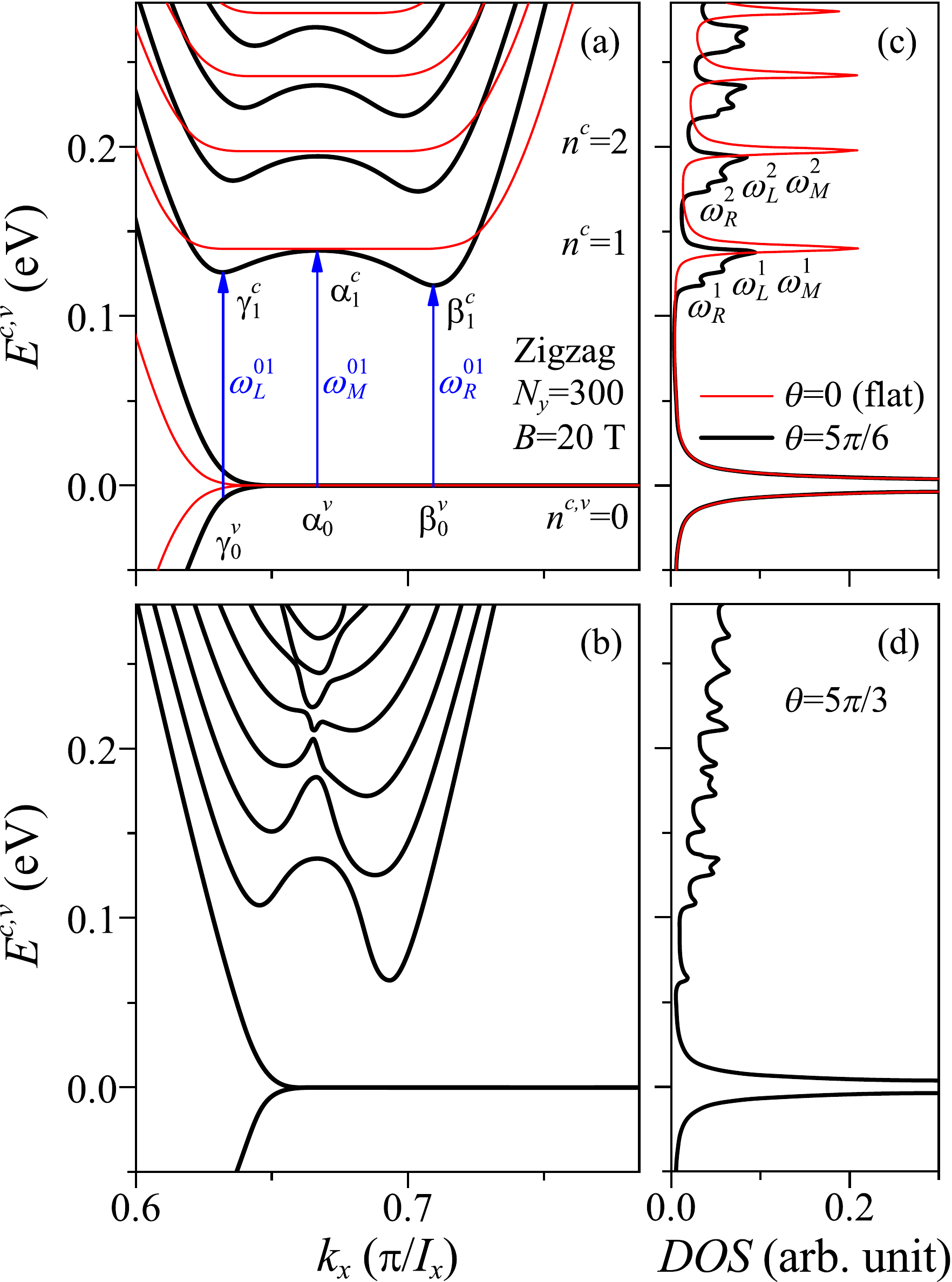}\\
  \caption[]{
  Low-energy magneto-electronic band structures of the $N_y = 300$ flat ZGNR (light red curves) and the curved ZGNR (heavy black curves) at $B = 20$ T for (a) $\theta=5\pi/6$ and (b) $\theta=5\pi/3$; (c) and (d): their corresponding DOSs.
  }
  \label{fig:BSandDOSofCurvedZGNR}
\end{center}
\end{figure}

The magneto-electronic structures of curved GNRs are drastically affected by the effectively non-uniform magnetic field (Fig.~\ref{fig:BSandDOSofCurvedZGNR}(a)).
For a moderate curvature, the dispersionless QLLs are converted into the oscillatory Landau subbands, which reveals that the electronic states hardly congregate at the same energy under the influence of curvature.
Each oscillatory Landau subbands has three band-edge states.
Such states would induce the peaks in the DOS and thus the absorption peaks.
As for the lower $n^{c,v}$ subbands, electronic states in the vicinity of the middle band edges ($k_x = 2/3$) still belong to the Landau states, while those around the other two band edges are drastically changed by the lateral confinement.
With the increment of subband index $n^{c,v}$, the energy spacing between the oscillatory Landau subband and QLL is enlarged due to the reduction of magnetic quantization.
This also leads to a slight decrease in the amplitude of oscillatory Landau subbands.
At higher-energy levels, the band structure exhibits the 1D parabolic subbands.
The aforementioned results directly reflect the difficulty in aggregation of electronic states in a curved GNR.
The described phenomena become more obvious for a curved GNR with larger central angle, where the effective magnetic field changes more rapidly along the $y$-direction.
The large $\theta$ enlarges the subband oscillation and the energy difference between the right and left band edges, while the number and the wavevector range of oscillatory Landau subbands are reduced (Fig.~\ref{fig:BSandDOSofCurvedZGNR}(b)).
In addition, the band-mixing phenomena around $k_x = 2/3$ happen, which result in extra optical transition channels.
On the other hand, the curvature effects only slightly reduce the dispersionless region of the subband at the Fermi energy, since the structure of zigzag edge remains.

Curvature significantly change the structure of DOS.
In the case of a moderate curvature (Fig.~\ref{fig:BSandDOSofCurvedZGNR}(c)), the peak at frequency $\omega = 0$ remains the symmetric peak, since the flat subbands at the Fermi energy are hardly affected.
The other symmetric peaks split into asymmetric ones with lower heights, and the frequencies are related to the extra band-edge states of oscillatory Landau subbands.
The peaks can be categorized into three groups.
$\omega_M^{n^c}$, $\omega_L^{n^c}$, and $\omega_R^{n^c}$ correspond to the energies of the middle, left, and right band-edge states in the $n^c$ oscillatory Landau subband, respectively.
With the increase of the central angle, the peak height is lowered (Fig.~\ref{fig:BSandDOSofCurvedZGNR}(d)).

\begin{figure}
\begin{center}
  \includegraphics[width=\linewidth, keepaspectratio]{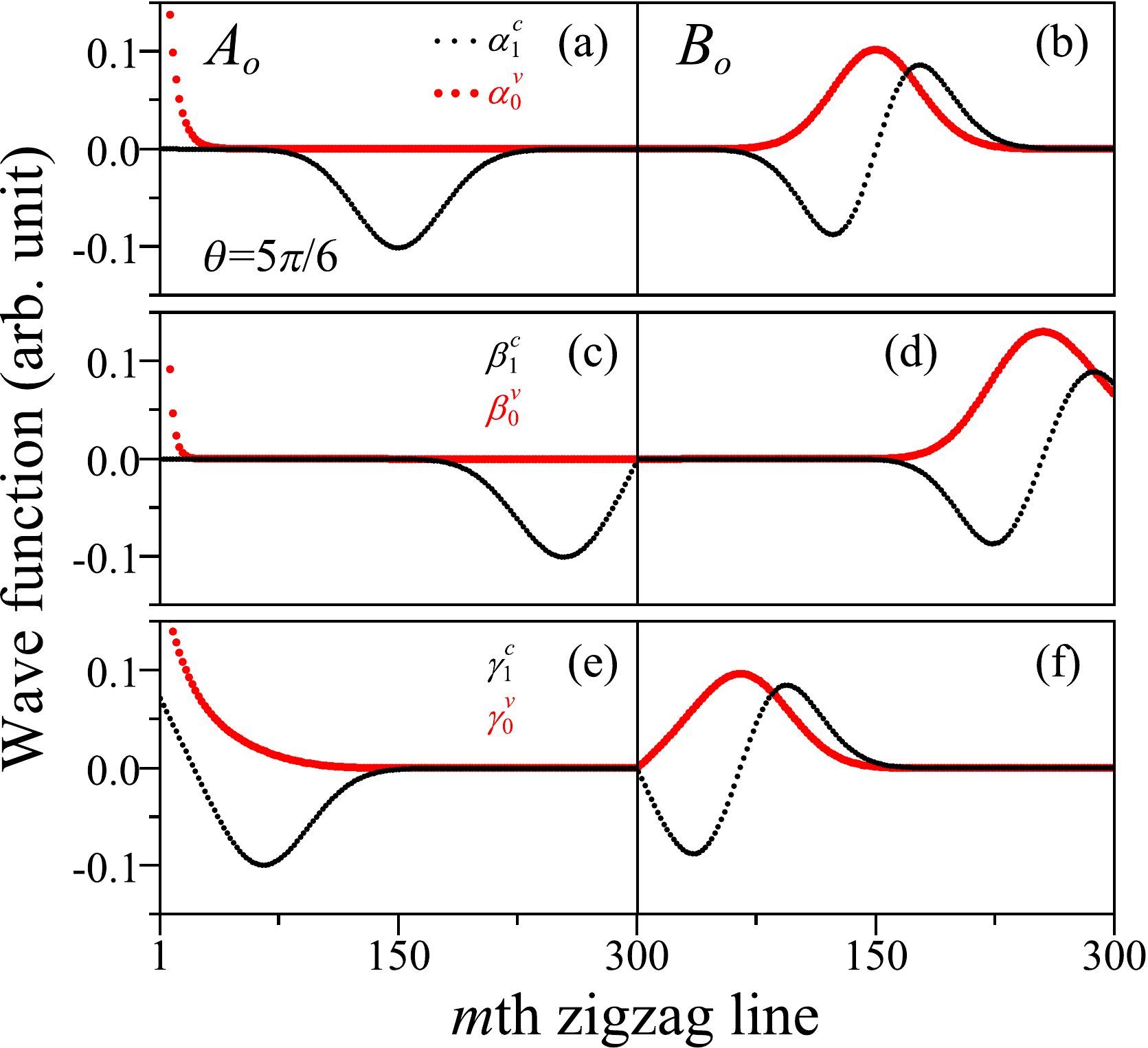}\\
  \caption[]{
  Wave functions of the $n^v=0$ and $n^c=1$ oscillatory Landau subbands for the states marked in Fig.~\ref{fig:BSandDOSofCurvedZGNR}.
  }
  \label{fig:WFofCurvedZGNR}
\end{center}
\end{figure}

The spatial distribution of wave functions, which dominates the selection rules and absorption peaks, strongly depends on the curvature and field strength.
In Fig.~\ref{fig:WFofCurvedZGNR}, the representative wave functions of the $n^v=0$ and $n^c=1$ oscillatory Landau subbands at the middle edge ($\alpha_0^v$, $\alpha_1^c$), right edge ($\beta_0^v$, $\beta_1^c$), and left edge ($\gamma_0^v$, $\gamma_1^c$) are given.
The relationships between the wave functions of the $A$ and $B$ sublattices in different subbands play critical roles in determining the optical transitions.

For the middle band-edge states (Fig.~\ref{fig:WFofCurvedZGNR}(a) and (b)), the wave functions still exhibit the spatial symmetry of Landau wave functions, and the proportional relationship is presented between the wave functions of the $B$ sublattice for the $n^v=0$ subband and the wave functions of the $A$ sublattice for the $n^c=1$ subband.
However, for states deviating from the oscillating center $k_x = 2/3$, the spatial symmetry of Landau wave functions is broken.
As to the right band-edge states (Fig.~\ref{fig:WFofCurvedZGNR}(c) and (d)), the wave functions are localized at one of the ribbon edges, and the distortion and truncation of wave functions are carried out by the open boundaries.
The relationship of wave functions between the $B$ sublattice for the $n^v=0$ subband and the $A$ sublattice for the $n^c=1$ subband still has a strong overlap, and further such overlap causes the nonzero velocity matrix element and related optical transition peak.
For the left band-edge states (Fig.~\ref{fig:WFofCurvedZGNR}(e) and (f)), the wave functions exhibit a similar phenomenon except that the distortions of wave functions occur at the other side of GNR.
These results suggest that the optical transitions between the $n^v = 0$ valence subband and the $n^c = 1$ conduction subband are still alive; however, the oscillation in Landau subbands leads to the multi-peak structure for the transition between two specific subbands.

\subsection{Magneto-electronic properties of CNTs}
\label{sec:MagnetoElectronicPropertiesOfCNTs}

Since a CNT possesses a closed cylindrical structure, its electronic states are characterized by the angular momentum quanta, $J$'s,\cite{Phys.Rev.B46(1992)1804R.Saito, Phys.Rev.B50(1994)17744M.F.Lin, Phys.Rev.B51(1995)7592M.F.Lin, Phys.Rev.B52(1995)8423M.F.Lin, Phys.Rev.Lett.78(1997)1932C.L.Kane} an essential difference in contrast to GNRs with open boundaries.
When an axial magnetic field is applied, a shift of $J$ leads to the metal-semiconductor transition as a result of Aharonov-Bohn effect.
On the other hand, a transverse magnetic field induces a coupling of the independent angular-momentum states, making energy subbands become less dispersive.
Recently, experimental observations on QLLs of CNTs were achieved via magneto-transport measurements in a CNT-based Fabry-Perot resonator under an extremely strong magnetic field.\cite{Phys.Rev.Lett.101(2008)046803B.Raquet, Phys.Rev.Lett.103(2009)256801S.Nanot}

As the polarization of the incident light is parallel to the tube axis, the allowed transitions are those between the occupied and unoccupied states with the same angular momentum, $\Delta J = 0$, which has been theoretically predicted\cite{PhysicaB201(1994)349H.Ajiki, J.Phys.Soc.Jpn.66(1997)3294M.F.Lin, Phys.Rev.B57(1998)9301S.Tasaki, Phys.Rev.B62(2000)13153M.F.Lin,  Phys.Rev.B62(2000)16092S.Roche, Carbon42(2004)3169J.Jiang, SuperlatticesMicrostruct.43(2008)399M.E.Portnoi} and experimentally verified.\cite{Science312(2006)554M.Y.Sfeir, Nat.Commun.4(2013)2542J.C.Blancon}
When an axial magnetic field is applied, the peak splitting and the periodicity of Aharonov-Bohn oscillations,\cite{Phys.Rev.115(1959)485Y.Aharonov, Phys.Rev.Lett.54(1985)2696R.A.Webb, Phys.Rev.Lett.56(1986)792A.Tonomura, Phys.Rev.A34(1986)815N.Osakabe} which are theoretically predicted in magneto-absorption spectra of CNTs\cite{J.Phys.Soc.Jpn.62(1993)1255H.Ajiki, PhysicaB201(1994)349H.Ajiki, J.Phys.Soc.Jpn.65(1996)505H.Ajiki, J.Phys.Soc.Jpn.66(1997)3294M.F.Lin, Phys.Rev.B62(2000)13153M.F.Lin, Phys.Rev.B62(2000)16092S.Roche, Phys.Rev.B67(2003)045405F.L.Shyu} have been verified experimentally,\cite{Nature397(1999)673A.Bachtold, SuperlatticesMicrostruct.34(2003)413S.Zaric, Science304(2004)1129S.Zaric, PhysicaE29(2005)469S.Zaric} whereas the selection rule remains.
On the other hand, perpendicular magnetic fields cause the coupling of the independent angular-momentum states.
This might trigger available optical transitions between various valence and conduction subbands.

\begin{figure}
\begin{center}
  \includegraphics[width=\linewidth, keepaspectratio]{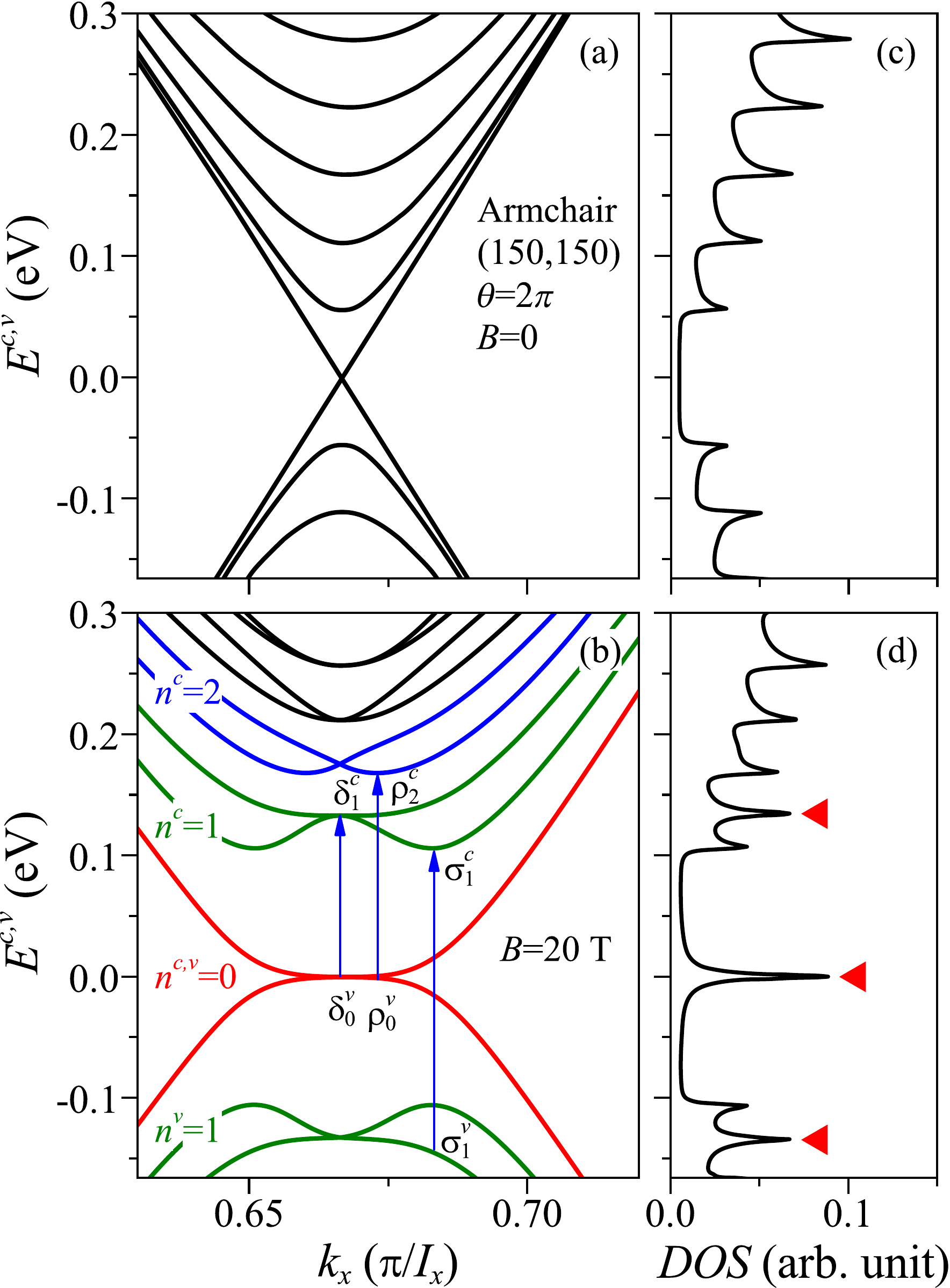}\\
  \caption[]{
  Low-energy magneto-electronic band structures for the ($150$, $150$) ACNT in (a) the absence of external fields, and (b) the presence of a uniform magnetic field $B = 20$ T; (c) and (d): their corresponding DOSs.
  }
  \label{fig:BSandDOSofACNT}
\end{center}
\end{figure}

In the case of $\theta=2\pi$, a zipped curved ZGNR corresponds to a ACNT, named according to the shape of circular cross section.
As shown in Fig.~\ref{fig:BSandDOSofACNT}(a), the band structure, symmetric about $k_x=2/3$, is composed of two linear subbands intersecting at the Fermi energy and several doubly degenerate parabolic subbands.
The closed boundary condition makes the edge-localized states at the Fermi energy disappear.
Also, the distinct magneto-electronic structures presented by the curved GNR and CNT are mainly due to the different boundary conditions.
The perpendicular magnetic field causes the coupling of states, converts the linear subbands into parabolic ones, and lifts the degeneracy in CNTs (Fig.~\ref{fig:BSandDOSofACNT}(b)).
Intersecting at the Fermi energy, the lowest conduction subband and the highest valence subband are labeled by the subband indices $n^c = 0$ and $n^v = 0$, respectively.
Furthermore, the pairs of conduction subbands (valence subbands) intersecting at $k_x = 2/3$ are denoted as $n^c$ ($n^v$) $=1$, $2$, $3$,..., ascending in accordance with the energies.
Instead of containing single angular momentum, the electronic state possesses multiple angular momenta of adjacent subbands.
It is also noted that $n^{c,v} = 0$ and $n^{c,v} = 1$ are the only ones that possess a small dispersionless region under the reduced magnetic quantization in CNTs.
The parabolic subbands are the main structures of the energy spectrum.
By increasing the diameter of CNT and the strength of magnetic field, the Landau subbands of higher energies will gradually form with the elongation of dispersionless region.
These important differences between CNTs and curved GNRs indicate that the boundary conditions play a crucial role in the electronic properties.
In the absence of external fields, the DOS of CNT contains lots of asymmetric peaks and a finite density near the Fermi energy (Fig.~\ref{fig:BSandDOSofACNT}(c)).
In magnetic fields, the three additional peaks around $\omega = 0$ and $\omega = \pm 0.12$ eV (indicated by triangles) are induced by the magnetic quantization, exhibiting a symmetric form (Fig.~\ref{fig:BSandDOSofACNT}(d)).

\begin{figure}
\begin{center}
  \includegraphics[width=\linewidth, keepaspectratio]{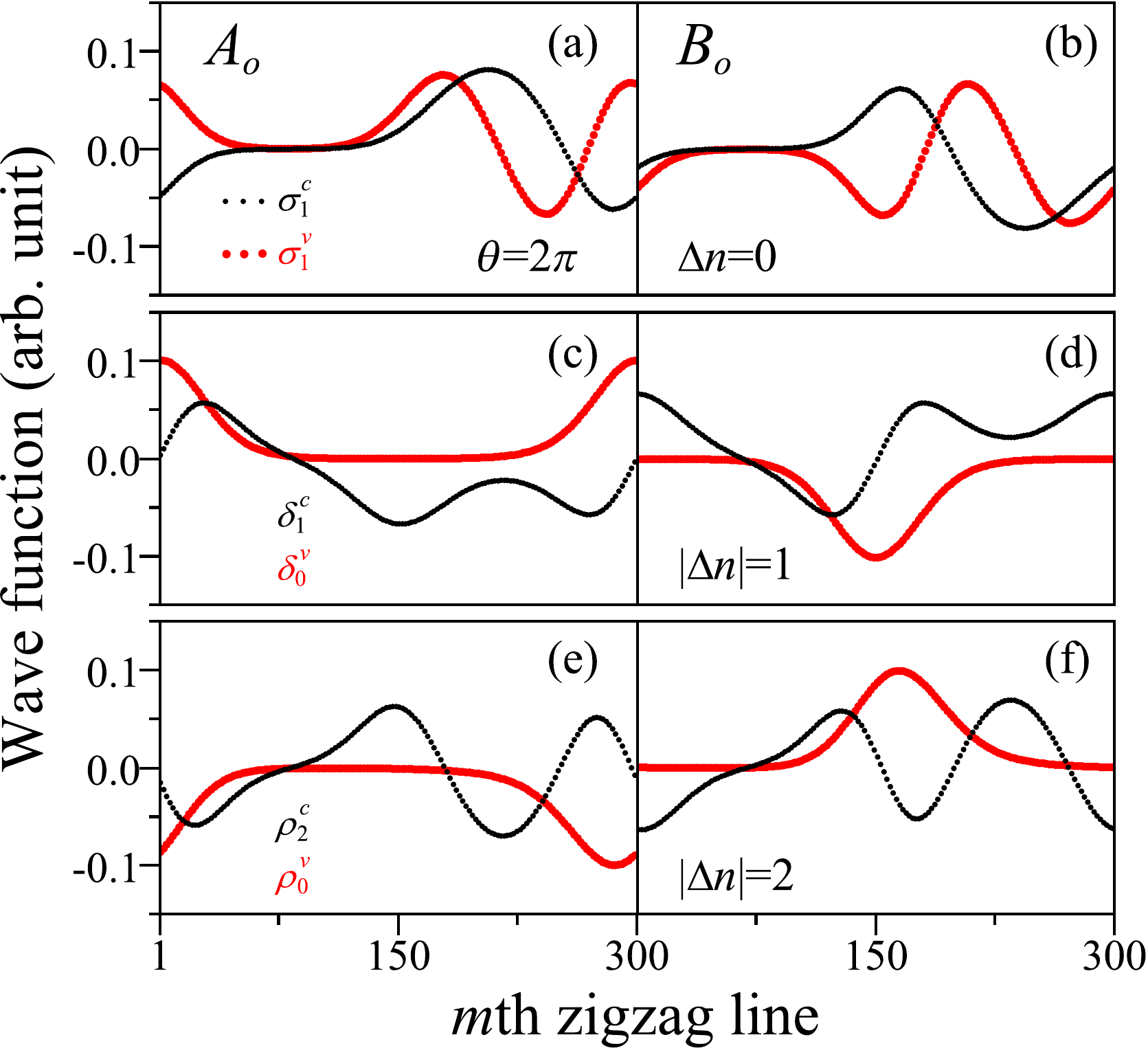}\\
  \caption[]{
  Wave functions of the $n^{c,v} = 0$, $1$, and $2$ oscillatory Landau subbands for the states marked in Fig.~\ref{fig:BSandDOSofACNT}.
  }
  \label{fig:WFofACNT}
\end{center}
\end{figure}

In a CNT, the wave function with an angular momentum $J$ is characterized by two linearly independent functions, $\sin(J\Phi_m)$ and $\cos(J\Phi_m)$, where $0\leq\Phi_{m}\leq2\pi$ indicates the azimuthal coordinate of a carbon atom.
As the polarization of incident light is parallel to the tube axis, the allowed transitions are only those satisfying the selection rule of $\Delta J = 0$.\cite{PhysicaB201(1994)349H.Ajiki, J.Phys.Soc.Jpn.66(1997)3294M.F.Lin, Phys.Rev.B62(2000)16092S.Roche}
On the other hand, the $J$-coupled states in a perpendicular magnetic field can be expressed as a combination of different $J$'s.
For example, as shown in Fig.~\ref{fig:WFofACNT}, $\delta_0^v$ consists of $J = 0$ and 1 components; $\sigma_1^v$ reveals the combination of $J = 0$, 1, and 2 components; ditto for the other $J$th-coupled states.
This gives rise to the available transitions between two states possessing the same $J$ component.

The selection rule of $\Delta n = 0$ is validated by many transitions, \emph{e.g.} the one between states $\sigma_1^v$ and $\sigma_1^c$ (Fig.~\ref{fig:WFofACNT}(a) and (b)).
Because of the coupling of adjacent angular momenta, the additional transition between $\sigma_0^v$ and $\sigma_1^c$ is also allowed ($|\Delta n| = 1$).
The corresponding wave functions shown in Fig.~\ref{fig:WFofACNT}(c) and (d), are mainly distributed around the locations of 1st and 150th zigzag line, \emph{i.e.} the bottom and top of nanotube, where the penetration of the magnetic flux is at a maximum.
This results in a significant overlap of wave functions, and makes transitions between $\delta_0^v$ and $\delta_1^c$ possible.
Once the Landau subbands are formed, the absorption spectrum exhibits more prominent peaks arising from channels obeying the selection rule of $|\Delta n| = 1$.
Furthermore, in addition to $\Delta n = 0$ and $|\Delta n| = 1$, the selection rule of $|\Delta n| = 2$ is introduced, such as the transition from state $\rho_0^v$ to state $\rho_2^c$ (Fig.~\ref{fig:WFofACNT}(e) and (f)).
It is thus deduced that when the field strength or tube diameter is increased, the stronger angular momentum coupling leads to further selection rules and absorption peaks.

The field strength and tube diameter are the crucial roles in controlling the magneto-electronic properties of CNTs.
For a narrow CNT under a weak magnetic field, the parabolic subbands in the energy spectrum remain, and the related asymmetric peaks are the main structures in the DOS.
For a CNT in a sufficiently large field, curvature-reduced magnetic quantization causes Landau subbands and symmetric DOS peaks.
For a wide CNT in a strong field, the dispersionless regions of Landau subbands enlarge, the Landau subbands with high energies appear, and the related symmetric peaks become the prominent structures in the DOS.
These predicted evolution in the magneto-electronic structures of CNTs can be verified by the STS, which has successfully been used in realizing the electronic structures of CNTs.\cite{Nature391(1998)59J.W.G.Wildoer, Nature391(1998)62T.W.Odom, Phys.Rev.Lett.82(1999)1225P.Kim}
The formation of Landau subband is a key feature of the magnetic quantization in CNTs.
Some signature of the Landau subband formation is first reported
by Kanda~\emph{et. al.}.\cite{PhysicaB323(2002)246A.Kanda}
Recent observations on the Landau states of CNTs are achieved via magneto-transport measurements in a CNT-based Fabry-Perot resonator in a very large magnetic field.\cite{Phys.Rev.Lett.101(2008)046803B.Raquet, Phys.Rev.Lett.103(2009)256801S.Nanot}

\subsection{Magneto-optical properties}

Magneto-optical properties of curved GNRs and CNTs are particularly interesting in terms of exploring the curvature effects during zipping process and the optical responses through the transformation of an open boundary into a closed cylindrical boundary system.
When a GNR is bent into a curved one, the prominent absorption peaks contributed by inter-QLL transitions split into tri-subpeaks, among which the spacing becomes separated from one another with an increment of the curvature.
However, as the curvature exceeds a certain critical value, some peaks vanish and extra peaks appear.
It should be noticed that as a zipped CNT is formed, the transition peaks and  optical selection rules are correlated to the degree of angular-momentum coupling.
They continuously change with a variation of the diameter and field strength.

\begin{figure}
\begin{center}
  \includegraphics[width=\linewidth, keepaspectratio]{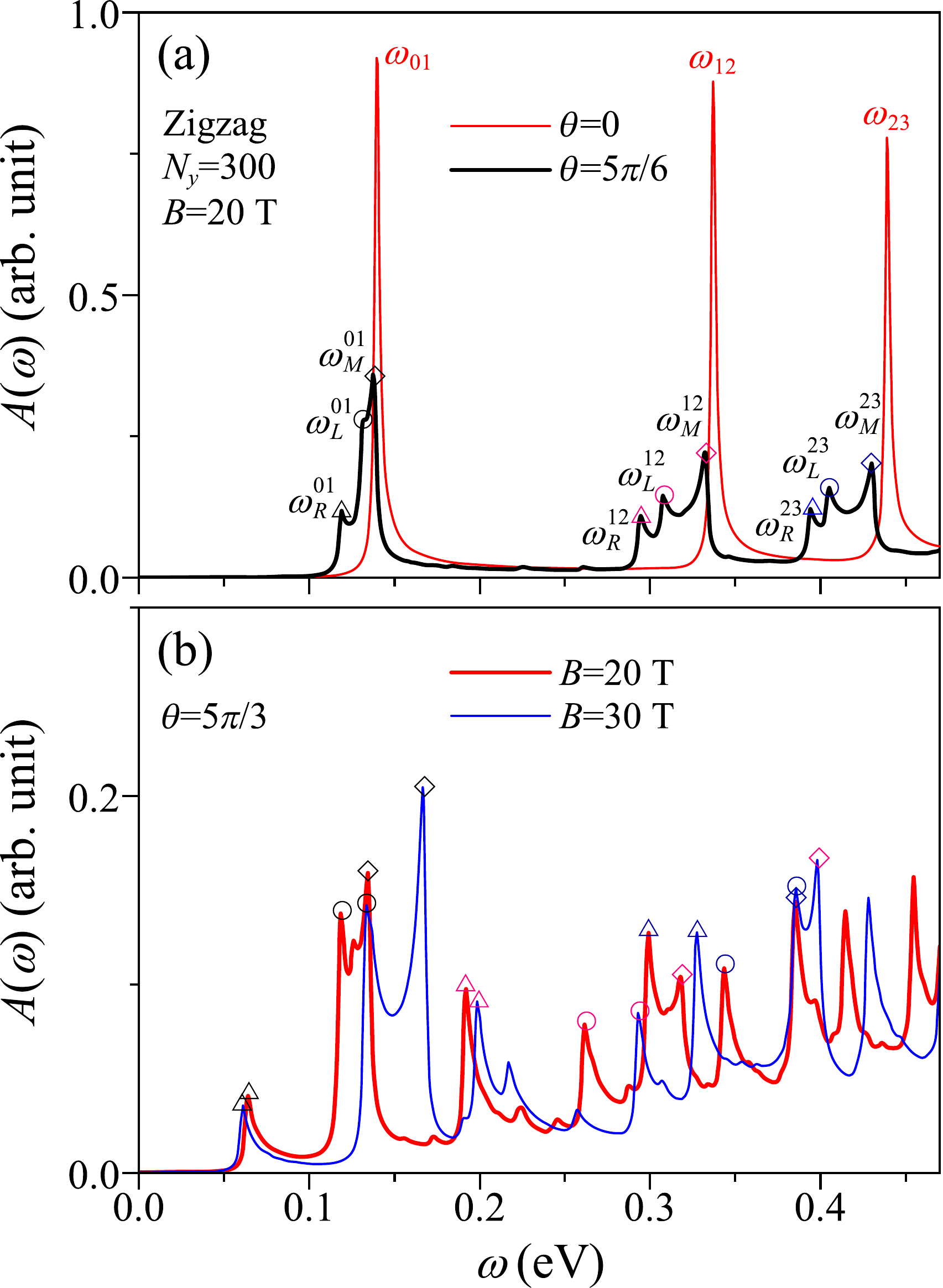}\\
  \caption[]{
  Optical absorption spectra of the $N_y = 300$ ZGNR for (a) different central angles ($0$ and $5\pi/6$) under $B = 20$ T; for (b) $5\pi/3$ under different field strengths ($B = 20$ T and $30$ T).
  The three subpeaks of each group following the selection rule of $|\Delta n| = 1$ are denoted by triangles, circles and diamonds.
  The subscripts $R$, $L$, and $M$ are used to identify the transitions from the right, left, and middle band-edge states of Landau subbands in Fig.~\ref{fig:BSandDOSofCurvedZGNR}(a).
  }
  \label{fig:ABSofCurvedZGNR}
\end{center}
\end{figure}

The direct transitions for a flat GNR (red curve in Fig.~\ref{fig:ABSofCurvedZGNR}(a)) that satisfy the selection rule of $|\Delta n|=1$ between valence and conduction QLLs contribute to symmetric delta-function-like absorption peaks, which are spaced approximately by a proportional relationship $\sqrt{B}(\sqrt{n^{c}}-\sqrt{n^{v}})$.
However, the curvature significantly changes the features of magneto-optical spectra, such as the number, peak structure, height and frequency of absorption peak.
For a moderate central angle of $5\pi/6$, the absorption spectrum, responsible for the peak splittings in the DOS, demonstrates more low-intensity asymmetric peaks (heavy black curve in Fig.~\ref{fig:ABSofCurvedZGNR}(a)).
It is still the selection rule of $|\Delta n|=1$ that dominates the magneto-optical spectrum, because the features of Landau wave functions are retained.
The oscillatory QLLs with three band-edge states result in tri-peaks for the same kind of transition channels, labeled as $\omega_{R}^{n(n+1)}$ (triangle), $\omega_{L}^{n(n+1)}$ (circle), and $\omega_{M}^{n(n+1)}$ (diamond), where the subscripts $R$, $L$, and $M$ are used to identify the interband transitions from the right, left, and middle band-edge states.
The spacing of a tri-peak structure is determined by the Landau subband amplitude.
The $\omega_{M}^{n(n+1)}$ subpeak can be regarded as from inter-QLL transitions (Section~\ref{sec:MagnetoElectronicPropertiesOfCurvedGNR}); it is closest to the inter-QLL excitation peak of the flat GNR and has the highest intensity among the tri-peaks.
The other two lower subpeaks, away from the Landau absorption peak, are under weaker $B_{eff}$ and significantly affected by the lateral confinement.

With an increment of the central angle, the subpeaks are red-shifted and broadened; moreover, additional subpeaks of $|\Delta n| \neq 1$ can be induced (heavy red curve in Fig.~\ref{fig:ABSofCurvedZGNR}(b)).
These subpeaks remain sharp in the low frequency region, while become obscure and distorted at high frequencies.
This indicates that the curvature effects on the magneto-optical properties is stronger at higher energies.
On the other hand, a stronger magnetic quantization causes the blue shift and  intensity enhancement of subpeaks (light blue curve in Fig.~\ref{fig:ABSofCurvedZGNR}(b)).
Consequently, the appearance or absence of the $|\Delta n| \neq 1$ additional subpeaks mainly results from the complex relationship between the geometric structure and magnetic field.

\begin{figure}
\begin{center}
  \includegraphics[width=\linewidth, keepaspectratio]{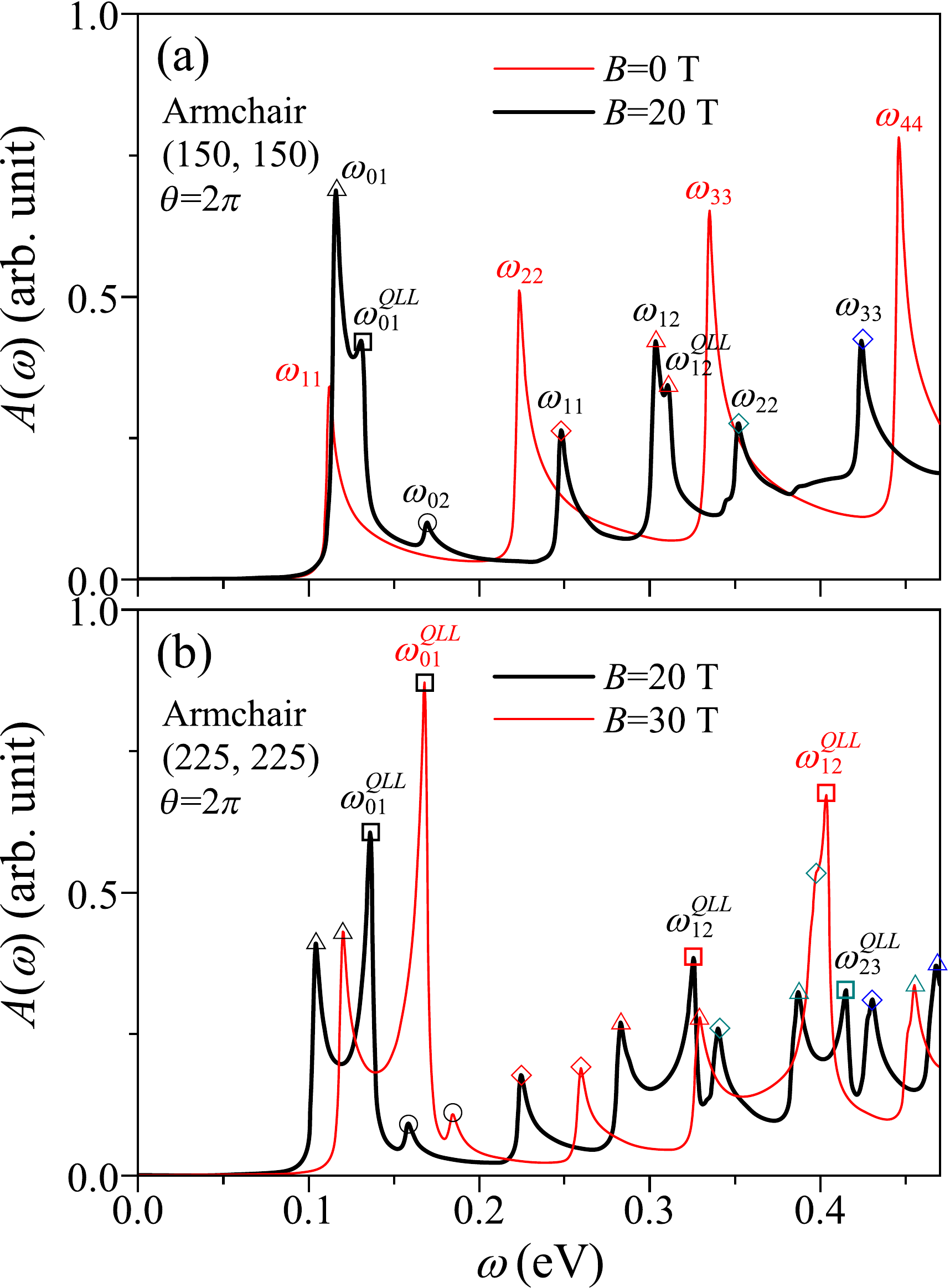}\\
  \caption[]{
  Optical absorption spectra for the (a) ($150$, $150$) ACNT at $B = 0$ and $B = 20$ T; (b) ($225$, $225$) ACNT at $B = 20$ T and $B = 30$ T.
  The absorption peaks satisfying the selection rules of $\Delta n = 0$, $|\Delta n| = 1$, and $|\Delta n| = 2$ are denoted by the diamonds, triangles and squares, as well as circles, respectively.
  }
  \label{fig:ABSofACNT}
\end{center}
\end{figure}

Due to the periodic cylindrical symmetry, the selection rules of CNTs are different from those of curved GNRs.
In the absence of external fields (light red curve in Fig.~\ref{fig:ABSofACNT}(a)), each asymmetric peak $\omega_{nn}$ is associated with the direct optical transitions between the parabolic valence and conduction subbands characterized by the same angular momentum (permitted by the selection rule of $|\Delta n| = 0$), as shown by the light red curve in Fig.~\ref{fig:ABSofACNT}(a).
However, additional selection rules, \emph{e.g.} $|\Delta n| = 1$ and $|\Delta n| = 2$, are induced through $J$ coupling in the presence of perpendicular magnetic fields.
At $B=20$ T (heavy black curve in Fig.~\ref{fig:ABSofACNT}(a)), the blue-shifted asymmetric peaks, $\omega_{nn}$'s, are suppressed, and the extra peaks, $\omega_{01}$, $\omega_{12}$, and $\omega_{02}$ derived from $|\Delta n| = 1$ and $|\Delta n| = 2$ take place.
The $\omega_{01}^{QLL}$ and $\omega_{12}^{QLL}$ peaks indicate the formation of low-lying Landau states.

At a fixed magnetic field, the angular-momentum coupling is more significant for a CNT with a large diameter, leading to a rich and diversified magneto-absorption spectrum (heavy black curve in Fig.~\ref{fig:ABSofACNT}(b)).
The spectrum contains pairs of peaks induced by the selection rule $|\Delta n| = 1$, where, in particular, the peaks $\omega_{01}^{QLL}$, $\omega_{12}^{QLL}$ and $\omega_{23}^{QLL}$, contributed by the transitions around $k_{x} = 2/3$, are formed as a consequence of the low-lying Landau subbands.
With an increase of diameter or field strength (Fig.~\ref{fig:ABSofACNT}(b)), the most noticeable peaks are mainly contributed by the inter-QLL transitions, that is to say, the optical excitations that obey the selection rule $|\Delta n| = 1$ dominate the major structure of the absorption spectrum, in contrast to the minor ones from $|\Delta n| = 0$ and $|\Delta n| = 2$.

During the variation of $\theta$ from $0$ to $2\pi$, the main characteristics of the absorption spectra, such as the number, intensity, frequency and prominent structure of absorption peaks are thoroughly altered.
Various peaks are exclusively formed for a flat structure and a curved structure with either open or closed boundary.
The inter-QLL excitation peaks of a curved GNR split to form tri-peak structures under the influence of a non-uniform effective magnetic field.
An increment of the curvature gives rise to an increase of the tri-peak spacing and a continuous decrease of the intensity and frequency, however, which are interrupted for a CNT ($\theta=2\pi$) where a cylindrical symmetry condition is given.
It is shown that the magnetic-field-induced angular-momentum coupling allows various excitation channels following the selection rules of $|\Delta n| = 0$, $1$, and $2$.
These predicted magneto-optical properties can be verified by the optical experiments, such as the optical transmission/absorption/scattering measurements providing a wide frequency spectrum for epitaxial GNR arrays\cite{Phys.Rev.Lett.110(2013)246803J.M.Poumirol} and CNTs.\cite{SuperlatticesMicrostruct.34(2003)413S.Zaric, Science304(2004)1129S.Zaric, PhysicaE29(2005)469S.Zaric}
The magneto-optical features affected by the respective boundary conditions, namely open and periodic ones, provide insight into the nature of the magneto-electronic properties in curved quasi-1D systems.

\section{Bilayer systems}
\label{BilayerSystem}

Monolayer GNRs exhibit diverse electronic and optical properties under external fields (Section~\ref{sec:MonolayerSystem}).
Especially, the width-dependent energy gap\cite{Phys.Rev.Lett.98(2007)206805M.Y.Han, Science319(2008)1229X.L.Li, Nat.Nanotechnol.3(2008)397L.Tapaszto} and high carrier mobility\cite{IEEEElectronDeviceLett.28(2007)282M.C.Lemme} trigger extensively both experimental and theoretical studies to focus on integrating GNRs into potential electronic devices, such as GNR-based FETs,\cite{Science306(2004)666K.S.Novoselov, Appl.Phys.Lett.88(2006)142102B.Obradovic, J.Appl.Phys.102(2007)054307G.Liang, IEEEElectronDeviceLett.28(2007)282M.C.Lemme, NanoLett.8(2008)1819G.Liang, Science319(2008)1229X.L.Li} resonant tunneling diodes,\cite{J.Appl.Phys.105(2009)084317H.Teong, J.Phys.D-Appl.Phys.43(2010)215101G.Liang, IEEEJ.ElectronDevicesSoc.2(2014)118F.Al-Dirini} quantum dot devices,\cite{Phys.Rev.Lett.98(2007)016802P.G.Silvestrov, Appl.Phys.Lett.91(2007)053109Z.F.Wang, NanoLett.8(2008)2378C.Stampfer, Appl.Phys.Lett.93(2008)212102J.Guttinger, Phys.Rev.Lett.103(2009)046810J.Guttinger, Mater.Today13(2010)44T.Ihn, Phys.Rev.B84(2011)041401D.Prezzi} negative differential resistance devices,\cite{Appl.Phys.Lett.94(2009)173110H.Ren, J.Appl.Phys.107(2010)063705V.N.Do, Appl.Phys.Lett.98(2011)192112K.M.M.Habib, Chem.Phys.Lett.554(2012)172P.Zhao, Appl.Phys.Lett.102(2013)043114Y.Khatami} and nanoelectromechanical contact switches.\cite{J.Am.Chem.Soc.131(2009)11147D.C.Wei}
Recently, few-layer GNRs have been synthesized by cutting graphene,\cite{NanoLett.9(2009)2083J.W.Bai, Science319(2008)1229X.L.Li} unzipping multi-wall CNTs,\cite{Nature458(2009)872D.V.Kosynkin, Nature458(2009)877L.Jiao} and CVD\cite{J.Am.Chem.Soc.131(2009)11147D.C.Wei} (Section~\ref{sec:Introduction}).
These GNRs are predicted to possess quite different physical properties compared with monolayer ones, owing to the number of layers, stacking configurations, and interlayer atomic interactions.
Bilayer GNRs can be served as a model system in understanding the essential properties of few-layer ones.

The stacking configurations and interlayer atomic interactions play important roles in the electronic properties of bilayer GNRs.
There are two groups of energy subbands and each group includes conduction and valence subbands.
The initial energies for AB and AA stackings are totally different because of the distinct interlayer atomic interactions.
The energy gap in AB stacking is determined by the conduction and valence subbands of the first group.
Gap modulation can be simply reached by electric fields, providing the flexibility in the design of top- or side-gated nanoelectronic devices.
Concerning the AA stacking with higher symmetry, the interlayer atomic interactions induce the significant overlap between the valence subbands of the first group and the conduction subbands of the second group.
The metallic property remains even in the presence of electric fields.
Magnetic field flocks the neighboring electronic states of each subband, thus leading to two groups of QLLs with the initial energies similar to those of zero-field energy subbands.
These highly degenerate states corresponding to large DOS are predicted to have intense optical transitions.
Electric fields competes against the magnetic quantization; therefore, energy dispersions are significantly altered.
The transverse one results in the tilt of QLLs and even the collapse of Landau states, while the perpendicular one lifts state degeneracy and thus induces QLL crossings and anti-crossings.

Bilayer AB and AA stackings exhibit distinct magneto-optical spectra.
Four groups of QLL-dependent optical transitions are revealed in the former, including two intragroup and two intergroup ones.
However, there exist only two intragroup transitions in the AA stacking, as a result of the special relations among the well-behaved Landau wave functions on different layers.
The electric fields severely distort the Landau wave functions, so that the magneto-absorption spectra will be dramatically altered.
These optical responses can be observed from the magneto-optical transmission and magneto-Raman spectroscopy, where the optical excitations between Landau levels in monolayer and few-layer graphene have been identified.\cite{Phys.Rev.Lett.97(2006)266405M.L.Sadowski, Phys.Rev.Lett.98(2007)197403Z.Jiang, Phys.Rev.B76(2007)081406R.S.Deacon, Phys.Rev.Lett.100(2008)087401P.Plochocka, NanoLett.14(2014)4548S.Berciaud}

The organization of this section is stated as follows.
A detail discussion on the magneto-electronic properties are in the first subsection, including the stacking-dependent QLL spectra, DOS, and Landau wave functions, as well as the competition between the magnetic quantization and electric fields.
Meanwhile, the zero-field energy spectra and electric-field-induced gap modulation are also provided.
The magneto-optical properties covering the QLL-dependent transitions of different stackings and effects of transverse and perpendicular electric fields are presented in the second subsection.
The comparisons are made between the theoretical predictions and recent experimental observations.
The edge-dependent features in the magneto-electronic and optical properties, such as the degeneracy of QLLs, extra band-edge states, and additional absorption subpeaks, have been discussed in Section~\ref{sec:MonolayerSystem}.
In order to focus on the main differences owing to the stacking configurations, interlayer atomic interactions, and external fields, the ZGNRs are chosen for simplicity.

\subsection{Magneto-electronic properties}
\label{MagnetoElectronicPropertiesOfBilayerSystem}

\begin{figure}
\begin{center}
  \includegraphics[width=\linewidth, keepaspectratio]{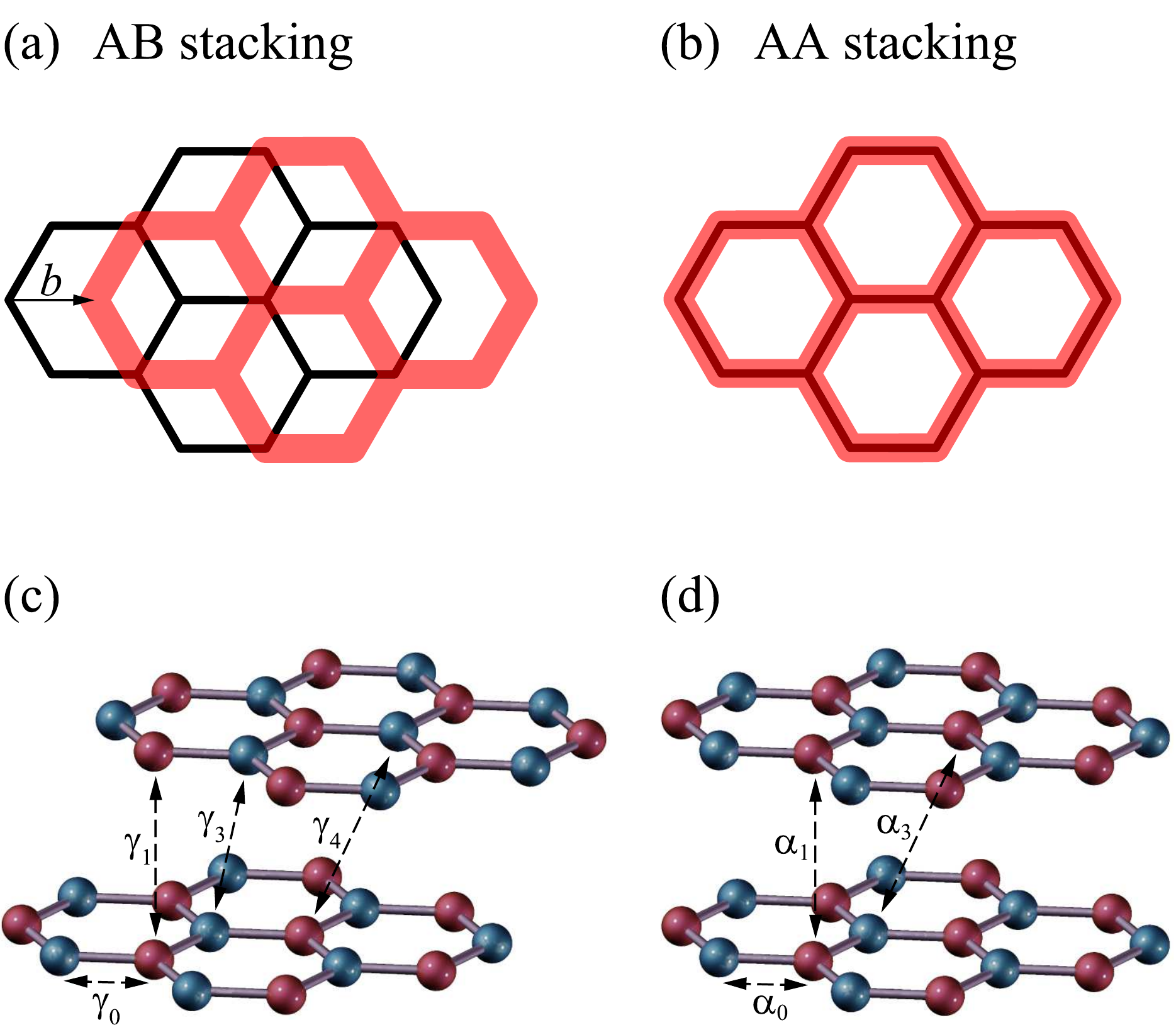}\\
  \caption[]{
  Top views of the (a) AB and (b) AA stackings with the corresponding interlayer atomic interactions in (c) and (d).
  $A^i$ and $B^i$ sublattices are indicated by the red and blue balls, respectively.
  }
  \label{fig:AB_AA_stacking}
\end{center}
\end{figure}

The low-lying electronic structures strongly depend on the stacking configurations and interlayer atomic interactions.
There are two typical stackings for bilayer GNRs with layer spacing of $c = 3.35$~{\AA}.
From the top view, the hexagonal lattices of two layers have a horizontal displacement of $b$ for the AB stacking (Fig.~\ref{fig:AB_AA_stacking}(a)), while they fully overlap for the AA stacking with a higher symmetry (Fig.~\ref{fig:AB_AA_stacking}(b)).
According to the Slonczewski-Weiss-McClure (SWMcC) model,\cite{Phys.Rev.71(1947)622P.R.Wallace, Phys.Rev.109(1958)272J.C.Slonczewski, Phys.Rev.108(1957)612J.W.McGlure} the atomic interactions in $2p_z$ orbitals for the AB stacking include the intralayer one between the nearest carbon atoms ($\gamma_0 = 2.598$ eV), interlayer ones between $A^1$ and $A^2$ atoms ($\gamma_1 = 0.364$ eV), $B^1$ and $B^2$  atoms ($\gamma_3 = 0.319$ eV), $A^{1}$ and $B^{2}$ ($A^{2}$ and $B^{1}$) atoms ($\gamma_4 = 0.177$ eV), as well as the chemical shift ($\gamma_6 = -0.026$ eV)\cite{Phys.Rev.B43(1991)4579J.C.Charlier} (Fig.~\ref{fig:AB_AA_stacking}(c)).
The superscripts 1 and 2 denote the top and bottom layers, respectively.
Such interactions in the AA stacking are the intralayer one between the nearest atoms ($\alpha_0 = 2.569$ eV), and interlayer ones between $A^1$ and $A^2$ ($B^1$ and $B^2$) atoms ($\alpha_1 = 0.361$ eV) \& $A^{1}$ and $B^{2}$ ($A^{2}$ and $B^{1}$) atoms ($\alpha_3 = -0.032$ eV)\cite{Phys.Rev.B44(1991)13237J.C.Charlier, Phys.Rev.B46(1992)4531J.C.Charlier} (Fig.~\ref{fig:AB_AA_stacking}(d)).

\subsubsection{AB stacking.}

\begin{figure}
\begin{center}
  \includegraphics[width=\linewidth, keepaspectratio]{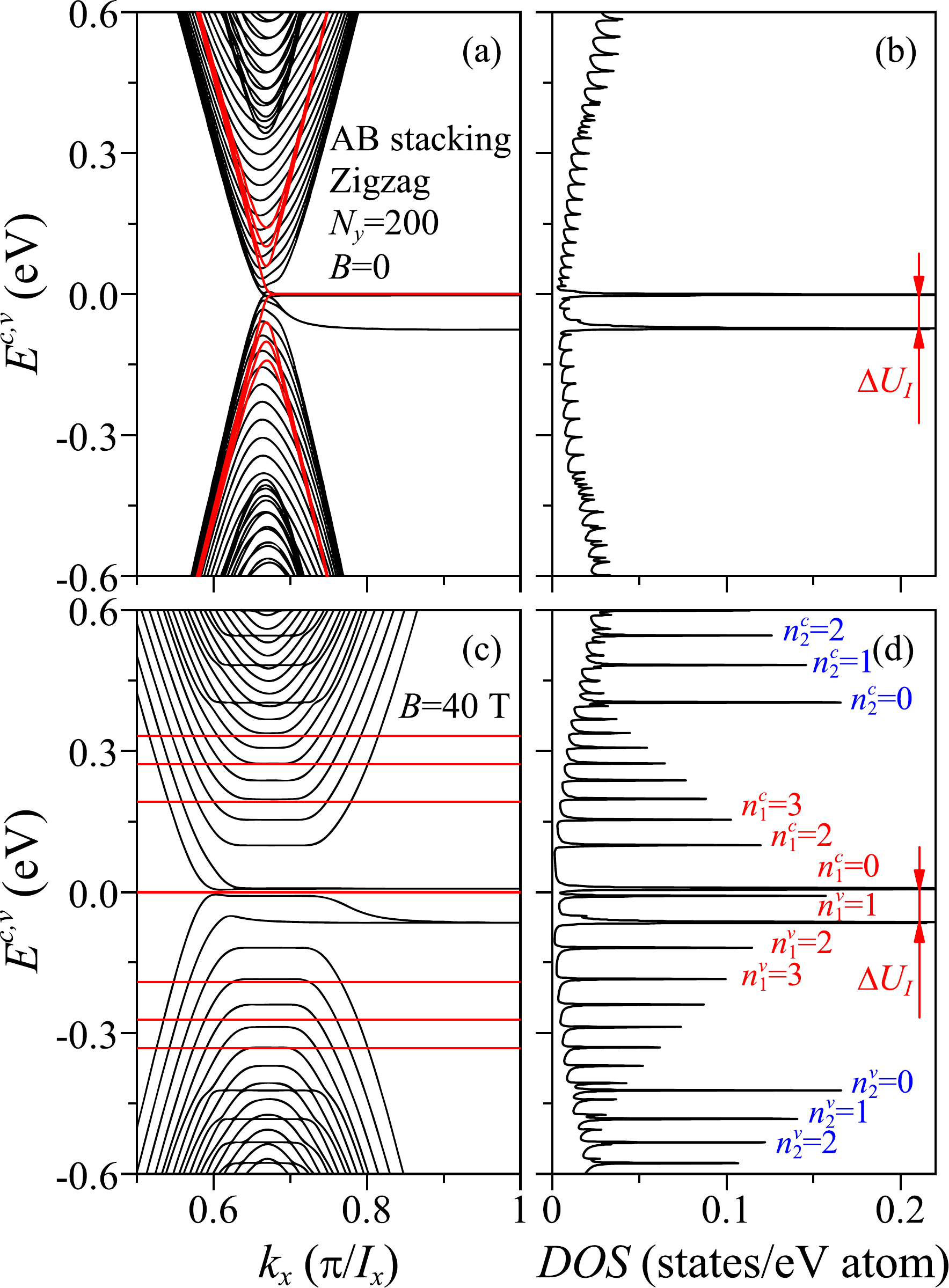}\\
  \caption[]{
  Low-energy band structure and DOS of the $N_y = 200$ bilayer AB-stacked ZGNRs at $B=0$ and $40$ T.
  The red curves (lines) indicate the low-lying parabolic subbands (QLL energies) of monolayer system.
  }
  \label{fig:BS_DOSofBi_AB_ZGNRs_inB}
\end{center}
\end{figure}

The low-lying energy dispersions for bilayer AB-stacked ZGNRs are composed of two groups of conduction and valence subbands, where they have parabolic dispersions except the two pairs of partial flat ones near the Fermi level (Fig.~\ref{fig:BS_DOSofBi_AB_ZGNRs_inB}(a)).
The subband symmetry about $E_F=0$ no longer exists because of the interlayer atomic interactions.
The parabolic subbands of the first group start to form near $E_F=0$, while those of the second group are initiated from $E^{c,v} \sim \pm\gamma_1$.
The energy spacings between the band-edge states are smaller than those of monolayer system (red curves).
Moreover, there is an energy difference, $\Delta U_I \sim 0.08$ eV, in the flat dispersions at the zone boundary.
DOS exhibits two intense symmetric and many asymmetric peaks related to the band-edge states of partial flat and parabolic subbands, respectively (Fig.~\ref{fig:BS_DOSofBi_AB_ZGNRs_inB}(b)).
For electronic states of $|E^{c,v}| > \gamma_1$, two groups of asymmetric peaks overlap, and thus a complex energy spectrum appears.

Bilayer AB-stacked ZGNRs have two groups of QLLs with the formation center at $k_x = 2/3$ and initial energies about $E_F$ and $\pm \gamma_1$ (Fig.~\ref{fig:BS_DOSofBi_AB_ZGNRs_inB}(c)).
With an increasing state energy, the QLL widths of the first group near $E_F$ gradually shrink, and parabolic subbands come to exist after the vanished QLLs.
Such subbands will overlap with the initial QLLs of the second group at higher energy.
Many high-intensity symmetric DOS peaks appearing at $|E^{c,v}| > \gamma_1$ are related to the second-group QLLs, in which each peak is surrounded by several low-intensity ones of the first group (Fig.~\ref{fig:BS_DOSofBi_AB_ZGNRs_inB}(d)).
This result is expected to have drastic changes in the differential conductance.
In other words, the STS measurements\cite{Science324(2009)924D.L.Miller, Nature467(2010)185Y.J.Song, Nat.Commun.4(2013)1744G.Li} are useful to determine the vertical interlayer atomic interaction of $\gamma_1$.
In addition, the dispersionless regions of the four partial flat subbands are extended.
According to the wave functions near $k_x = 2/3$, half of them belong to the  edge-localized or Landau states.

The competition among the interlayer atomic interactions, magnetic fields, and finite widths results in more complicated magneto-electronic structures, especially for the $B$-field dependence of QLL energies.
Two groups of QLLs are formed as the magnetic length is smaller than the ribbon width (Fig.~\ref{fig:BS_DOSofBi_AB_ZGNRs_inB}(c)).
The first (second) group of QLLs is labeled by $n_1^c = 0$, $2$, $3$,... and $n_1^v = 1$, $2$, $3$,... ($n_2^{c,v} = 0$, $1$, $2$,...) according to the state energy measured from $E_F$.
Considering the QLLs with $|E^{c,v}| < \gamma_1$, their energies roughly obey the linear $B$-dependence.\cite{Phys.Rev.B78(2008)115422Y.C.Huang}
With respect to the rise of their energies, QLLs gradually deviate from the linear $B$-dependence, but satisfy the $\sqrt{B}$-dependence that is an identifiable feature in the monolayer system.
The evolution of the $B$-dependent QLL energies can be identified by experimental measurements of tunneling current\cite{Science324(2009)924D.L.Miller} and cyclotron resonance.\cite{Phys.Rev.Lett.104(2010)067404E.A.Henriksen}
The similar $B$-field dependence has also been observed in LLs of bilayer graphene experimentally\cite{Phys.Rev.Lett.100(2008)087403E.A.Henriksen} and theoretically.\cite{Phys.Rev.Lett.96(2006)086805E.McCann, Phys.Rev.B76(2007)115419J.M.Pereira, Phys.Rev.B77(2008)085426Y.H.Lai}

\begin{figure}
\begin{center}
  \includegraphics[width=\linewidth, keepaspectratio]{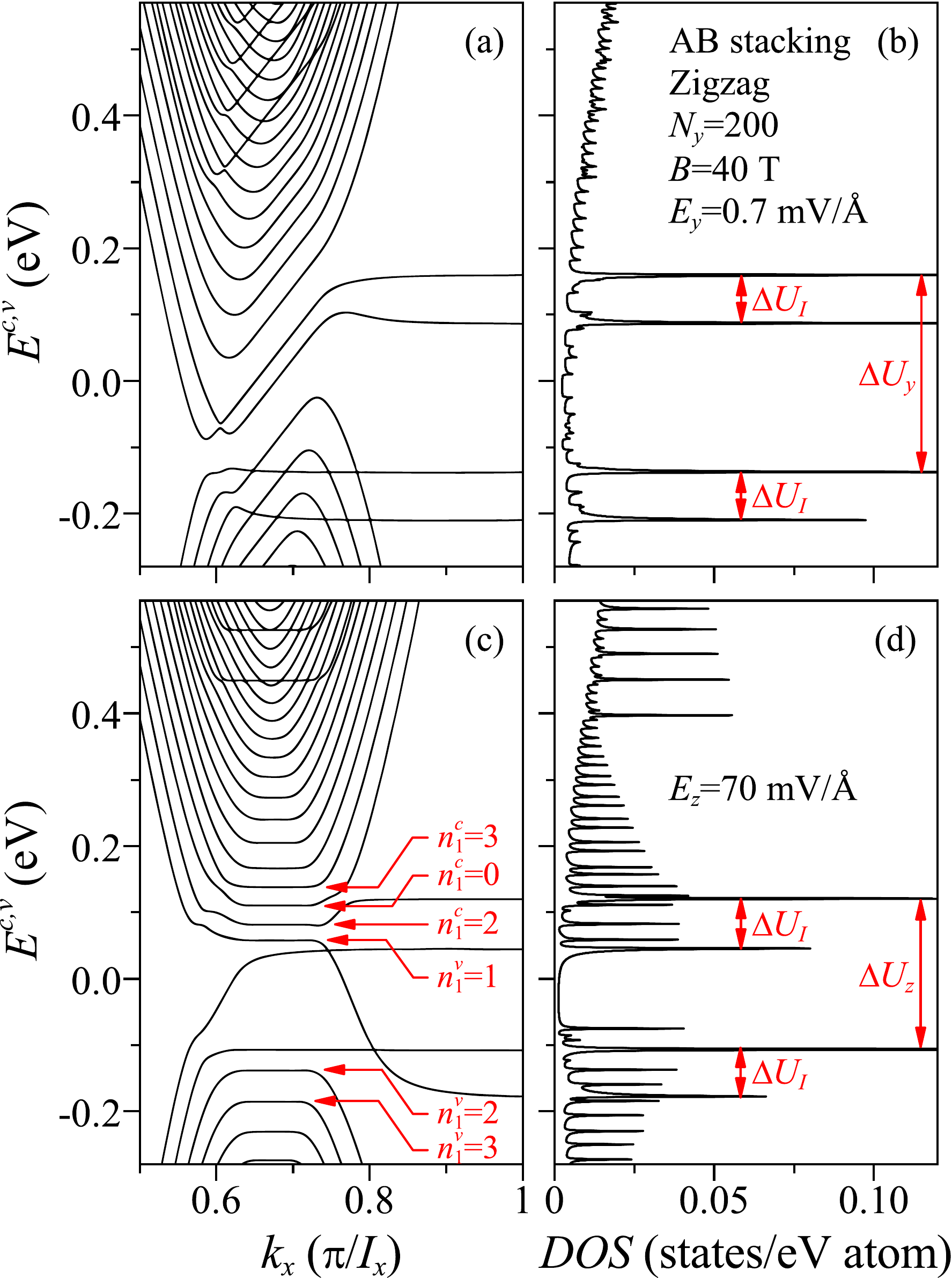}\\
  \caption[]{
  Low-energy band structures and DOS of the $N_y = 200$ bilayer AB-stacked ZGNRs at $B=40$ T; (a) \& (c) ($E_y = 0.7$ mV/{\AA}, $E_z = 0$), and (b) \& (d) ($E_y = 0$, $E_z = 70$ mV/{\AA}).
  }
  \label{fig:BS_DOSofBi_AB_ZGNRs_inB_inEyEz}
\end{center}
\end{figure}

Transverse and perpendicular electric fields can suppress the magnetic quantization and thus diversify the magneto-electronic structures.
The site energies of $A^i$ and $B^i$ sublattices have smooth variations across a nanoribbon under a transverse electric field.
Both groups of QLLs are tilted with the same angle (Fig.~\ref{fig:BS_DOSofBi_AB_ZGNRs_inB_inEyEz}(a)).
All the QLL symmetric peaks in the DOS are destroyed (Fig.~\ref{fig:BS_DOSofBi_AB_ZGNRs_inB_inEyEz}(b)); that is, the high-intensity symmetric peaks are turned into the lower asymmetric ones.\cite{Philos.Mag.94(2014)1859H.C.Chung, Phys.Rev.B91(2015)155409B.Ostahie}
It should be noted that the energy spacings between tilted QLLs quickly decline and the collapse takes place after $E_y$ reaches the critical value (Section~\ref{sec:Effects_of_Composite_Fields_Monolayer}).
On the other hand, a site energy difference exists between two layers in the perpendicular electric field.
Although the QLLs remain dispersionless, their energies are shifted with abnormal ordering for small quantum numbers (Fig.~\ref{fig:BS_DOSofBi_AB_ZGNRs_inB_inEyEz}(c)).
The QLL peaks in the DOS are shifted to higher frequencies (Fig.~\ref{fig:BS_DOSofBi_AB_ZGNRs_inB_inEyEz}(d)).\cite{Phys.Chem.Chem.Phys.15(2013)868H.C.Chung}
In addition, electric fields also lift the degeneracy of partial flat subbands, where the splitting energies, $\Delta U_y = -e E_y W_{zig}$ and $\Delta U_z = -e E_z c$, correspond to the transverse and perpendicular ones, respectively.

\begin{figure}
\begin{center}
  \includegraphics[width=\linewidth, keepaspectratio]{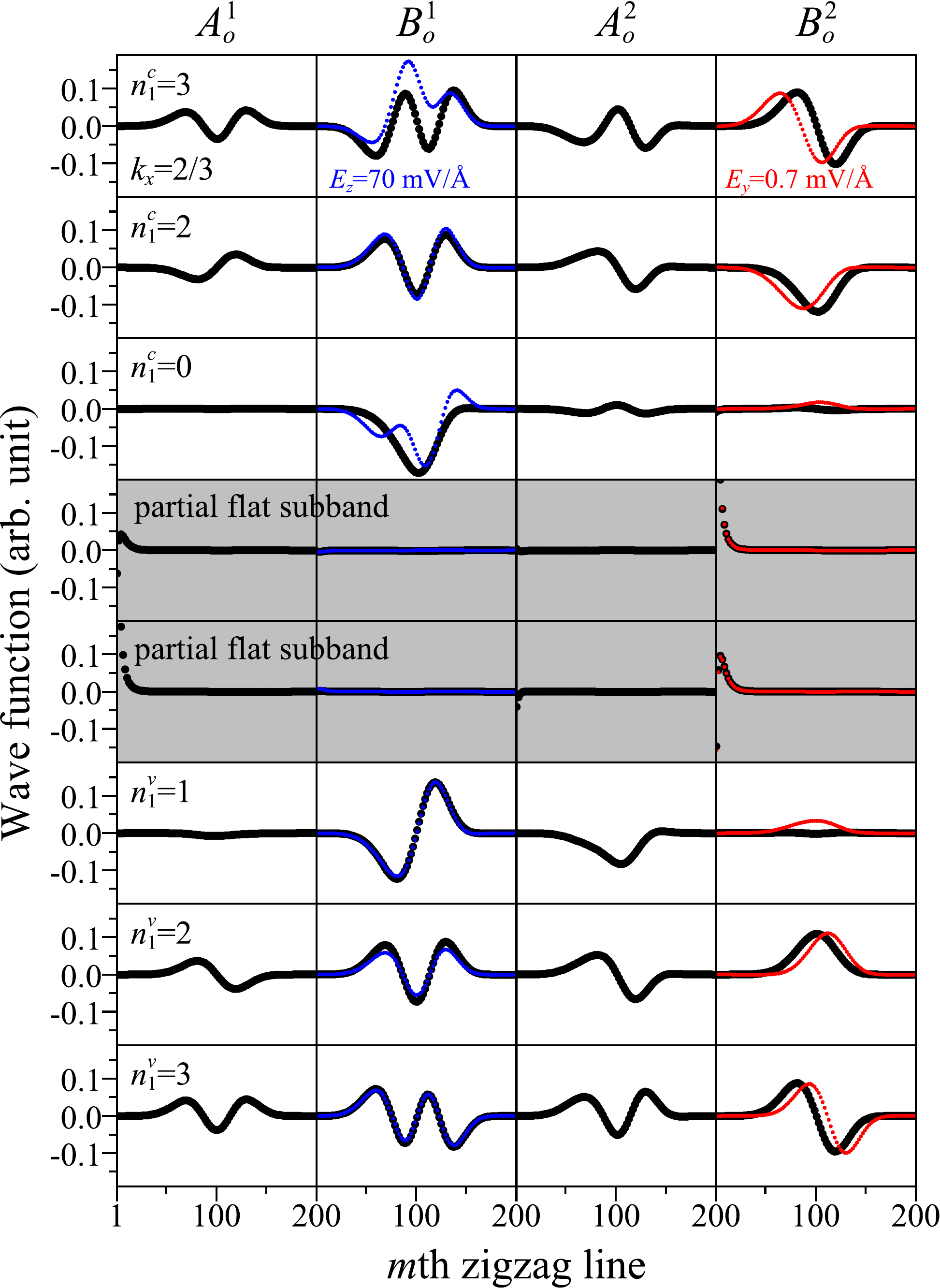}\\
  \caption[]{
  Low-energy wave functions of the $N_y = 200$ AB-stacked bilayer ZGNR at $k_x = 2/3$ in $B = 40$ T (heavy black dots), ($B = 40$ T, $E_y = 0.7$ mV/{\AA}) (light red dots), and ($B = 40$ T, $E_z = 70$ mV/{\AA}) (light blue dots).
  }
  \label{fig:WF_Bi_AB_ZGNR_in_B_Ey_Ez}
\end{center}
\end{figure}

The spatial distributions of QLL wave functions, including the amplitude, spatial symmetry, and node number, are not well behaved under the influence of interlayer interactions and electric fields.
In the first group, the wave functions at $k_x = 2/3$ are somewhat distorted and the amplitudes are different for the four sublattices (Fig.~\ref{fig:WF_Bi_AB_ZGNR_in_B_Ey_Ez}).
Apparently, the amplitude of $B^1$ sublattice is the dominating one among the four ones, so that the corresponding node number is characterized as a quantum number.
The node numbers differ by one for $A^i$ and $B^i$ sublattices in each layer, and they are identical (different) for $A^1$ and $A^2$ ($B^1$ and $B^2$) sublattices depending on the ($x$, $y$) projection.
For a $n_1^{c,v} = n$ ($\geq 2$) QLL, the number of nodes are, respectively, $n-1$, $n$, $n-1$, and $n-2$ in sublattices $A^1$, $B^1$, $A^2$, and $B^2$.
Especially, the electronic states of the four partial flat subbands are in sharp contrast to the mixed states in monolayer system\cite{J.Phys.Soc.Jpn.80(2011)044602H.C.Chung} and AA stacking (Fig.~\ref{fig:WF_Bi_AA_ZGNR_Ny=200_in_B}).
The $n_1^c = 0$ and $n_1^v = 1$ subbands belong to the Landau states, but the other two remain the edge-localized states (gray zone).
The similar quantum modes are also found in the second-group QLLs (not shown), where the node number of the dominating $A^i$ sublattice is defined as a quantum number.
The transverse electric field leads to the opposite shifts for the spatial distributions of the conduction and valence wave functions (light red dots), in which they respectively exhibit electron- and hole-like motions.
The shift of localization center will suppress the QLL-dependent optical transitions.
On the other hand, the node number is dramatically changed in the perpendicular electric field, \emph{e.g.} the $n_1^c = 0$ and $3$ QLLs (light blue dots).
The state mixing between $n^{c,v}$ and $n^{c,v} \pm 3$ QLLs is responsible for this result (discussed below).

\begin{figure}
\begin{center}
  \includegraphics[width=\linewidth, keepaspectratio]{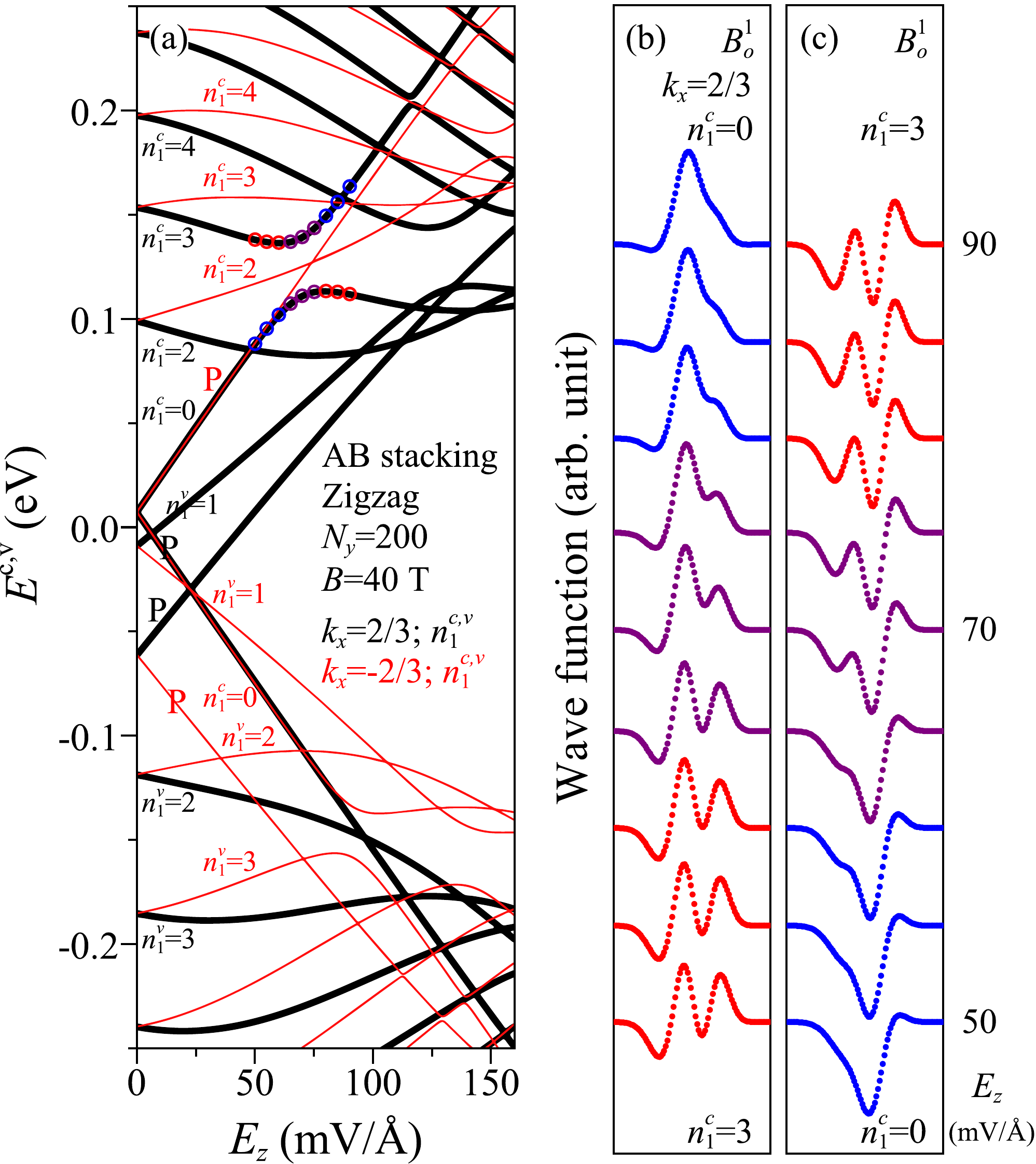}\\
  \caption[]{
  (a) $E_z$-dependent energy spectrum at $k_x = \pm 2/3$, including QLLs ($n_1^{c,v}$, $\textcolor{red}{n_1^{c,v}}$) and partial flat subbands (P, \textcolor{red}{P}).
  (b) and (c), respectively, indicate the spatial transformations in the dominating $B_o^1$ sublattice of the upper and lower QLL branches at $k_x = 2/3$ in $50 < E_z < 90$ mV/{\AA}.
  }
  \label{fig:Ez_dependent_QLL_Energies_of_Bi_AB_ZGNR}
\end{center}
\end{figure}

The $E_z$-dependent energy spectra exhibit the rich crossings and anti-crossings, as clearly indicated in Fig.~\ref{fig:Ez_dependent_QLL_Energies_of_Bi_AB_ZGNR}.
The QLLs are split at $k_x > 0$ (heave black curves) and $k_x < 0$ (light red curves) because of the destruction of inversion symmetry (Fig.~\ref{fig:Ez_dependent_QLL_Energies_of_Bi_AB_ZGNR}(a)).
As $E_z$ increases from zero, the QLL energies will grow or decline.
They have the non-linear $E_z$-dependence with crossings and anti-crossings.
The QLL crossings occur frequently in the absence of state mixings.
Specifically, the QLL anti-crossings are revealed in certain $E_z$-ranges.
Two $E_z$-dependent QLLs repel each other instead of direct crossing, and an exchange of QLL characteristics takes place.
For instance, the $n_1^c = 0$ and $3$ QLLs of $k_x = 2/3$ have an anti-crossing near $E_z = 70$ mV/{\AA} (purple circles in Fig.~\ref{fig:Ez_dependent_QLL_Energies_of_Bi_AB_ZGNR}(a))
The spatial distribution of the upper QLL branch is transformed from $n_1^c = 3$ (red dots) to $n_1^c = 0$ (blue dots) with the increase of $E_z$ (Fig.~\ref{fig:Ez_dependent_QLL_Energies_of_Bi_AB_ZGNR}(b)).
In contrast, the lower QLL branch undergoes an inverse transformation (Fig.~\ref{fig:Ez_dependent_QLL_Energies_of_Bi_AB_ZGNR}(c)).
Both branches in the anti-crossing possess the mixed characteristics of the $n_1^c = 0$ and 3 QLLs (purple dots in Fig.~\ref{fig:Ez_dependent_QLL_Energies_of_Bi_AB_ZGNR}(b) and (c)).
Such anti-crossings originate from the cooperation between the perpendicular electric field and interlayer atomic interactions.
Each $n^{c,v}$ wave function is composed of the main ($n^{c,v}$) and side ($n^{c,v} \pm 3$) modes, owing to the interlayer interactions between $B^1$ and $B^2$ sublattices.\cite{J.Phys.Soc.Jpn.17(1962)808M.Inoue, RSCAdv.4(2014)56552Y.P.Lin}
The weight of the latter is tiny at $E_z = 0$, while it is largely enhanced by the increasing $E_z$.
Two modes, with the spatial characteristics of $n^{c,v}$ and $n^{c,v} \pm 3$ nodes, become comparable during the QLL anti-crossing.
The relevant LL crossings and anti-crossings are also predicted in graphene-related systems.\cite{Phys.Rev.B87(2013)075417Y.H.Ho, RSCAdv.4(2014)56552Y.P.Lin}
Furthermore, the dramatically altered $E_z$-dependent DOS, with the QLL crossings and anti-crossings, can be explored through the STS measurements.

\begin{figure}
\begin{center}
  \includegraphics[width=\linewidth, keepaspectratio]{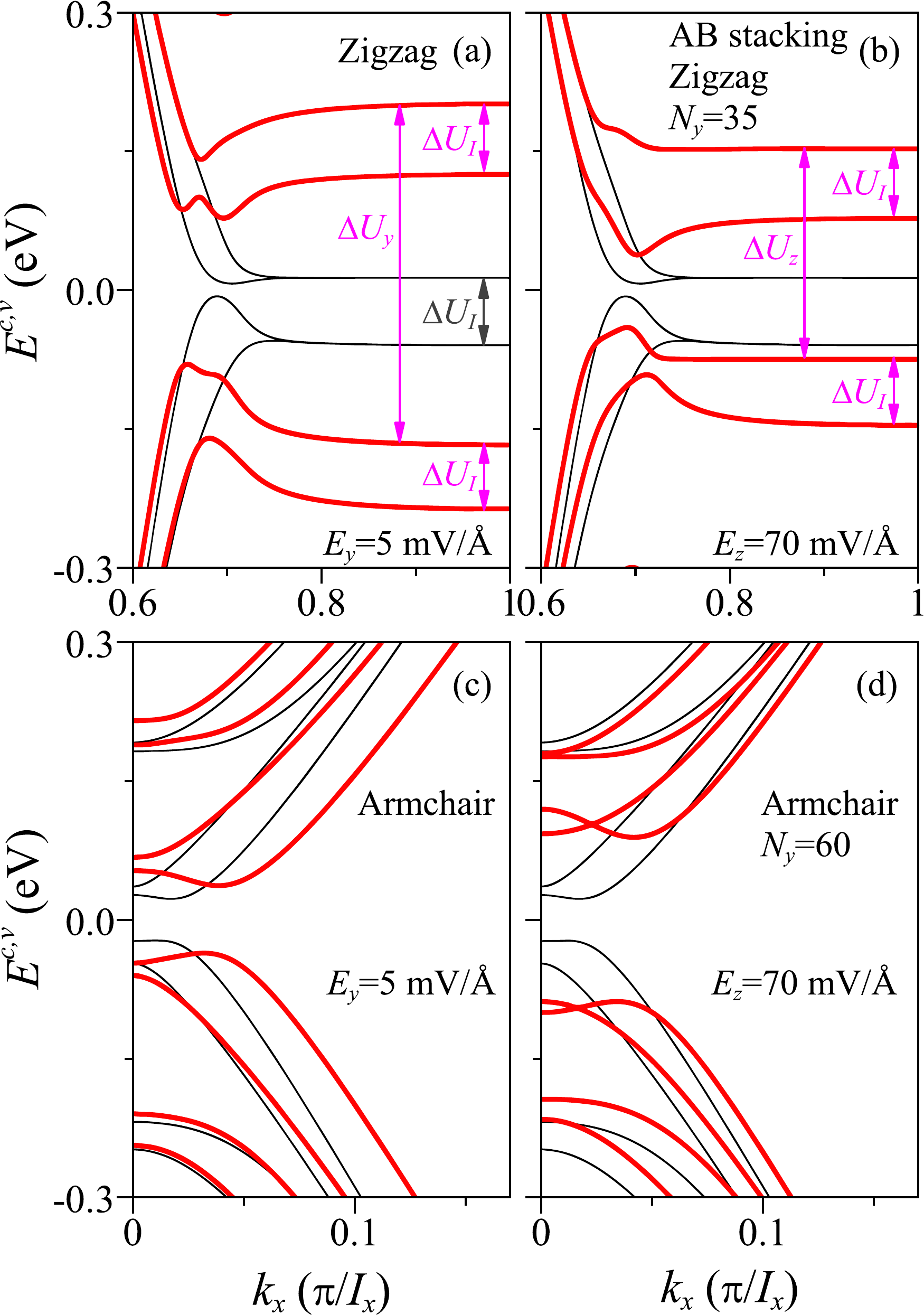}\\
  \caption[]{
  $E_y$- and $E_z$-induced gap openings for narrow bilayer AB-stacked systems: (a) \& (b) for the $N_y = 35$ ZGNRs, (c) \& (d) for the $N_y = 60$ AGNRs.
  Zero-field energy spectra are also illustrated by the light curves.
  }
  \label{fig:BSofBiAB_GNRs_Ny=35_60_B=0_Ey_Ez}
\end{center}
\end{figure}

The theoretical studies show that energy gaps of bilayer AB-stacked GNRs can be tuned by transverse\cite{Phys.Rev.Lett.99(2007)056802D.S.Novikov, J.Appl.Phys.104(2008)103714Y.C.Huang, Phys.Rev.B77(2008)245434H.Raza, J.Appl.Phys.109(2011)073704S.B.Kumar, Appl.Phys.Lett.98(2011)263105S.B.Kumar, Carbon77(2014)1031S.L.Chang} and perpendicular\cite{Phys.Rev.B78(2008)045404B.Sahu, J.Appl.Phys.104(2008)103714Y.C.Huang, J.Appl.Phys.110(2011)044309S.B.Kumar} electric fields.
Zigzag systems have an observable energy gap determined by partial flat subbands for the sufficiently narrow width ($N_y < 90$). 
Such subbands can be further separated by $E_y$ and $E_z$ (Fig.~\ref{fig:BSofBiAB_GNRs_Ny=35_60_B=0_Ey_Ez}(a) and (b)), so that the energy gap is enlarged or reduced.
Apparently, the modulation effects are relatively strong in the presence of $E_y$.
On the other hand, $E_g$ of armchair systems are determined by the energy spacing between two parabolic subbands nearest to $E_F$.
$E_y$ and $E_z$ make parabolic dispersions become oscillatory ones and thus change $E_g$ (Fig.~\ref{fig:BSofBiAB_GNRs_Ny=35_60_B=0_Ey_Ez}(c) and (d)).
The $E_y$- and $E_z$-dependent energy gaps belong to the direct or indirect ones.
The experimental verifications have been done on GNR-related systems, including the $E_y$-induced enhancement of $E_g$ in few-layer GNRs\cite{IEEETrans.ElectronDevices61(2014)3329L.T.Tung} and gap modulation in top-gated bilayer graphenes\cite{Nature459(2009)820Y.Zhang, Phys.Rev.Lett.99(2007)216802E.V.Castro, Science313(2006)951T.Ohta, Phys.Rev.Lett.102(2009)256405K.F.Mak, NanoLett.10(2010)4521J.Yan, NanoLett.10(2010)715F.Xia, NanoLett.11(2011)4759W.J.Yu} and GNRs.\cite{Sci.Rep.3(2013)1248W.J.Yu}
Especially, the predicted $E_g$ enhancements at weak $E_y$ and $E_z$ are in accord with the experimental measurements.\cite{IEEETrans.ElectronDevices61(2014)3329L.T.Tung, Sci.Rep.3(2013)1248W.J.Yu}

\subsubsection{AA stacking.}

\begin{figure}
\begin{center}
  \includegraphics[width=\linewidth, keepaspectratio]{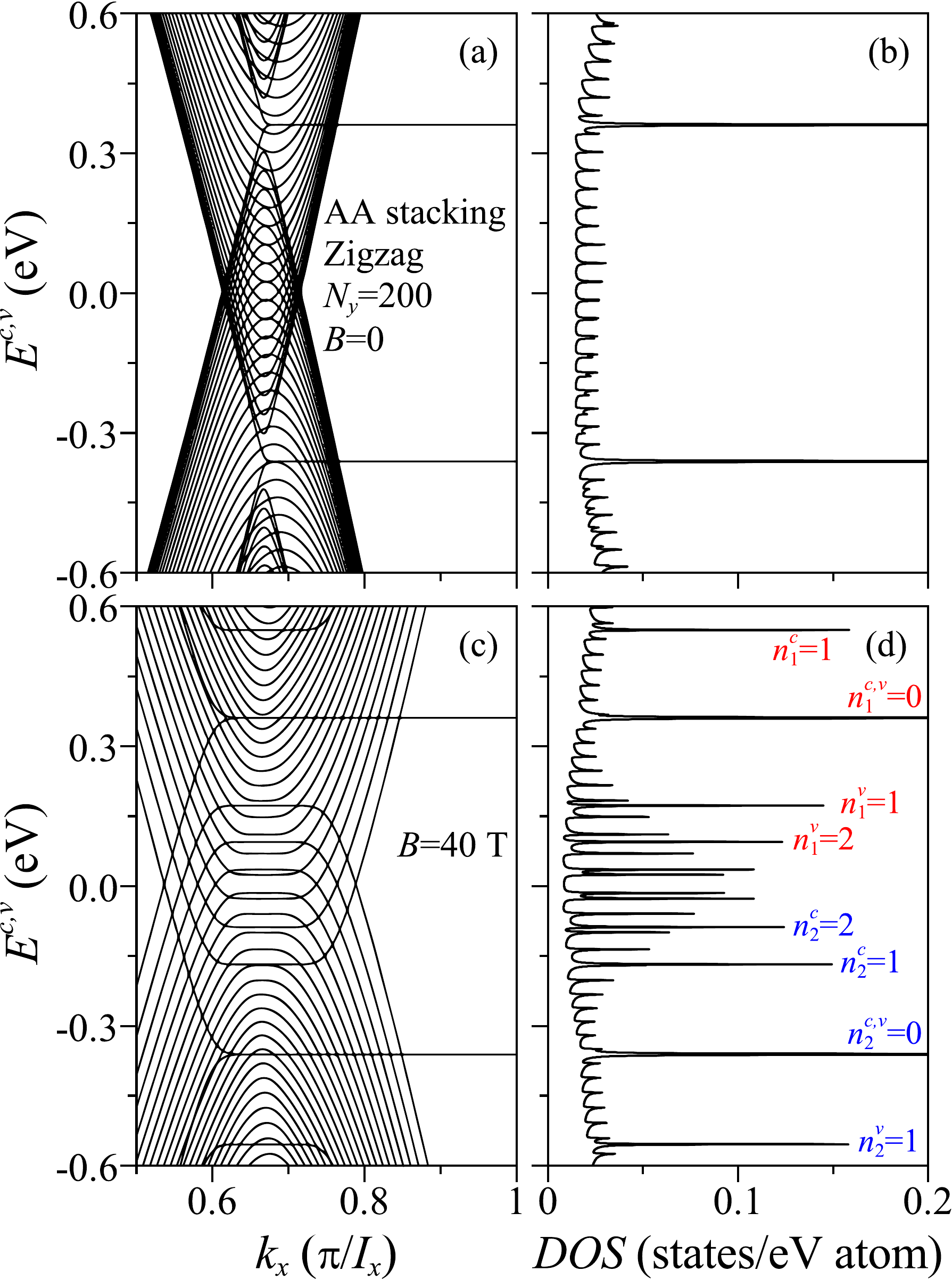}\\
  \caption[]{
  Low-energy band structure and DOS of the $N_y = 200$ bilayer AA-stacked ZGNRs at $B=0$ and 40 T.
  }
  \label{fig:BS_DOSofBi_AA_ZGNRs_inB}
\end{center}
\end{figure}

The low-energy electronic structure of bilayer AA-stacked ZGNRs consists of two groups of conduction and valence subbands (Fig.~\ref{fig:BS_DOSofBi_AA_ZGNRs_inB}(a)).
Each group, which is similar to the energy dispersions of the monolayer system (Section~\ref{sec:ElectronicStructureOfGrapheneInNoBandE}), has many parabolic dispersions and a pair of degenerate partial flat subbands.
The first and second ones, respectively, start to form at $\alpha_1$ and $-\alpha_1$, and they overlap each other without subband mixing, owing to the vertical interlayer atomic interactions.
The metallic property in the AA stacking can not be altered by transverse and perpendicular electric fields.
The DOS exhibits two highest symmetric peaks at $\pm \alpha_1$ related to partial flat subbands and many asymmetric ones of parabolic subbands, \emph{i.e.} it can be regarded as a superposition of two shifted monolayer DOSs (Fig.~\ref{fig:BS_DOSofBi_AA_ZGNRs_inB}(b)).

Two groups of QLLs are revealed in the presence of magnetic field (Fig.~\ref{fig:BS_DOSofBi_AA_ZGNRs_inB}(c)).
The first and second groups are, respectively, initiated at $\alpha_1$ and $-\alpha_1$.
The main features are similar to the monolayer system, including the $\sqrt{B}$-dependent energy spacing and reduced QLL widths with the increasing energy.
The DOS possesses two groups of primary symmetric and secondary asymmetric peaks (Fig.~\ref{fig:BS_DOSofBi_AA_ZGNRs_inB}(d)).
The former and the latter are associated with the QLLs and parabolic subbands, respectively.
Especially, there are two high-intensity peaks at $\pm \alpha_1$ surrounded by many asymmetric peaks.
The STS measurements\cite{Science324(2009)924D.L.Miller, Nature467(2010)185Y.J.Song, Nat.Commun.4(2013)1744G.Li} on them can determine the vertical interlayer atomic interaction of $\alpha_1$.
In addition, $E_y$ will result in the tilt and collapse of each QLL group, while there is no $E_z$-induced QLL splitting for $k_x > 0$ \& $k_x <0$, and the $E_z$-dependent QLL spectra only show crossings (not shown).

\begin{figure}
\begin{center}
  \includegraphics[width=\linewidth, keepaspectratio]{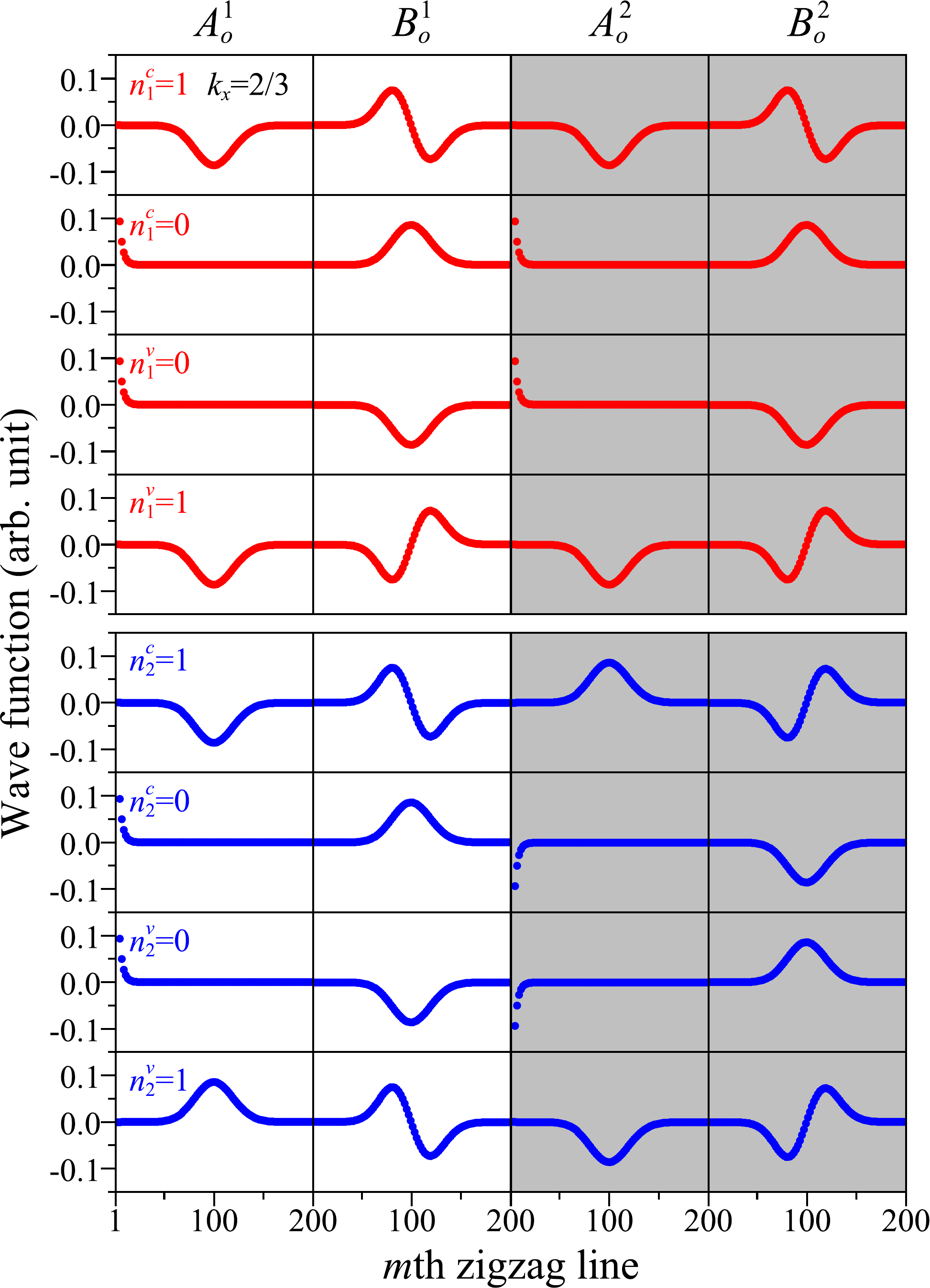}\\
  \caption[]{
  Wave functions of the $N_y = 200$ AA-stacked bilayer ZGNR at $B = 40$ T and $k_x = 2/3$.
  }
  \label{fig:WF_Bi_AA_ZGNR_Ny=200_in_B}
\end{center}
\end{figure}

The spatial distributions of Landau wave functions are well behaved, as a result of the highly symmetric stacking.
The amplitude, spatial symmetry, and node number of $A^i$ and $B^i$ sublattices in each layer are similar to those of the monolayer system.
Only the $n_1^{c,v} = 0$ and $n_2^{c,v} = 0$ QLLs have the characteristics of Landau and edge-localized states (Fig.~\ref{fig:WF_Bi_AA_ZGNR_Ny=200_in_B}).
In both QLL groups, the node numbers of $A^1$, $B^1$, $A^2$, and $B^2$ sublattices of the $n$th QLL are, respectively, $n-1$, $n$, $n-1$, and $n$.
There exists a special relation between the wave functions of $A^1$ and $A^2$ ($B^1$ and $B^2$) sublattices, in which they are in phase for the first group and out of phase for the second group.
In other words, the first- and second-group wave functions are only the symmetric and anti-symmetric superpositions of the tight-binding functions on different layers, respectively.
This will result in the absence of intergroup QLL transitions in optical spectra.

\subsection{Magneto-optical properties}
\label{sec:MagnetoOpticalPropertiesOfBilayerGNRs}

\begin{figure}
\begin{center}
  \includegraphics[width=\linewidth, keepaspectratio]{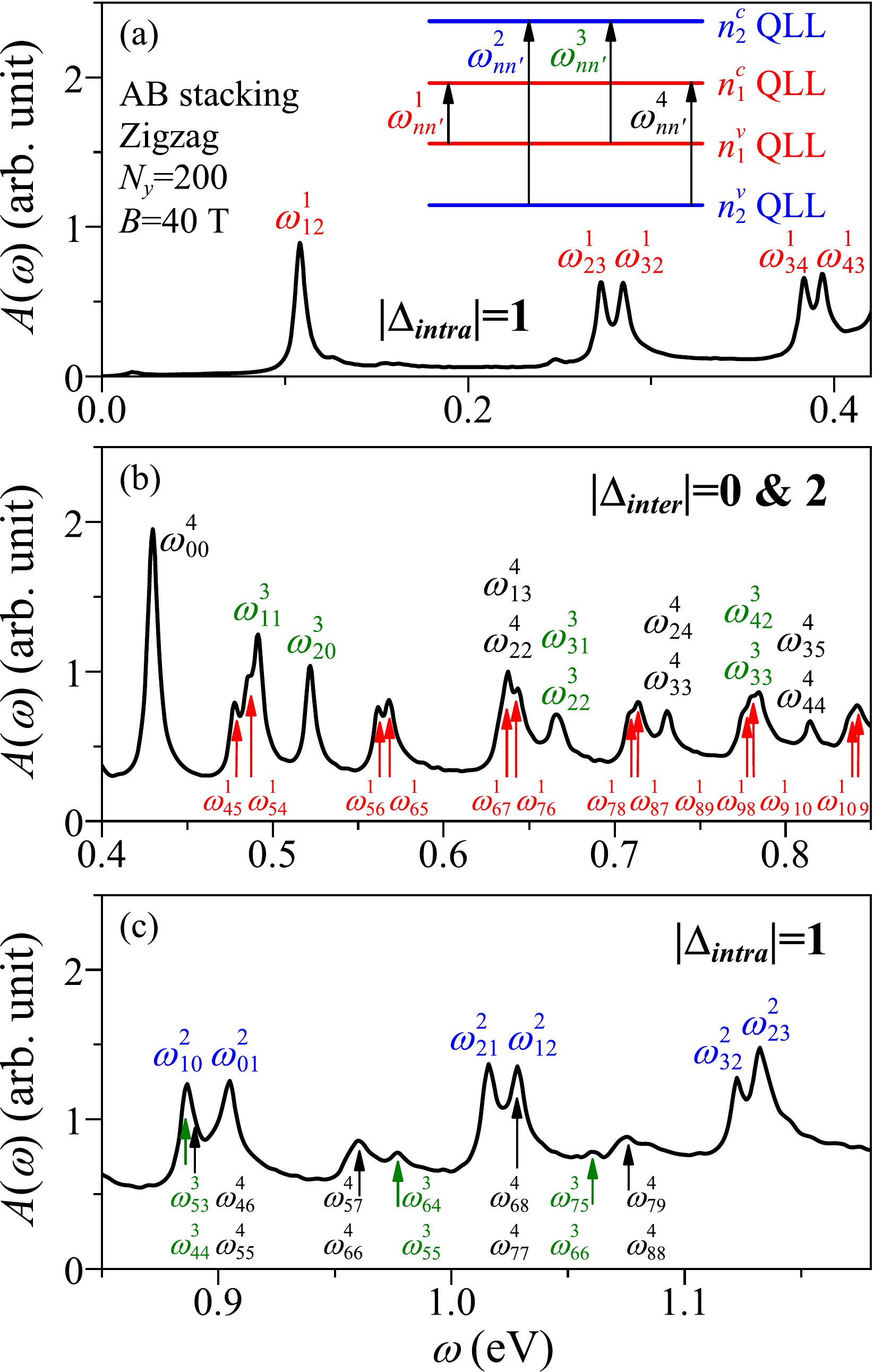}\\
  \caption[]{
  Low-energy magneto-absorption spectrum of the $N_y = 200$ bilayer AB-stacked ZGNR at $B = 40$ T.
  }
  \label{fig:ABS_Bi_AB_ZGNR_inB}
\end{center}
\end{figure}

Two groups of QLLs in bilayer AB-stacked ZGNR exhibit four categories of inter-QLL absorption peaks, where half of them belong to intragroup or intergroup excitations.
$\omega_{nn'}^1$, $\omega_{nn'}^2$, $\omega_{nn'}^3$, and $\omega_{nn'}^4$ indicate the inter-QLL transitions of $n^v_1 \rightarrow n'^c_1$, $n^v_2 \rightarrow n'^c_2$, $n^v_1 \rightarrow n'^c_2$, and $n^v_2 \rightarrow n'^c_1$, respectively.
The threshold absorption frequencies are $\omega \sim 0$ for $\omega_{nn'}^1$ (Fig.~\ref{fig:ABS_Bi_AB_ZGNR_inB}(a)), $\omega \sim 2 \gamma_1$ for $\omega_{nn'}^2$ (Fig.~\ref{fig:ABS_Bi_AB_ZGNR_inB}(c)), and $\omega \sim \gamma_1$ for $\omega_{nn'}^3$ and $\omega_{nn'}^4$ (Fig.~\ref{fig:ABS_Bi_AB_ZGNR_inB}(b)).
The intragroup and intergroup optical transitions, respectively, obey the selection rules of $| \Delta n_{intra} | = 1$ and $| \Delta n_{inter} | = 0$ \& $2$, since the Landau modes are the same on the $i$th layer for the $A^i$ ($B^i$) sublattice of the initial state and the $B^i$ ($A^i$) sublattice of the final states.\cite{Phys.Rev.B78(2008)115422Y.C.Huang}
The intragroup absorption spectrum has prominent twin-peak structures for $\omega_{n(n-1)}^i$ and $\omega_{(n-1)n}^i$ except the threshold transition peak in the first category.
However, the intergroup spectrum presents strong single peaks.
The twin- and single-peak structures are determined by the asymmetric QLL spectrum and quantum mode.
Magneto-absorption spectra can be further changed by transverse and perpendicular electric fields.
The tilted and collapsed QLLs in the transverse one significantly suppress the excitation channels.
The splitting QLLs in the perpendicular one result in more low-intensity absorption peaks.

\begin{figure}
\begin{center}
  \includegraphics[width=\linewidth, keepaspectratio]{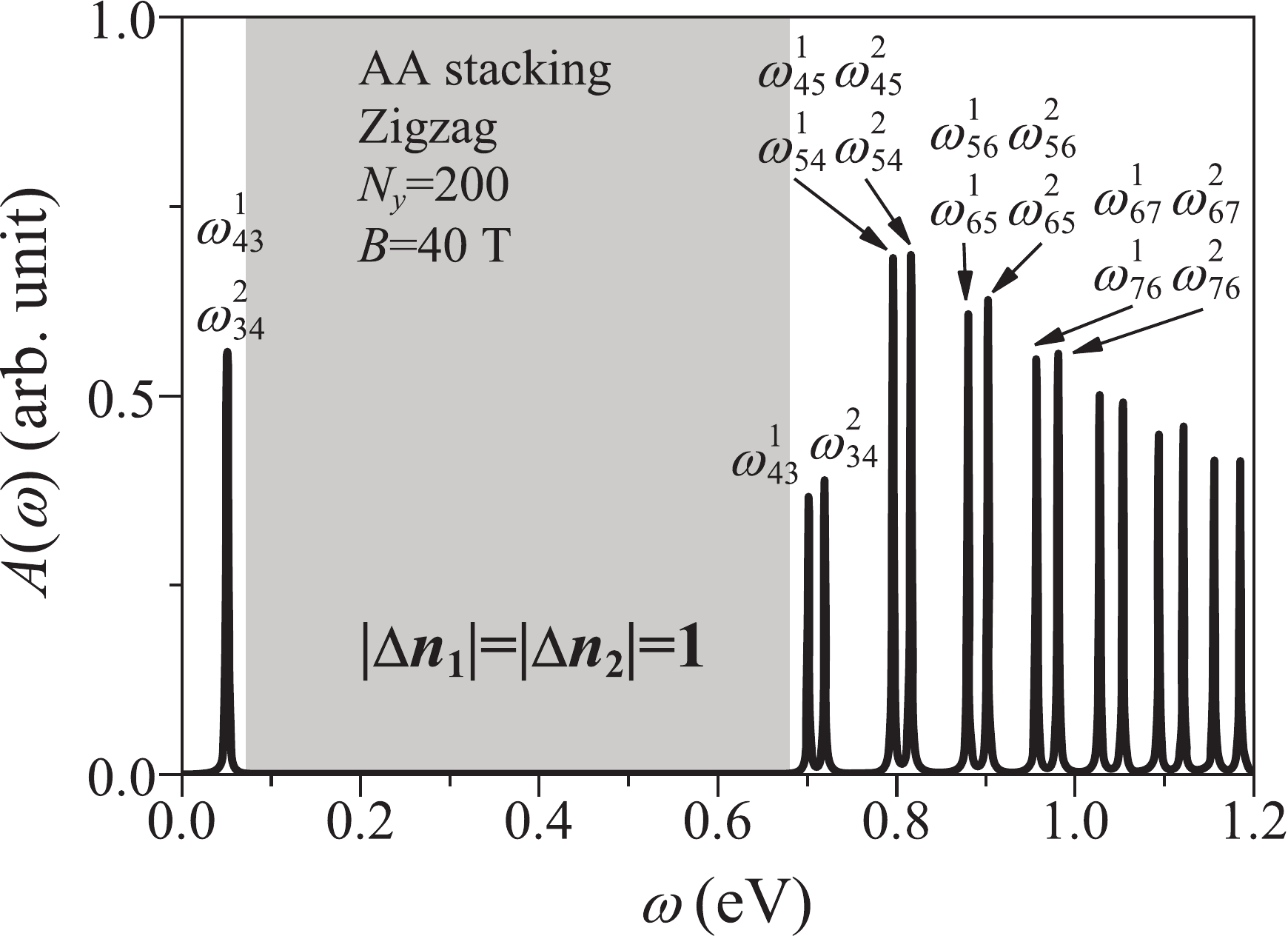}\\
  \caption[]{
  Low-energy magneto-absorption spectrum of the $N_y = 200$ bilayer AA-stacked ZGNR at $B = 40$ T.
  }
  \label{fig:ABS_Bi_AA_ZGNR_inB}
\end{center}
\end{figure}

Bilayer AA-stacked ZGNRs demonstrate peculiar magneto-absorption spectra based on the simple relations in QLL wave functions and Fermi-Dirac distribution.
The first- and second-group wave functions are, respectively, the symmetric and anti-symmetric superpositions of the Landau modes on different layers, so that the velocity matrix elements are finite only for the intragroup transitions.
Two categories of intragroup QLL peaks obeying the selection rules of $|\Delta n_1| = |\Delta n_2| = 1$ are clearly illustrated in Fig.~\ref{fig:ABS_Bi_AA_ZGNR_inB}.
The threshold absorption peak is composed of the valence-to-valence optical transition of $\tilde{n}^v_1 \rightarrow \tilde{n}^v_1-1$ and the conduction-to-conduction one of $\tilde{n}^c_2 \rightarrow \tilde{n}^c_2 + 1$, since the Fermi level is located in the middle of the two related QLLs.
However, the other peaks come from the valence-to-conduction excitations of $n^v_1 \rightarrow n^c_1=n^v_1 \pm 1$ and $n^v_2 \rightarrow n^c_2=n^v_2 \pm 1$, where $n^v_1 \geq \tilde{n}^v_1$ and $n^v_2 \geq \tilde{n}^v_2$.
Each single peak consists of $\omega^i_{n(n+1)}$ and $\omega^i_{(n+1)n}$ with the same frequency except that two initial ones ($\tilde{n}^v_1 \rightarrow n^c_1=\tilde{n}^v_1 - 1$ and $\tilde{n}^v_2 \rightarrow n^c_2=\tilde{n}^v_2 + 1$) are limited by the Fermi-Dirac distribution.
The valence-to-conduction excitation frequencies are much higher than the threshold one, thus leading to an optical gap of $2\alpha_1$ (gray zone).

The inter-QLL transitions and selection rules can be explored through the experimental measurements, as done for few-layer graphene-related systems by magneto-optical transmission\cite{Phys.Rev.Lett.97(2006)266405M.L.Sadowski, Phys.Rev.Lett.98(2007)197403Z.Jiang, Phys.Rev.B76(2007)081406R.S.Deacon, Phys.Rev.Lett.100(2008)087401P.Plochocka, Phys.Rev.Lett.110(2013)246803J.M.Poumirol} and magneto-Raman spectroscopy.\cite{NanoLett.14(2014)4548S.Berciaud}
Theoretical predictions indicate that the important differences in AB and AA stackings include excitation categories, structures, frequencies, and numbers of absorption peaks.
Bilayer GNR arrays, which have been recently synthesized,\cite{NanoLett.14(2014)4581H.Yan} can provide strong-intensity optical responses.
Measurements on inter-QLL transitions are useful in the identification of stacking configuration.
Furthermore, the vertical interlayer atomic interactions ($\gamma_1$ and $\alpha_1$) can be examined from the initial energies in distinct excitation categories.

\section{Non-uniform GNRs and GNR-CNT hybrids}
\label{NonUniformGNRs}

More complicated carbon-related systems can be synthesized by the assembly of distinct subsystems.
They display rich essential properties, and most of them only have the weak van der Waals interactions.
Non-uniform few-layer GNRs can be fabricated by the mechanical exfoliation,\cite{Phys.Rev.B79(2009)235415C.P.Puls, Phys.Rev.B88(2013)125410J.Tian, NanoLett.10(2010)562X.Xu} cutting of graphene,\cite{Nat.Nanotechnol.3(2008)397L.Tapaszto, NanoLett.8(2008)1912S.S.Datta, NanoLett.9(2009)457N.Severin, NanoRes.2(2009)695F.Schaffel, J.Am.Chem.Soc.132(2010)10034S.Fujii} chemical unzipping of CNT,\cite{Nature458(2009)872D.V.Kosynkin, Nature458(2009)877L.Jiao, Nanoscale3(2011)3876U.K.Parashar, ACSNano5(2011)968D.V.Kosynkin, Nat.Nanotechnol.5(2010)321L.Jiao} and CVD.\cite{NanoLett.8(2008)2773J.Campos-Delgado, J.Am.Chem.Soc.131(2009)11147D.C.Wei}
Some theoretical studies on them are chiefly focused on the electronic,\cite{Appl.Phys.Lett.97(2010)263114G.Kim} magnetic,\cite{Phys.Rev.B82(2010)205436M.Koshino, Philos.Mag.94(2014)2812M.H.Lee} and transport\cite{Phys.Rev.B76(2007)165416J.Nilsson, Phys.Rev.B81(2010)195406J.W.Gonzalez} properties.
Moreover, the coupling of carbon allotropes can form various hybridized systems with unique geometric configurations, such as the graphene-nanotube hybrids,\cite{Adv.Mater.20(2008)1706D.Cai, NanoLett.9(2009)1949V.C.Tung, Chem.Commun.46(2010)8279Q.Su, J.Phys.Chem.Lett.1(2010)467D.Yu, Carbon49(2011)3597M.Y.Yen, J.Mater.Chem.22(2012)9949B.P.Vinayan, Energy53(2013)282J.Y.Jhan, Synth.Met.203(2015)127H.K.Park} GNR networks,\cite{Nanoscale3(2011)5156X.Dong} carbon nanopeapods (a fullerene chain within a CNT),\cite{Nature396(1998)323B.W.Smith, Phys.Rev.Lett.85(2000)5384K.Hirahara, Chem.Phys.Lett.337(2001)48S.Bandow, Nature415(2002)1005J.Lee, Science295(2002)828D.J.Hornbaker, NanoLett.4(2004)2451K.Urita, Phys.Rev.B73(2006)075406H.Shiozawa, Appl.Phys.Lett.92(2008)183115Y.F.Li, Chem.-Eur.J.19(2013)14061T.Iwamoto} and carbon-nanobud hybrids.\cite{Nat.Nanotechnol.2(2007)156A.G.Nasibulin, Chem.Phys.Lett.446(2007)109A.G.Nasibulin, J.Am.Chem.Soc.130(2009)7188Y.Tian}
Extensive theoretical researches have been performed on the electronic structure,\cite{Phys.Rev.Lett.86(2001)3835S.Okada, ACSNano2(2008)1459X.Wu, J.Phys.Chem.C113(2009)20822H.Y.He, Phys.Chem.Chem.Phys.13(2011)3925C.H.Lee, ACSNano6(2012)5539S.Osella} magnetic quantization,\cite{J.Phys.-Condes.Matter21(2009)435302T.S.Li} optical spectra,\cite{J.Phys.Chem.C119(2015)5679F.Fergani} and quantum conductance.\cite{J.Appl.Phys.107(2010)063714T.S.Li}

The magneto-electronic properties of bilayer non-uniform GNRs can be easily tuned by the geometric structures, such as the widths of the top and bottom layers, relative position between them, and inner boundaries of narrow ribbon.
Four kinds of magneto-electronic energy spectra are revealed, including the monolayer-like, bilayer-like, and coexistent QLLs, as well as the distorted energy subbands.
Especially, the coexistent one can survive when the monolayer and bilayer regions are much wider than the magnetic length.
The preliminary evidences for the coexistence of QLL spectra are revealed from the conductance measurements.\cite{Phys.Rev.B79(2009)235415C.P.Puls, Phys.Rev.B88(2013)125410J.Tian}
The QLL wave functions are well behaved, being responsible for many strong absorption peaks with two specific selection rules.
As to the fourth kind, the wave functions present serious variations between two layers and severe distortions with piecewise continuity.
Consequently, there are many low-intensity absorption peaks without any selection rules.

A GNR-CNT hybrid, a coupling of GNR with open boundaries and CNT with a closed boundary, is chosen to comprehend the hybridization effects.
The relative position between GNR and CNT can change the interactions and enrich the magneto-electronic properties, such as the breaking of symmetric energy spectrum, partial destruction of QLLs, more extra band-edge states, disruption of DOS peaks, and piecewise distorted wave functions.
Moreover, the prominent magneto-absorption peaks are largely reduced and surrounded by additional subpeaks.

The organization of this section is stated as follows.
The first subsection describes the magneto-electronic properties of non-uniform bilayer GNRs and GNR-CNT hybrids, including the geometry-dependent energy spectra, DOS, and wave functions.
The effects of atomic interactions between two subsystems are discussed in detail.
The feature-rich optical properties, with different kinds of magneto-absorption spectra, are presented in the next subsection.
Some theoretical predictions are roughly consistent with the experimental evidences.

\subsection{Magneto-electronic properties}
\label{MagnetoElectronicPropertiesOfNonUniformAndHybridSystems}

\begin{figure}
\begin{center}
  \includegraphics[width=\linewidth, keepaspectratio]{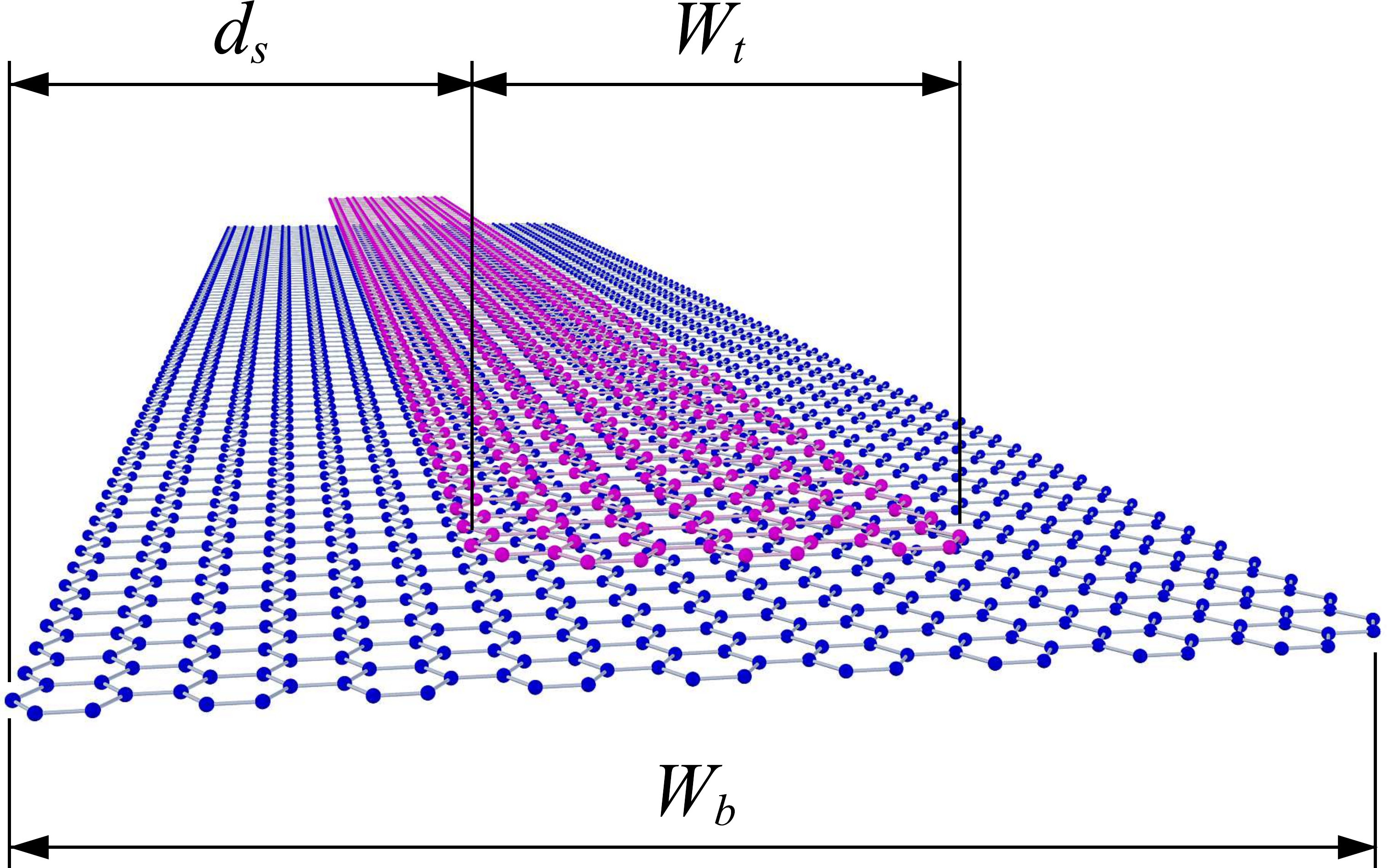}\\
  \caption[]{
  Geometric structure of a non-uniform bilayer AB-stacked ZGNR.
  The top layer of width $W_t$ is shifted by a distance of $d_s$ from the edge of the wide bottom layer of width $W_b$.
  }
  \label{fig:GeometricStructure_Bi_AB_ZGNR_unequalWidth}
\end{center}
\end{figure}

A non-uniform bilayer AB-stacked ZGNR is illustrated to explore the effects of geometric structures on essential properties.
The narrow top and wide bottom layers have widths $W_t$ and $W_b$, respectively (Fig.~\ref{fig:GeometricStructure_Bi_AB_ZGNR_unequalWidth}).
The former will be shifted from the left-hand side of the latter, where $d_s$ is the distance between their boundaries, and the AB stacking is fixed in the theoretical calculations.
The tuning of the relative position and ribbon widths, leading to obvious variations in the interlayer atomic interactions and boundary conditions, can diversify the magneto-electronic and magneto-optical properties.

\begin{figure*}
\begin{center}
  \includegraphics[width=\linewidth, keepaspectratio]{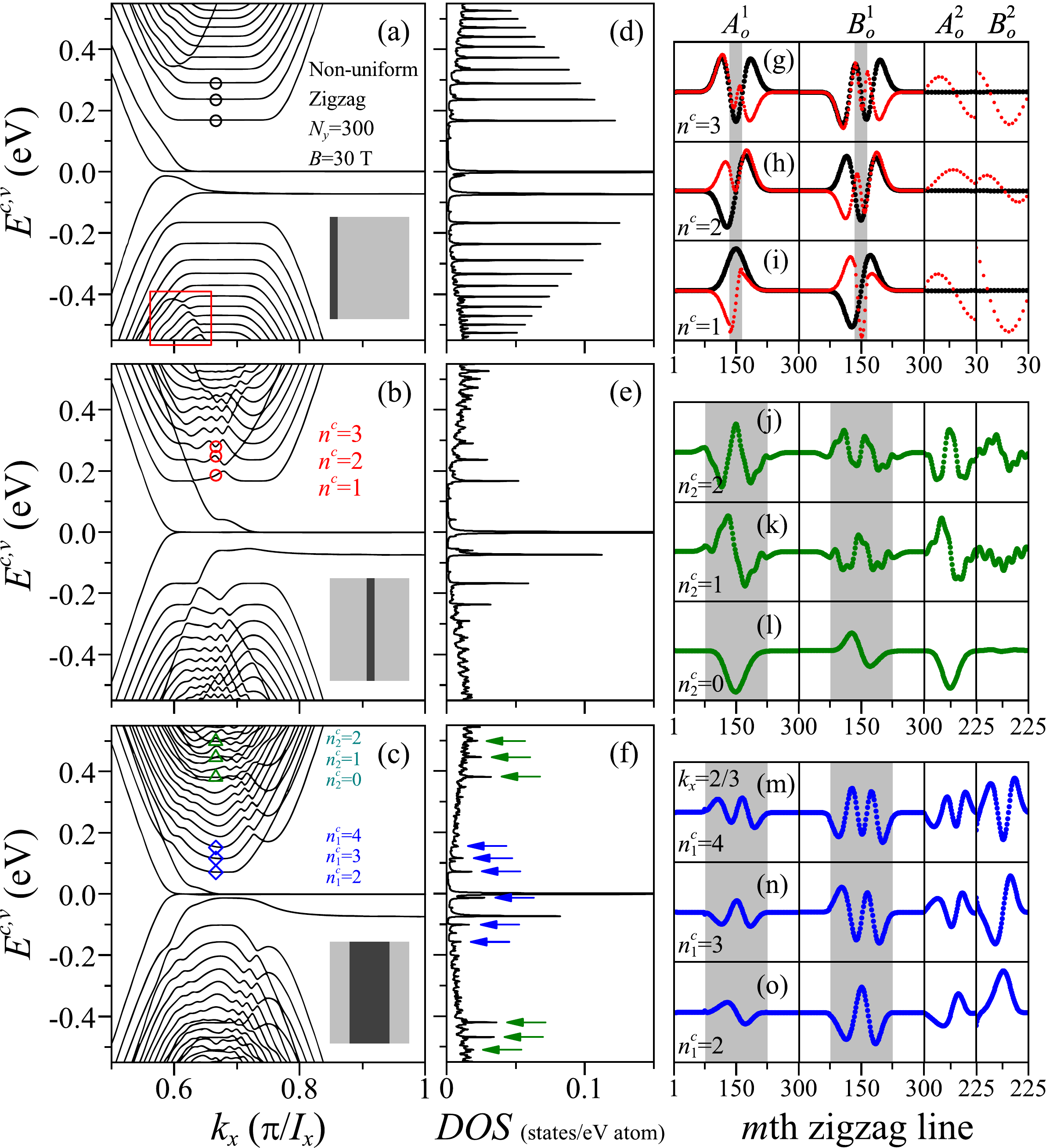}\\
  \caption[]{
  Low-energy magneto-electronic structures of the $N_y=300$ non-uniform bilayer AB-stacked ZGNRs at $B = 30$ T for (a) ($W_t = W_b/10$, $d_s = b/2$); (b) ($W_t = W_b/10$, $d_s = (W_b-W_t+b)/2$); (c) ($W_t = W_b/2$, $d_s = (W_b-W_t+b)/2$).
  Also shown is the top view of each geometric configuration.
  The corresponding DOSs are shown in (d), (e), and (f), respectively.
  The Landau wave functions at the positions marked by the black and red circles, green triangles, and blue diamonds are demonstrated in (g)--(o) with the same color.
  The gray zones indicate the top layer.
  }
  \label{fig:BS_DOS_WF_Bi_AB_ZGNR_unequalWidth_inB}
\end{center}
\end{figure*}

The magneto-electronic structure of a non-uniform system is roughly composed of two monolayer-like ones, when the top layer is much narrower than the bottom one (Fig.~\ref{fig:BS_DOS_WF_Bi_AB_ZGNR_unequalWidth_inB}(a) and (b)).
At $d_s = b/2$, the parabolic subbands of the top layer are slightly mixed with the monolayer-like QLLs of the bottom layer (Fig.~\ref{fig:BS_DOS_WF_Bi_AB_ZGNR_unequalWidth_inB}(a)).
There are some subband mixings between the parabolic and Landau subbands, \emph{e.g.} $E^{c,v} < -0.4$ eV as indicated by the red rectangle.
More band-edge states due to them are revealed in the subpeak structures surrounding the high-intensity QLL peaks in the DOS (Fig.~\ref{fig:BS_DOS_WF_Bi_AB_ZGNR_unequalWidth_inB}(d)).
The dispersionless features of low-lying QLLs are almost unaffected, and so do the wave functions.
For instance, it is obvious that the $n^c = 1$--$3$ wave functions at $k_x = 2/3$ (black circles in Fig.~\ref{fig:BS_DOS_WF_Bi_AB_ZGNR_unequalWidth_inB}(a)) possess Landau modes in $A^1$ and $B^1$ sublattices similar to those of monolayer GNR (heavy black dots in Fig.~\ref{fig:BS_DOS_WF_Bi_AB_ZGNR_unequalWidth_inB}(g)--(i)).

The mixing of energy spectra is largely enhanced, as the top layer is gradually shifted to the center of the bottom layer.
Especially at $d_s = (W_b-W_t+b)/2$, many dispersionless QLLs are turned into oscillatory ones with a lot of extra band-edge states (Fig.~\ref{fig:BS_DOS_WF_Bi_AB_ZGNR_unequalWidth_inB}(b)).\cite{Philos.Mag.94(2014)2812M.H.Lee}
The QLL peaks in the DOS are seriously suppressed and most of them are replaced by the lower asymmetric peaks (Fig.~\ref{fig:BS_DOS_WF_Bi_AB_ZGNR_unequalWidth_inB}(e)).
Also, they induce drastic alternations in the wave functions.
For example, at $k_x = 2/3$ (red circles in Fig.~\ref{fig:BS_DOS_WF_Bi_AB_ZGNR_unequalWidth_inB}(b)), the wave functions become piecewise continuous functions of differentiability class $C^0$ or have two jump discontinuities in $A^1$ and $B^1$ sublattice at two edge positions of the top ribbon (edges of gray zone in Fig.~\ref{fig:BS_DOS_WF_Bi_AB_ZGNR_unequalWidth_inB}(g)--(i)), so that the normal Landau modes are absent.
Moreover, the spatial distributions of the top layer turn out to be observable, showing the distorted standing waves.

The monolayer-like QLL spectrum can be transformed into the bilayer-like one, when the width of the top layer is larger than the magnetic length.
For $W_t = W_b/2~(\gg l_B)$, the monolayer-like QLLs evolves into bilayer-like ones (Fig.~\ref{fig:BS_DOS_WF_Bi_AB_ZGNR_unequalWidth_inB}(c)).
The second group of QLLs are initiated at the energies of the vertical interlayer atomic interactions ($\pm \gamma_1$).
The energy spacings between QLLs are shrunk, and there is no $\sqrt{B}$-dependence in the QLL energies.
It should be noted that the two groups of QLLs are also mixed with the extra parabolic subbands.
The latter are related to the non-overlapping region of the bottom layer, whose width is insufficiently wide for the magnetic quantization.
Two groups of QLL peaks are indicated by the blue and green arrows in Fig.~\ref{fig:BS_DOS_WF_Bi_AB_ZGNR_unequalWidth_inB}(f).
The low-intensity asymmetric peaks are contributed by the extra band-edge states.
The QLL wave functions at $k_x = 2/3$ are similar to those of bilayer ZGNRs.
The Landau modes and the dominating sublattices for each group keep the same, \emph{e.g.} the low-lying QLLs of the second and first groups as shown in Fig.~\ref{fig:BS_DOS_WF_Bi_AB_ZGNR_unequalWidth_inB}(j)--(l) and Fig.~\ref{fig:BS_DOS_WF_Bi_AB_ZGNR_unequalWidth_inB}(m)--(o), respectively.
However, the Landau wave functions of the former also have the extra component of the standing waves due to parabolic subbands, presenting slightly distorted distributions (Fig.~\ref{fig:BS_DOS_WF_Bi_AB_ZGNR_unequalWidth_inB}(j) and (k)).

\begin{figure}
\begin{center}
  \includegraphics[width=\linewidth, keepaspectratio]{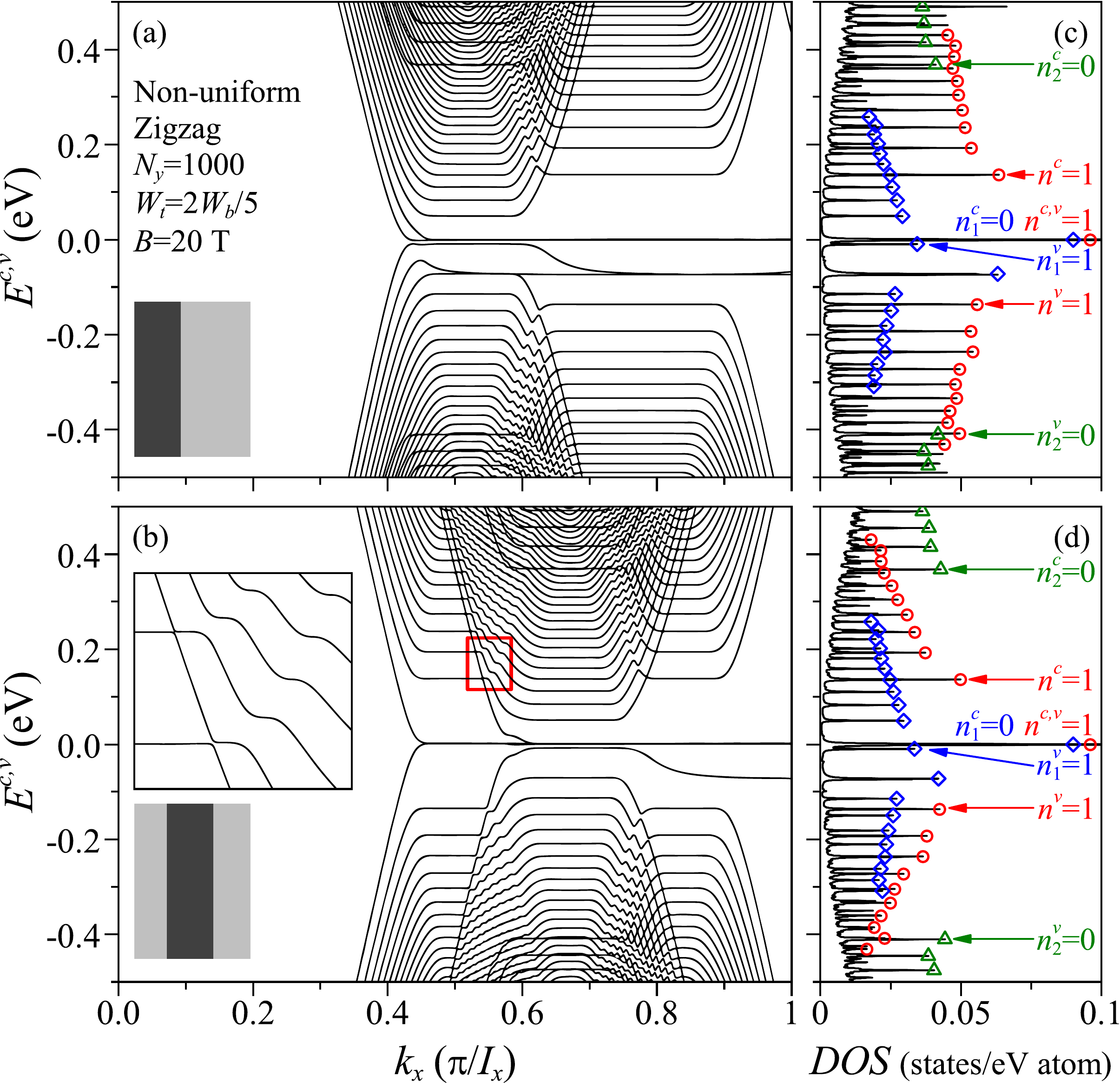}\\
  \caption[]{
  Monolayer- and bilayer-like QLL spectra in the $N_y = 1000$ non-uniform GNRs of two different geometric configurations: (a) ($W_t = 2W_b/5$, $d_s = b/2$) and (b) ($W_t = 2W_b/5$, $d_s = (W_b-W_t+b)/2$).
  The top view of each configuration is illustrated by the gray zones.
  Inset magnifies the intermediate QLLs within the red box.
  The corresponding DOSs are shown in (c) and (d), where the red circles, blue diamonds, and green triangles indicate the first few monolayer-like QLL peaks and bilayer-like QLL peaks of the first and second groups, respectively.
  }
  \label{fig:TheCoexistenceOfMonolayerAndBilayerQLL}
\end{center}
\end{figure}

The monolayer- and bilayer-like QLL spectra can survive simultaneously, as the overlapping and non-overlapping regions are sufficiently wide to quantize electronic states.
For $W_t = 2W_b/5$ and $d_s = b/2$ (Fig.~\ref{fig:TheCoexistenceOfMonolayerAndBilayerQLL}(a)), the bilayer- and monolayer-like QLLs, respectively, appear at $k_x < 2/3$ and $k_x > 2/3$.
Also, there is an extra group of intermediate Landau subbands strongly mixing with QLLs.
The wave functions are distributed in the piecewise form near the interface between two regions, where the distorted bilayer and monolayer Landau wave functions are joined.\cite{Phys.Rev.B82(2010)205436M.Koshino}
But for $W_t = 2W_b/5$ and $d_s = (W_b-W_t+b)/2$ (Fig.~\ref{fig:TheCoexistenceOfMonolayerAndBilayerQLL}(b)), there are two monolayer-like QLLs, one bilayer-like QLLs, and two extra groups of intermediate Landau subbands, reflecting the geometric structure with two interfaces.
The monolayer-like QLL peaks of wider spacings (red circles) and bilayer-like QLL peaks of narrower spacings (blue diamonds and green triangles) are revealed in the DOS (Fig.~\ref{fig:TheCoexistenceOfMonolayerAndBilayerQLL}(c) and (d)), and the relative peak height is related to the geometric configurations.
For instance, the monolayer-like QLL peaks decline rapidly because of subband mixings, as illustrated in Fig.~\ref{fig:TheCoexistenceOfMonolayerAndBilayerQLL}(d).

\begin{figure}
\begin{center}
  \includegraphics[width=\linewidth, keepaspectratio]{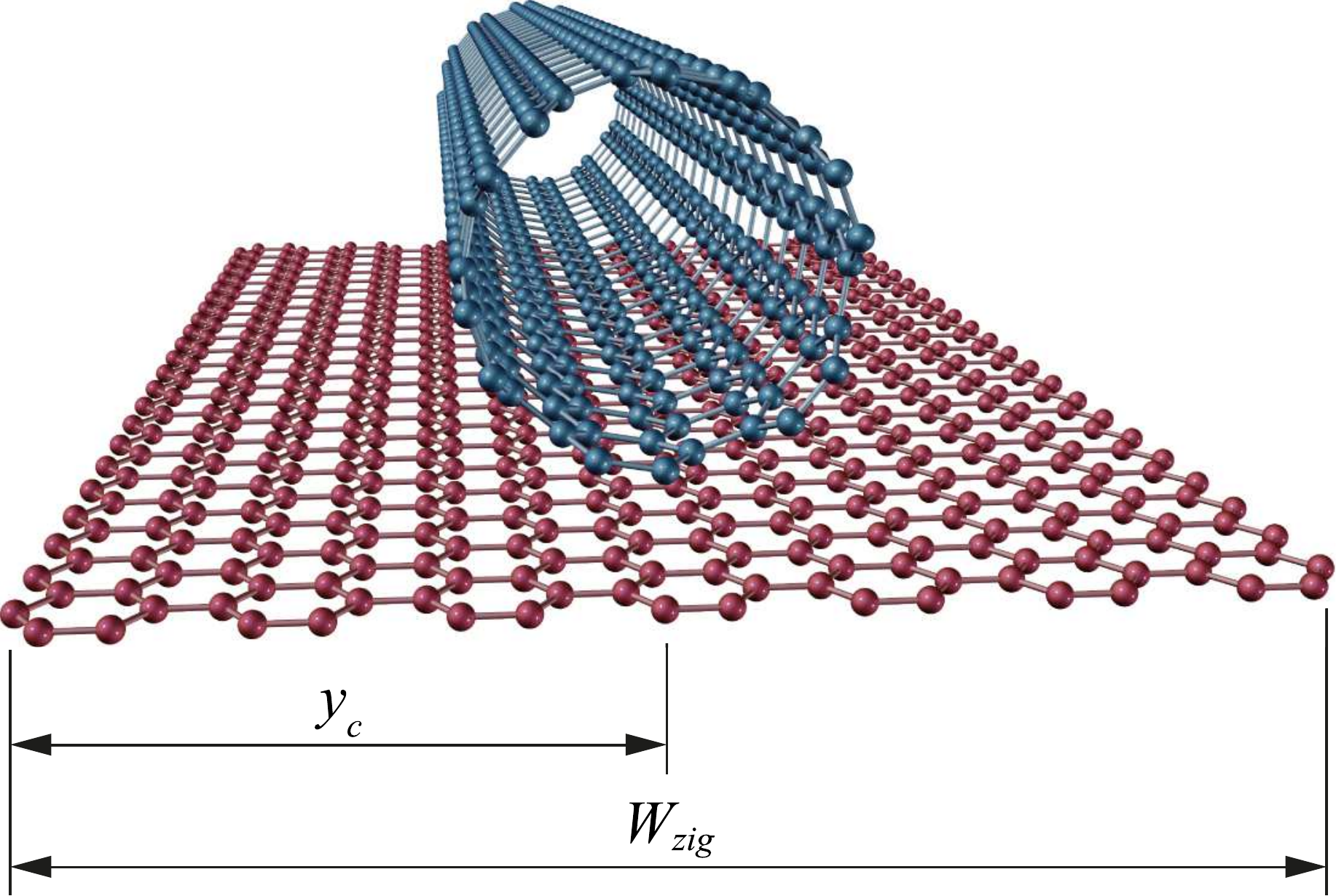}\\
  \caption[]{
  Geometric structure of a GNR-CNT hybrid.
  $y_c$ is the transverse distance between the CNT axis and ribbon edge, and $W_{zig}$ is the ribbon width.
  }
  \label{fig:GeometricStructure_of_ACNT_ZGNR_Hybrid}
\end{center}
\end{figure}

The GNR-CNT hybrids can be synthesized by the chemical unzipping of multi-wall CNTs.\cite{Nature458(2009)872D.V.Kosynkin, Carbon48(2010)2596F.Cataldo, Angew.Chem.-Int.Edit.52(2013)3996Z.Yang}
In the theoretical model, an ($m$, $m$) ACNT is located above a ZGNR (Fig.~\ref{fig:GeometricStructure_of_ACNT_ZGNR_Hybrid}), where they have the same period along the longitudinal direction.
There are $4m+2N_y$ carbon atoms in a primitive unit cell.
The AB stacking remains the same during the shift of CNT.
The Lennard-Jones potential is employed to find the optimal vertical distance corresponding to the minimal van der Waals interaction energy\cite{J.Chem.Phys.25(1956)693L.A.Girifalco}.
The inter-subsystem interactions in the Hamiltonian are assumed to decay exponentially with the interatomic distance.\cite{Phys.Rev.Lett.90(2003)026601K.H.Ahn, J.Phys.-Condes.Matter21(2009)435302T.S.Li}
The addition and shift of CNT strongly modifies the inter-subsystem interactions and essential properties, such as energy dispersions, DOS, wave functions, and magneto-absorption spectra.

\begin{figure}
\begin{center}
  \includegraphics[width=\linewidth, keepaspectratio]{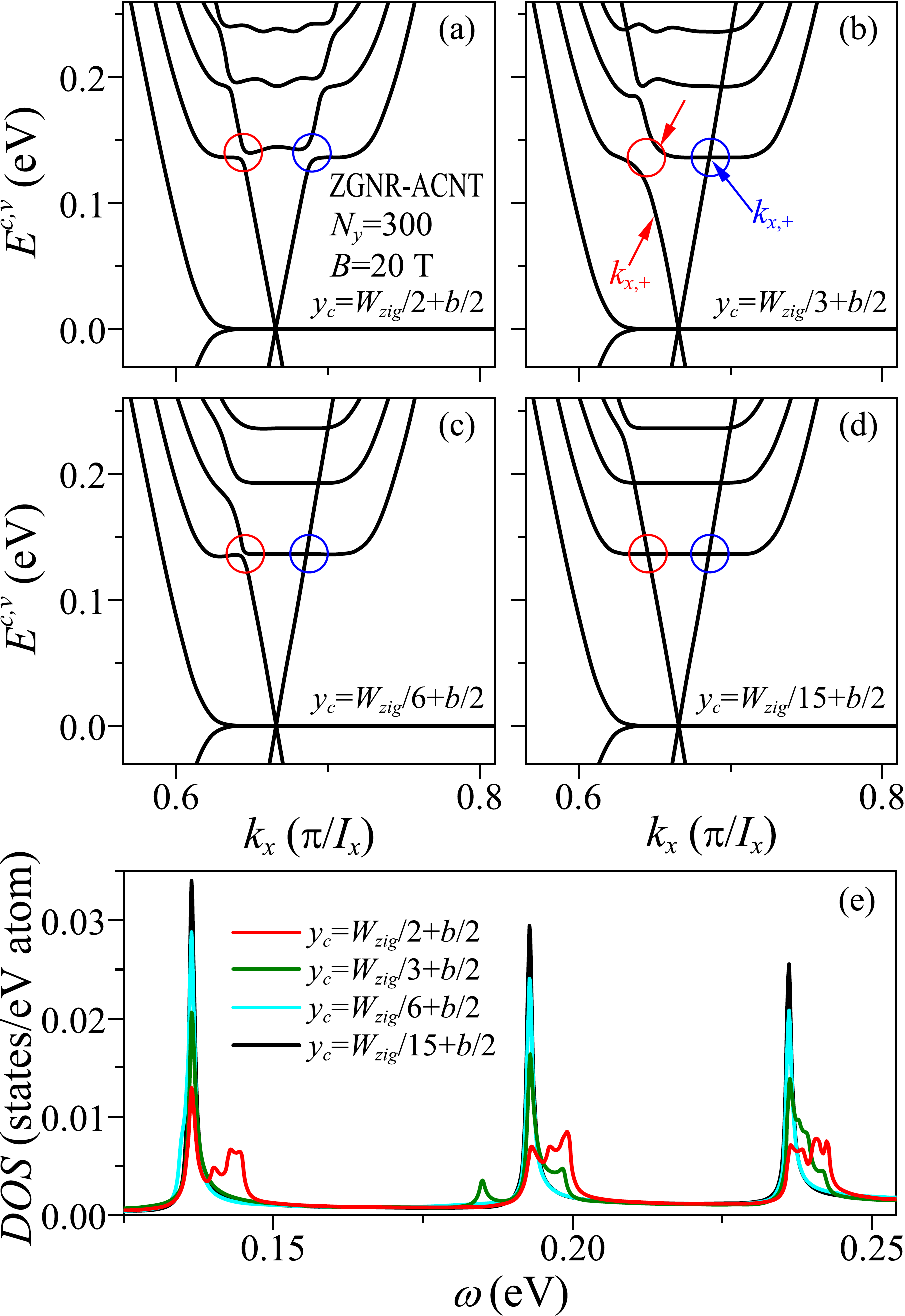}\\
  \caption[]{
  Low-energy magneto-electronic spectra of the ZGNR-ACNT hybrid at $B = 20$ T.
  The ($6$, $6$) ACNT is located at $y_c =$ (a) $W_{zig}/2+b/2$, (b) $W_{zig}/3+b/2$, (c) $W_{zig}/6 + b/2$, and (d) $W_{zig}/15 + b/2$.
  DOS is shown in (e).
  }
  \label{fig:BS_DOS_HybridSystem}
\end{center}
\end{figure}

The GNR-CNT interactions alter the magneto-electronic structures considerably.
The linear subbnads contributed by a metallic ($6$, $6$) ACNT mix with the low-lying QLLs, resulting in the breaking of symmetric spectrum about $E_F$, partial destruction of dispersionless QLLs, more extra band-edge states, and piecewise distorted wave functions.
These variations strongly depends on the shift of CNT, as shown in Fig.~\ref{fig:BS_DOS_HybridSystem}.
For the case of CNT at the ribbon center, the strong mixings between the linear subbands and QLLs occur at $k_{x,-}$ ($< 2/3$) and $k_{x,+}$ ($> 2/3$) (red and blue circles in Fig.~\ref{fig:BS_DOS_HybridSystem}(a)).
When the CNT gradually approaches to the ribbon edge, the subband mixings at $k_{x,+}$'s will almost disappear (blue circles in Fig.~\ref{fig:BS_DOS_HybridSystem}(b)--(d)).
These results are closely related to the DOS and spatial distributions of Landau wave functions.
The prominent QLL peaks in DOS are significantly lowered, and they are surrounded by more asymmetric peaks (Fig.~\ref{fig:BS_DOS_HybridSystem}(e)).

\begin{figure}
\begin{center}
  \includegraphics[width=\linewidth, keepaspectratio]{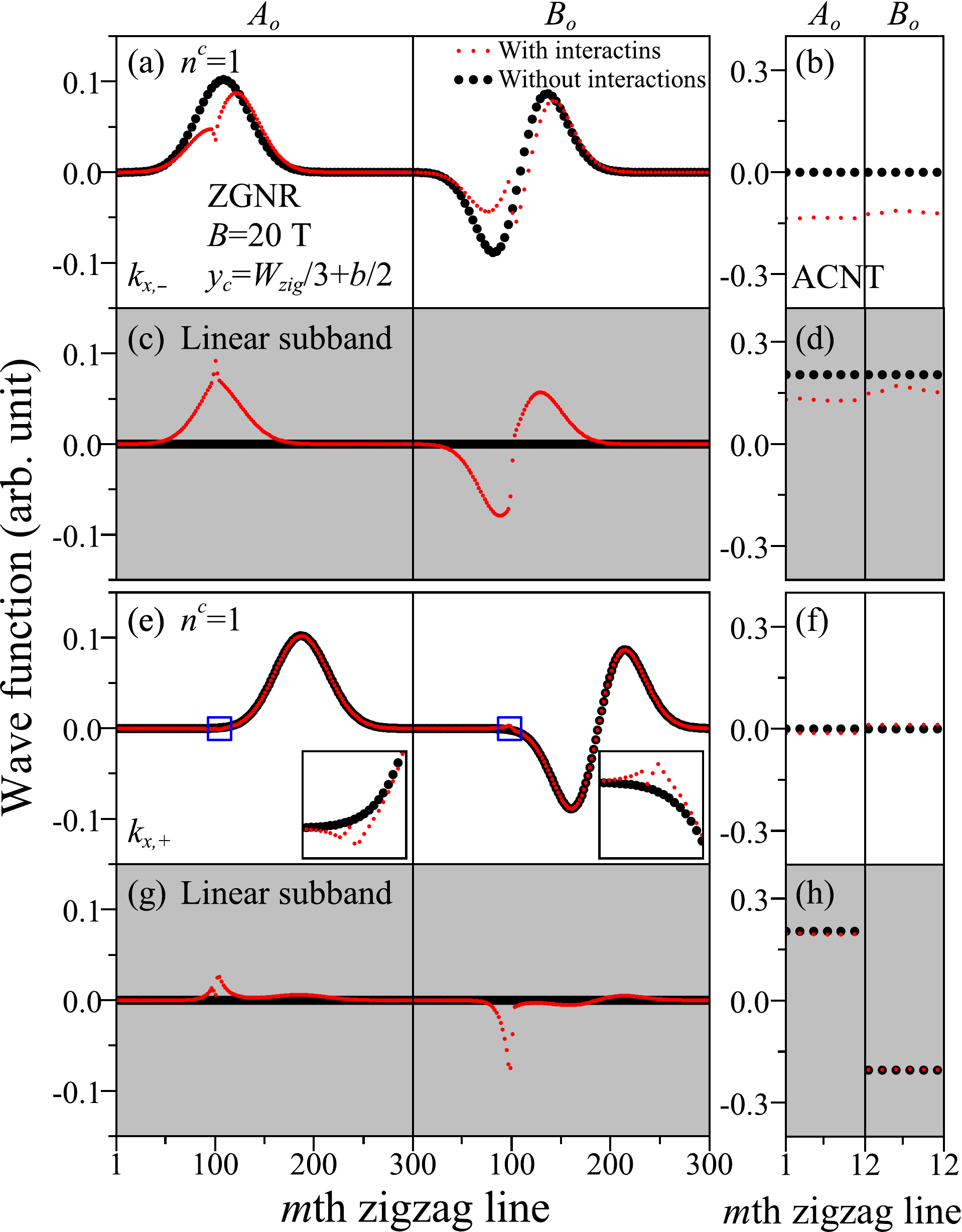}\\
  \caption[]{
  Wave functions of the ZGNR-ACNT hybrid at $B=20$ T, $y_c = W_{zig}/3 + b/2$, and $k_x= k_{x,-}$ \& $k_{x,+}$ (light red dots).
  Also shown are those of two independent systems (heavy black dots).
  Insets magnify the region near the CNT center.
  }
  \label{fig:WF_HybridSystem}
\end{center}
\end{figure}

The wave functions, depending on the wave vector and CNT location, present more information about the state mixings, as illustrated by the $n^c = 1$ QLL and linear subband (Fig.~\ref{fig:WF_HybridSystem}).
Two independent subsystems, respectively, exhibit the well-behaved Landau modes of $n^c = 1$ and the uniform spatial distributions along the azimuthal direction of CNT (heavy black dots in Fig.~\ref{fig:WF_HybridSystem}(a) and (d)).
The inter-subsystem interactions induce serious distortions in wave functions as  CNT is situated at the localization center of Landau wave function.
At $k_{x,-}$ (red arrows in Fig.~\ref{fig:BS_DOS_HybridSystem}(b)), the spatial distributions in CNT become non-uniform (light red dots in Fig.~\ref{fig:WF_HybridSystem}(d)).
Furthermore, the Landau wave function possesses the kink-form spatial distribution (Fig.~\ref{fig:WF_HybridSystem}(a)), indicating that the Landau state is highly suppressed by CNT.
More importantly, state mixing leads to the non-uniform azimuthal distribution in the $n^c = 1$ QLL (Fig.~\ref{fig:WF_HybridSystem}(b)) and the distorted Landau states in the linear subband (Fig.~\ref{fig:WF_HybridSystem}(c)).
But for $k_{x,+}$, there exist only weak mixings (blue arrow in Fig.~\ref{fig:BS_DOS_HybridSystem}(b)), so that the electronic states almost remain the same (Fig.~\ref{fig:WF_HybridSystem}(e)--(h)), and slight disruptions in Landau wave functions are revealed at the position of CNT.

The aforementioned four kinds of magneto-electronic energy spectra in the non-uniform GNRs can be identified by the STS measurements on the DOS.\cite{Phys.Rev.B71(2005)193406Y.Kobayashi, Phys.Rev.Lett.94(2005)226403T.Matsui, Phys.Rev.B73(2006)085421Y.Niimi, Phys.Rev.Lett.97(2006)236804Y.Niimi, Science324(2009)924D.L.Miller}
Especially, the coexistence of monolayer- and bilayer-like QLLs has been evidenced by conductance measurements.
Through adjusting the gate voltage, a dramatic suppression of the bilayer resistance oscillation is revealed, because of the coexistent monolayer and bilayer QLLs.\cite{Phys.Rev.B79(2009)235415C.P.Puls}
In order to understand the inter-subsystem interactions, the CNT-position-induced special structures in DOS require further experimental measurements.
Also, the spectroscopic-imaging STM\cite{Phys.Rev.Lett.101(2008)256802K.Hashimoto, Nat.Phys.6(2010)811D.L.Miller, Phys.Rev.Lett.109(2012)116805K.Hashimoto, Nat.Phys.10(2014)815Y.S.Fu} is appropriate to figure out the hybridization effects on the spatial distributions of wave functions.
The theoretical studies on the non-uniform GNRs and hybridized systems are related to those of the partly overlapped GNRs\cite{Phys.Rev.B81(2010)195406J.W.Gonzalez, Appl.Phys.Lett.98(2011)192112K.M.M.Habib, Carbon75(2014)411M.Berahman} and GNR array-graphene hybrids.\cite{Synth.Met.161(2011)489C.H.Lee, Diam.Relat.Mat.20(2011)1026C.H.Lee, J.Phys.Chem.C117(2013)7326J.H.Wong}
In addition, the partly overlapped structure between CVD-grown graphene grains has been recently synthesized.\cite{ACSNano5(2011)6610A.W.Robertson, Science336(2012)1143A.W.Tsen, Carbon80(2014)513R.Rao}

\subsection{Magneto-optical properties}

\begin{figure}[]
\begin{center}
  \includegraphics[width=\linewidth, keepaspectratio]{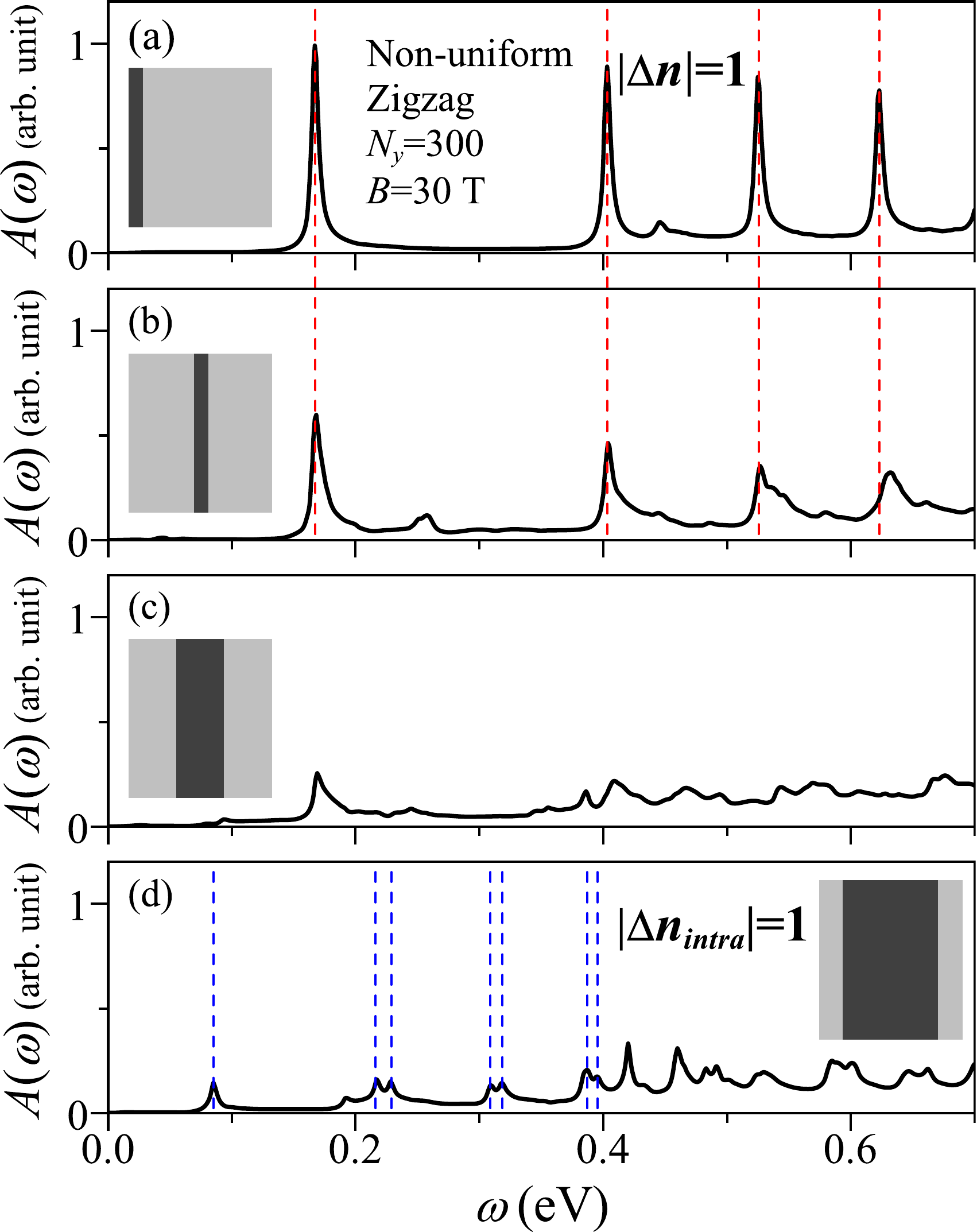}\\
  \caption[]{
  Low-frequency magneto-optical spectra of the $N_y=300$ non-uniform ZGNRs at $B = 30$ T for (a) ($W_t = W_b/10$, $d_s = b/2$); (b) ($W_t = W_b/10$, $d_s = (W_b-W_t+b)/2$); (c) ($W_t = W_b/3$, $d_s = (W_b-W_t+b)/2$); (d) ($W_t = 2W_b/3$, $d_s = (W_b-W_t+b)/2$).
  Gray rectangles demonstrate the top view of geometric configurations.
  Dashed red and blue lines indicate the frequencies of monolayer and bilayer QLL transition peaks, respectively.
  }
  \label{fig:ABS_NonUniformSystem}
\end{center}
\end{figure}

\begin{figure}
\begin{center}
  \includegraphics[width=\linewidth, keepaspectratio]{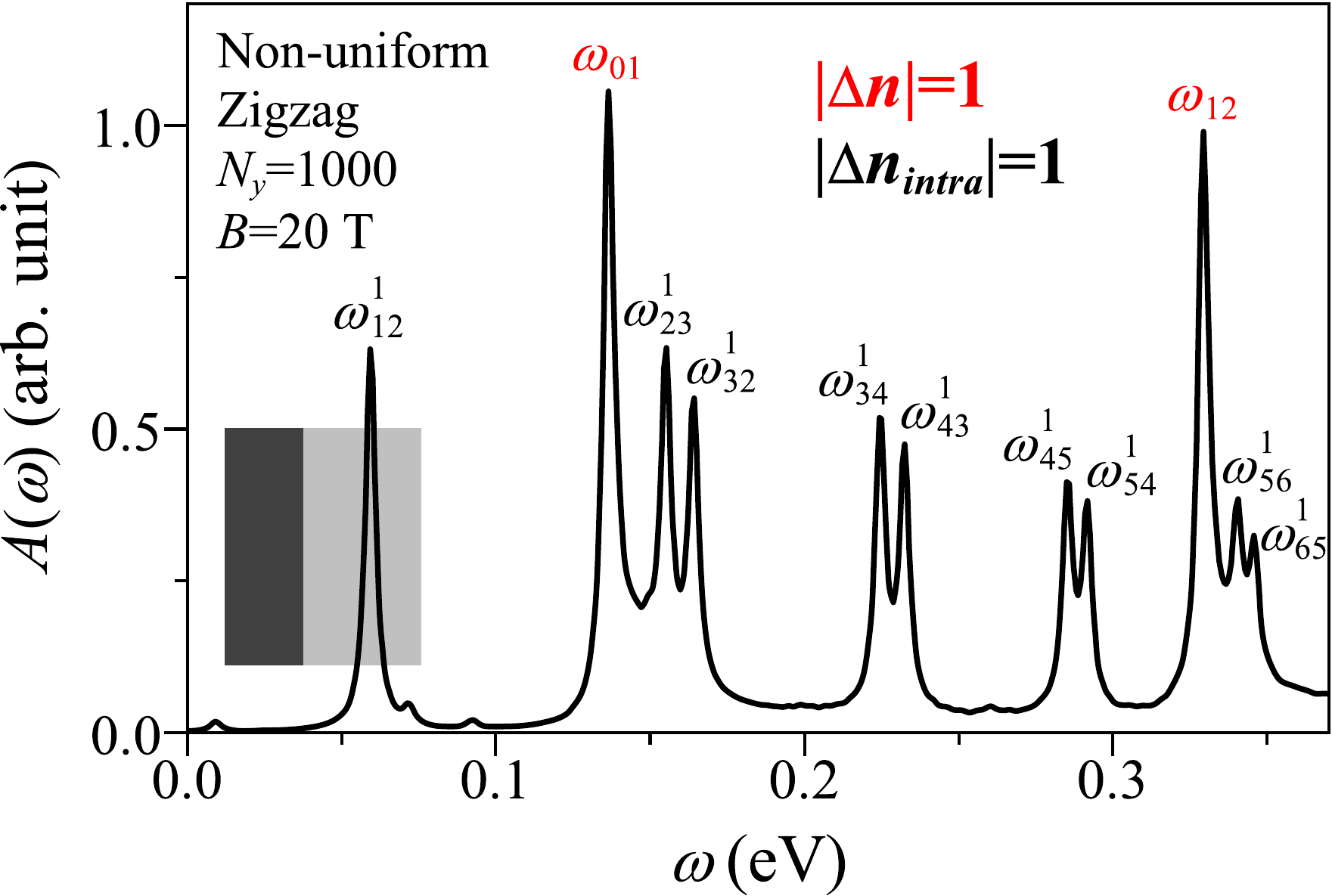}\\
  \caption[]{
  Magneto-optical spectrum of the $N_y=1000$ non-uniform ZGNR at $B = 30$ T, $W_t = 2W_b/5$, and $d_s = b/2$.
  The top view of geometric configuration is illustrated by the gray rectangles.
  }
  \label{fig:ABS_CoexistentSpectra_NonUniformSystem}
\end{center}
\end{figure}

Based on the geometric configurations, the non-uniform GNRs possess four kinds of magneto-absorption spectra, including the monolayer, bilayer and coexistent ones, as well as the irregular one.
The magneto-absorption spectrum is almost the same with that of monolayer (dashed red lines) except for few extra subpeaks, when the narrow top GNR is on the edge of the bottom one (Fig.~\ref{fig:ABS_NonUniformSystem}(a)).
The number, intensity, and frequency of absorption peaks are very sensitive to the position of the top layer, especially at $d_s = (W_b-W_t+b)/2$ (Fig.~\ref{fig:ABS_NonUniformSystem}(b)).
The intensity of prominent peaks is significantly reduced, but the selection rule of $|\Delta n| = 1$ keeps unchanged.
As the top ribbon becomes wider and covers the distribution region of Landau wave functions ($W_t = W_b/3$ in Fig.~\ref{fig:ABS_NonUniformSystem}(c)), the severe subband mixings cause the suppression of QLL transition peaks, the absence of selection rule, and many low-intensity peaks.
The main reason is that the dispersionless QLLs are absent and the spatial distributions of wave functions are seriously distorted.
However, two groups of Landau subbands are formed as the top layer width is sufficiently wide for the magnetic quantization, \emph{e.g.} $W_t = 2W_b/3 > l_B$ in Fig.~\ref{fig:ABS_NonUniformSystem}(d).
Consequently, the absorption spectrum is similar to the bilayer one.
The main peaks obey the same selection rule, and their frequencies are very close to those of bilayer GNRs (dashed blue lines).
Particularly, the monolayer and bilayer QLL transitions can survive simultaneously, when the non-overlapping and overlapping regions are wide enough for the magnetic quantization (Fig.~\ref{fig:ABS_CoexistentSpectra_NonUniformSystem}).
The absorption peaks of $|\Delta n| = 1$ and $|\Delta n_{intra}| = 1$ selection rules are clearly present, and the relative peak height between them is roughly proportional to the width ratio of two distinct regions.\cite{Non_uniform_arXiv1512.08415}

\begin{figure}
\begin{center}
  \includegraphics[width=\linewidth, keepaspectratio]{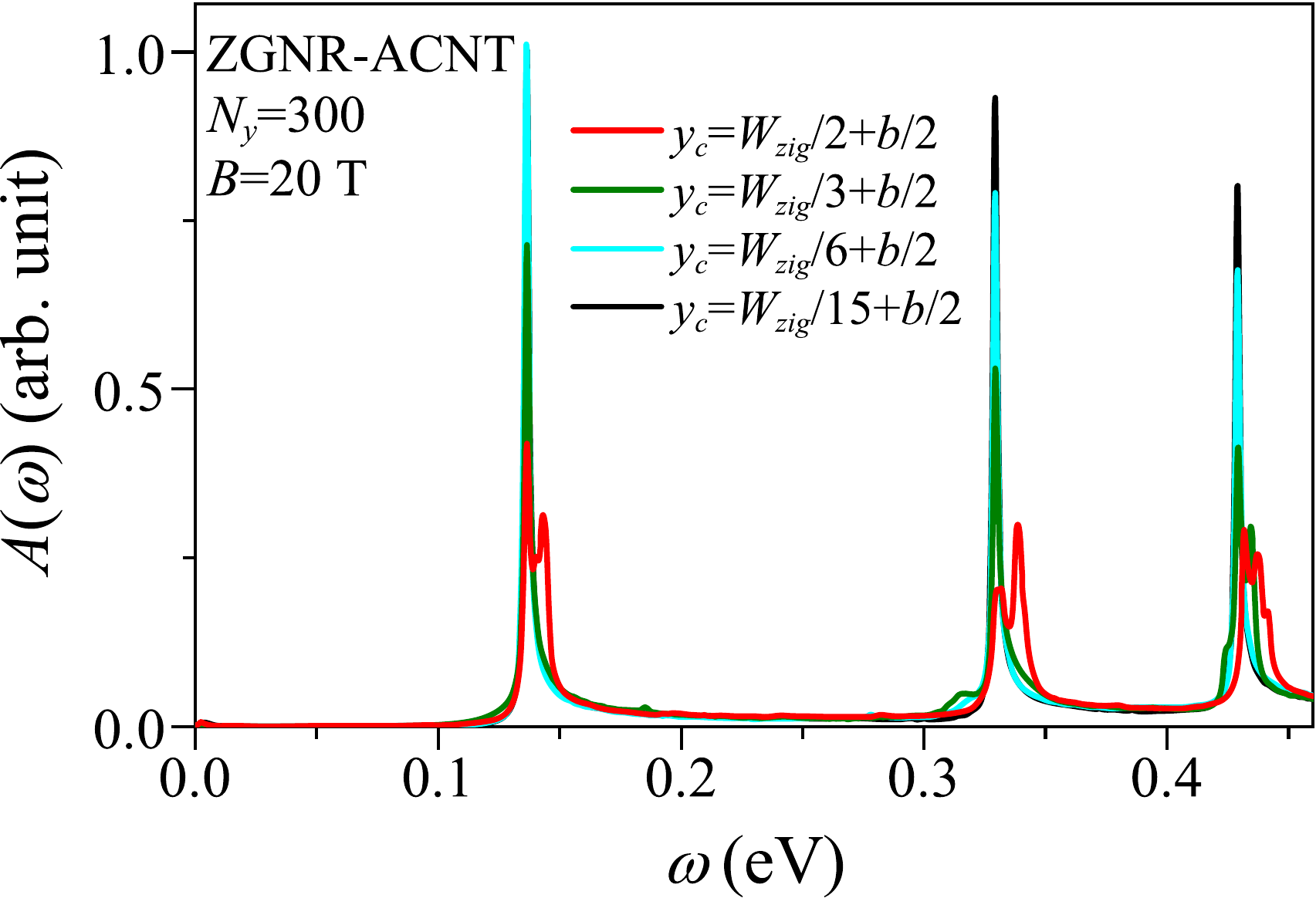}\\
  \caption[]{
  Magneto-absorption spectra of the ZGNR-ACNT hybrid at $B = 20$ T.
  The ($6$, $6$) ACNT is placed at $y_c =$ (a) $W_{zig}/2+b/2$, (b) $W_{zig}/3+b/2$, (c) $W_{zig}/6 + b/2$, and (d) $W_{zig}/15 + b/2$.
  }
  \label{fig:ABS_HybridSystem}
\end{center}
\end{figure}

The magneto-absorption spectra of GNR-CNT hybrids have substantial changes, owing to the inter-subsystem atomic interactions (Fig.~\ref{fig:ABS_HybridSystem}).
As the CNT is shifted from the ribbon edge to center, the monolayer QLL transition peaks are gradually lowered and become twin-peak structures (red curves).
These structures are closely related to the state mixings, since a specific QLL quantum mode exists in two hybridized subbands.
The hybridization effects are relatively easily observed in the wider GNRs and weaker magnetic fields; that is, there are more twin peaks under such conditions.

The predicted magneto-optical properties in non-uniform GNRs and GNR-CNT hybrids can be verified by the optical absorption and infrared transmission experiments,\cite{Phys.Rev.Lett.97(2006)266405M.L.Sadowski, Phys.Rev.Lett.98(2007)197403Z.Jiang} including the monolayer, bilayer, and coexistent QLL absorption spectra, magneto-optical selection rules, as well as single-, twin-, and irregular-peak structures.
They are quite sensitive to the geometric configurations.
Especially, the coexistent one is useful to understand the optical properties of other related structures, such as partly overlapped GNRs\cite{Appl.Phys.Lett.98(2011)192112K.M.M.Habib, Carbon75(2014)411M.Berahman, ACSNano5(2011)6610A.W.Robertson, Science336(2012)1143A.W.Tsen, Carbon80(2014)513R.Rao} and GNR-graphene superlattices.\cite{Synth.Met.161(2011)489C.H.Lee, Diam.Relat.Mat.20(2011)1026C.H.Lee, J.Phys.Chem.C117(2013)7326J.H.Wong}

\section{Conclusion and outlook}

The essential electronic and optical properties of GNR-related systems have been systematically reviewed.
By means of the generalized tight-binding model, the geometric configurations and external fields responsible for the wide diverse properties are explored in detail.
The former include the ribbon widths, edge structures, curvatures, stackings, as well as the hybrid structures; meanwhile, the latter cover the electric, magnetic, composite, and spatially modulated fields.
The tunable electronic structures give the flexibilities in the design of electronic devices.
Furthermore, the rich optical spectra with different selection rules can provide more opportunities for the applications in nanophotonics.
Above all, many presented results are compared with those from other theoretical methods and validated by the experimental measurements, while some predictions need further experimental verifications.
The developed theoretical framework shall advance the future studies on other quasi-1D materials and the related nanostructures.

Transverse lateral confinements lead to the 1D parabolic subbands and thus the asymmetric peaks in the DOS.
The energy gap is inversely proportional to the ribbon width.
The wave functions present standing waves with different wavelengths depending on the subband index.
The simple relations between the valence and conduction wave functions account for the edge-dependent optical selection rules, $|\Delta J| = odd$ for ZGNRs and $\Delta J = 0$ for AGNRs.
In further competition with the magnetic quantization, the low-lying dispersionless QLLs are formed, and hence the associated symmetric peaks are revealed in the DOS.
The edge-independent selection rule, $|\Delta n | = 1$, which is due to the spatial symmetry of Landau states, dominates the inter-Landau absorption peaks.
On the other hand, the transverse electric fields significantly alter the low-lying energy subbands and modulate the energy gap.
Many additional DOS peaks appear because of the complicated subband mixings.
The slightly distorted wave functions with additional relations lead to the suppression of original absorption peaks and appearance of extra peaks.
Furthermore, the uniform electric field is responsible for the tilt and collapse of  QLLs, while the modulated one results in the oscillatory Landau subbands.

Curvatures suppress the magnetic quantization and cause the electrons to experience a non-uniform effective magnetic field.
The energy and optical spectra are drastically changed by the competition between the curvature and magnetic field.
For a slightly curved GNR, the QLLs change into the oscillatory Landau subbands with several band edges, leading to the peak splitting in the DOS.
Each symmetric Landau absorption peak splits into multi-asymmetric ones with weaker intensity, where the selection rule of $|\Delta n | = 1$ remains unchanged.
As the GNR is severely curved, the subband mixings among the strongly oscillatory subbands create many band-edge states, which are reflected by the complex peak structures in the DOS and the $|\Delta n | \neq 1$ extra absorption peaks.
The crucial differences between CNTs and GNRs are revealed to result from the boundary conditions.
The zero-field absorption spectra of CNTs are dominated by the selection rule of $\Delta n  = 0$, whereas these rules are edge-dependent in GNRs.
The angular-momentum coupling induced by the perpendicular magnetic field contributes to the additional rules ($|\Delta n | = 1$ and $2$), of which more come to exist due to the increase of either (both) of the factors: the tube diameter and field strength.
Particularly, once the two factors exceed certain critical values, the optical spectra can reflect the Landau absorption peaks.
The reported features of the spectra provide insight to optical excitations for curved systems with either an open or a closed boundary condition.

Stacking configurations enrich the band structures and absorption spectra in the bilayer GNRs.
Apart from the transverse electric field, the energy gap of the AB-stacked bilayer GNRs can be modulated through the perpendicular electric field, while such gap modulation is unavailable in the metallic AA-stacked bilayer GNRs.
In the magneto-electronic structures, two groups of QLLs and the corresponding symmetric DOS peaks are distinguishable in either stacking configuration.
For the AB stacking, the first group initiates near the Fermi energy and the second group appears at energies $\sim \pm \gamma_1$; whereas, for the AA stacking, the first and second groups occur at $\sim \alpha_1$ and $\sim - \alpha_1$, respectively.
Transverse electric fields cause the tilt and collapse of QLLs, as well as turn the symmetric DOS peaks into the asymmetric ones.
Perpendicular electric fields lead to the crossings and anti-crossings of QLLs; the dramatic changes in the DOS.
The AB-stacked system exhibits four categories of Landau absorption peaks, including two intragroup and two intergroup ones.
Meanwhile, the intragroup transitions occurring near the Fermi level contribute to the prominent peaks in the low-energy spectrum.
On the other hand, only two categories of the intragroup transitions are permitted in the bilayer AA-stacked system.
More importantly, the Fermi-Dirac distribution is responsible for the forbidden absorption range of $\sim 0.6$ eV ($\sim 2\alpha_1$).
It is obvious that the magneto-absorption spectra of bilayer GNRs are quite distinct from those of monolayer ones, including the excitation categories, structures, frequencies, and numbers of absorption peaks.
Such features depend on the quantum modes, special relations in wave functions, and symmetric/asymmetric QLL spectra.
In general, the QLL transitions can survive when the Landau modes are identical on the same layer for the $A^i$ ($B^i$) sublattice of the initial state and the $B^i$ ($A^i$) sublattice of the final state.

The variations of non-uniform configurations and hybrid structures can dramatically tune the magneto-electronic and optical properties.
As the non-uniform configuration changes from monolayer to bilayer, there are four kinds of magneto-electronic spectra, including the monolayer-like, bilayer-like, and coexistent QLL spectra, and the distorted energy dispersions.
The various kinds of electronic structures also underline the features exhibited by the magneto-absorption spectra.
Especially, the inter-QLL transitions are mostly preserved in the disordered spectrum, and a lot of lower-intensity absorption peaks exist in the absence of selection rules.
The monolayer-like and bilayer-like absorption spectra can concurrently survive, when both the widths of monolayer and bilayer regions are sufficiently wide for the magnetic quantization.
The GNR-CNT interactions cause the partial destruction of QLLs, disruption of DOS peaks, and piecewise continuous wave functions; therefore, the hybridization lowers the prominent Landau absorption peaks with twin-peak structures.

To date, the experimental observations have confirmed the major characteristics of the electronic properties of monolayer GNRs, including the 1D parabolic subbands induced by the lateral confinement,\cite{ACSNano6(2012)6930P.Ruffieux} the exponentially decaying edge-localized states,\cite{Nat.Commun.5(2014)4311Y.Y.Li} and the width-dependent energy gap.\cite{Phys.Rev.Lett.98(2007)206805M.Y.Han, Science319(2008)1229X.L.Li, Nature514(2014)608G.Z.Magda}
The observed gradual disappearance of the Landau peaks in the local DOS near the ribbon edges is a direct consequence of the competition between the lateral confinement and magnetic quantization.\cite{Nat.Commun.4(2013)1744G.Li}
Further evidences of the competition are identified by the transport measurements, including the anomalous SdHO and the destruction of the Hall plateau of larger filling factor.\cite{Phys.Rev.Lett.107(2011)086601R.Ribeiro}
In the DOS, two groups of Landau peaks in the bilayer GNRs, the destruction of Landau peaks by the electric fields with different directions or spatial modulation, and the curvature-induced triple- or twin-peak structure could be verified by the STS experiments.
Also, the spectroscopic imaging STM serves as a powerful tool to visualize the zero-field standing waves and Landau wave functions.
Transport measurements have shown that the gap modulation in monolayer and bilayer GNRs can be achieved via the exertion of the electric fields.\cite{Sci.Rep.3(2013)1248W.J.Yu, IEEETrans.ElectronDevices61(2014)3329L.T.Tung}
However, owing to the edge roughness and the width controlling limitation, the detailed conditions about the metal-semiconductor and semiconductor-metal transitions require further investigations.
In terms of optical properties, the infrared transmission measurement on the monolayer GNR arrays has demonstrated the dependence of the threshold inter-Landau absorption frequency on the strength of magnetic field.\cite{Phys.Rev.Lett.110(2013)246803J.M.Poumirol}
On the other hand, there are many issues left for further verifications, including the zero-field edge-dependent selection rules in the monolayer GNRs, the splitting of Landau absorption peaks due to the curvature-induced effective magnetic field, the low-lying Landau and $J$-coupling absorption peaks in the CNTs, four (two) categories of transitions in the AB-stacked (AA-stacked) bilayer GNRs, and the coexistence of the monolayer and bilayer Landau absorption spectra in the non-uniform GNRs.

Recently, experimental and theoretical researches have unveiled other graphene-related honeycomb lattices of group-IV elements, such as silicene,\cite{Phys.Rev.Lett.102(2009)236804S.Cahangirov, Phys.Rev.B80(2009)155453H.Sahin, Appl.Phys.Lett.97(2010)223109B.Lalmi, Phys.Rev.Lett.108(2012)155501P.Vogt, Appl.Phys.Lett.102(2013)163106P.DePadova, Sci.Rep.3(2013)2399A.Resta} germanene,\cite{Phys.Rev.Lett.102(2009)236804S.Cahangirov, ACSNano5(2013)4414E.Bianco, NewJ.Phys.16(2014)095002M.E.Davila, J.Phys.-Condens.Matter26(2014)442001P.Bampoulis, Adv.Mater.26(2014)4820L.Li} and stanene.\cite{J.Phys.Chem.C115(2011)13242J.C.Garcia, Phys.Rev.B92(2015)081112Y.Xu, Nat.Mater.14(2015)1020F.F.Zhu}
Also found are the transition metal dichalcogenides (TMDs) consisting of hexagonal layers of transition-metal atoms (M) sandwiched between two layers of chalcogen atoms (X) with stoichiometry MX$_2$.\cite{Science331(2011)568J.N.Coleman, Nanotechnol.22(2011)125706M.M.Benameur}
TMDs have more than 40 different types,\cite{Adv.Phys.18(1969)193J.A.Wilson, Int.Rev.Phys.Chem.3(1983)177E.A.Marseglia} depending on the combination of chalcogen (S, Se; Te) and transition-metal atoms (Mo, W, Nb, Ta; Ti).
The layered 2D systems can also be formed via van der Waals interactions.
Furthermore, some theoretical studies focus on the strip-like nanostructures.\cite{J.Am.Chem.Soc.130(2008)16739Y.Li, Phys.Rev.B81(2010)195120S.Cahangirov, Phys.Rev.Lett.109(2012)055502M.Ezawa, NewJ.Phys.14(2012)033003M.Ezawa, Appl.Phys.Lett.102(2013)172103M.Ezawa, Phys.Rev.89(2014)195303S.Rachel, Phys.Chem.Chem.Phys.17(2015)6865J.Xiao}
The silicene,\cite{Appl.Phys.Lett.96(2010)183102B.Aufray, Appl.Phys.Lett.96(2010)261905P.DePadova} MoS$_2$,\cite{Nat.Commun.4(2013)1776X.Liu} and WS$_2$\cite{ACSNano7(2013)7311C.Nethravathi} nanoribbons have been successfully synthesized.
This work is useful in understanding the essential properties of GNR-related quasi-1D systems, such as the effects due to the lateral confinements, edge structures, curvatures, stackings, non-uniform and hybridized configurations; magnetic and electric fields.


\begin{acknowledgments}
We would like to thank all the contributors to this article for their valuable discussions and recommendations, especially Godfrey Gumbs, Jung-Chun Andrew Huang, Chung-Lin Wu, Shean-Jen Chen, Hsisheng Teng, Shih-Hui Chang, Dah-Chin Ling, Cheng-Hao Chuang, Bogdan Ostahie,  Alexandru Aldea, Luis E. F. Fo\`{a} Torres, To-Sing Li, Yu-Huang Chiu, Yen-Hung Ho, Yih-Jon Ou, Ming-Hsun Lee, Matisse Wei-Yuan Tu, Ping-Yuan Lo, and Yu-Ming Wang.
The authors thank Po-Hua Yang and Pei-Ju Chien for English discussions and corrections.
One of us (Hsien-Ching Chung) thanks Ming-Hui Chung and Su-Ming Chen for financial support.
This research received funding from the Headquarters of University Advancement at the National Cheng Kung University, which is sponsored by the Ministry of Education, Taiwan.
This work was supported in part by the National Science Council of Taiwan under grant number NSC 102-2112-M-006-007-MY3.





\end{acknowledgments}

\bibliography{Reference}
\bibliographystyle{apsrev4-1} 

\end{document}